\documentclass[useAMS,usenatbib]{mnras}

\usepackage[british]{babel}             
\usepackage{graphicx}

\usepackage{times}
\usepackage{amsmath}
\usepackage{amssymb}
\usepackage{color,graphicx}
\usepackage{multirow}
\usepackage{booktabs}
\usepackage{bm}
\usepackage[flushleft]{threeparttable}
\usepackage{longtable}     
\usepackage{rotating}
\usepackage{hyperref}

\newcommand{\lsim}{\raisebox{-.5ex}{$\,\stackrel{\textstyle <}{\sim}\,$}}
\newcommand{\gsim}{\raisebox{-.5ex}{$\,\stackrel{\textstyle >}{\sim}\,$}}

\def\apj{ApJ}
\def\apjs{ApJS}
\def\apjl{ApJL}
\def\aj{AJ}
\def\mnras{MNRAS}

\def\nat{nat}
\def\araa{ARA\&A}
\def\aap{A\&A}

\title[Newly identified N-rich giants]{Chemodynamics of newly identified giants with globular cluster like abundance patterns in the bulge, disk, and halo of the Milky Way}
\author[Jos\'e G. Fern\'andez-Trincado et al.]{Jos\'e G. Fern\'andez-Trincado$^{1,2}$\thanks{jose.fernandez@uda.cl, jfernandez@obs-besancon.fr, jfernandezt87@gmail.com}, 
	Timothy C. Beers$^{3}$,
	Baitian Tang$^{4}$,
	Edmundo Moreno$^{5}$,
	\newauthor
	Angeles P\'erez-Villegas$^{6}$ 
	and Mario Ortigoza-Urdaneta$^{1}$\\
$^{1}${Instituto de Astronom\'ia y Ciencias Planetarias, Universidad de Atacama, Copayapu 485, Copiap\'o, Chile}\\
$^{2}$Institut Utinam, CNRS UMR6213, Univ. Bourgogne Franche-Comt\'e, OSU THETA , Observatoire de Besan\c{c}on, \\BP 1615, 25010 Besan\c{c}on Cedex, France\\
$^{3}$Department of Physics and JINA Center for the Evolution of the Elements, University of Notre Dame, Notre Dame, IN 46556, USA\\
$^{4}$School of Physics and Astronomy, Sun Yat-sen University, Zhuhai 519082, China\\
$^{5}$Instituto de Astronom\'ia, Universidad Nacional Aut\'onoma de M\'exico, Apdo. Postal 70264, M\'exico D.F., 04510, M\'exico\\
$^{6}$Universidade de S\~ao Paulo, IAG, Rua do Mat\~ao 1226, Cidade Universit\'aria, S\~ao Paulo 05508-900, Brazil\\
}

\begin{document}

\maketitle

\label{firstpage}

\begin{abstract}
   The latest edition of the APOGEE-2/DR14 survey catalogue and the first \texttt{Payne} data release of APOGEE abundance determinations by Ting et al.  are examined. We identify 31 previously unremarked metal-poor giant stars with anomalously high levels of nitrogen in the chemical space defined by [Fe/H] and [N/Fe]. The APOGEE chemical abundance patterns of such objects revealed that these are chemically distinct from the Milky Way (MW) in most chemical elements. We have found all these objects have a [N/Fe]$\gsim+0.5$, and are thus identified here as nitrogen-rich stars. An orbital analysis of these objects revealed that a handful of them shares the orbital properties of the bar/bulge, and possibly linked to tidal debris of surviving globular clusters trapped into the bar component. 3 of the 31 stars are actually halo interlopers into the bulge area, which suggests that halo contamination is not insignificant when studying N-rich stars found in the inner Galaxy, whereas the rest of the N-rich stars share orbital properties with the halo population. Most of the newly identified population exhibit chemistry similar to the so-called \textit{second-generation} globular cluster stars (enriched in aluminum, [Al/Fe]$\gsim+0.5$), whereas a handful of them exhibit lower abundances of aluminum, [Al/Fe]$<+0.5$, which are thought to be chemically associated with the \textit{first-generation} of stars, as seen in globular clusters, or compatible with origin from a tidally disrupted dwarf galaxy. 
\end{abstract}

\begin{keywords}
stars: abundances, stars: chemically peculiar, Galaxy: abundances, Galaxy: bulge, globular clusters: general, Galaxy: halo
\end{keywords}

\section{INTRODUCTION}

 \label{section1}
 
 The advent of large spectroscopic surveys such as APOGEE \citep{Majewski2017} and its capability to measure the atmospheric composition of $\sim$176,000 giants distributed homogeneously over much of the Milky Way (bulge, disk, and halo), and especially designed to observe where extinction by dust is significantly higher, has opened up a new panoramic window on our Galaxy, providing the precise chemical abundance of more than 23 chemical species (e.g., C, N, O, Na, Mg, Al, Si, P, S, K, Ca, Ti, V, Cr, Mn, Co, Ni, Cu, Ge, Rb, Nd, Ce and Yb). 
 
 The detailed and unprecedented precision of most of the stellar elemental abundances provided in the \textit{H}-band ($\lambda$1.5--1.7$\mu$m) has enabled the discovery of giants with unusual abundance patterns throughout of the Galaxy. For example, \citet[][]{Martell2016, Fernandez-Trincado2016, Fernandez-Trincado2017L, Schiavon2017a, Reis2018, Koch2019} and \citet{Fernandez-Trincado2019} have identified many such stars through a simple technique known as "weak" chemical tagging \citep{Freeman2002, Ting2015, Ting2016, Hogg2016, Schiavon2017a}, i.e., chemically tagging stars born in the same type of stellar system characterised by an unique detailed abundance pattern. Most of the APOGEE spectra have been re-analyzed manually in the works above, with the conclusion that a significant fraction of giants fell within the bounds of the  chemically anomalous stars unique to Galactic and/or extragalactic globular cluster environments. Their number has become increased in recent years \citep[see][]{Recio-Blanco2017, Tang2019, Kemp2018}, thanks to other extensive spectroscopic surveys such as LAMOST, , which obtained low-resolution spectra \citep[][]{Zhao2012, Luo2012, Deng2012, Cui2012}, and the Gaia-ESO survey, with both medium- and high-resolution spectra \citep[][]{Gilmore2012, Randich2013}. So far, these results demonstrate that those inhomogeneities appear to occur in other star-formation environments. Such stars have received significant attention in recent years, primarily because they are considered as evaporated from stellar clusters, and as such, play an important role in deciphering the early history of the Galactic formation process \citep{Martell2010, Martell2011, Carollo2013, Fernandez-Trincado2015a, Fernandez-Trincado2015b, Fernandez-Trincado2016b, Helmi2018, Khoperskov2018, Minniti2018, Ibata2019}, as well as providing clues on the mechanism responsible for the ejection from stellar clusters and its relation with chemical peculiarity \citep[e.g.,][]{Pereira2017, Pereira2019}.
 
 Often, stars with "anomalous chemistry" have been qualitatively linked to the so--called \textit{second-generation}\footnote{Here, we refer to second-generation to the groups of stars showing enhanced N and Al, and depleted C and O abundances, with respect to other field stars at the same metallicity [Fe/H].} globular cluster stars, which clearly exhibit enhanced N, Na, and Al and depleted C, Mg and O abundances with respect to field stars at the same metallicity \citep[e.g.,][]{Bastian2018}, however, this terminology is ambiguous in multiple ways. The implicit assumption, that the unusual abundance patterns result after the chemically mundane stars and have been enriched by a previous generation, has not been fully demonstrated. Here, we refer to the stars with peculiar chemical composition as enriched, and the stars having field-like abundances as primordial. 
 
 Furthermore, it is now firmly established that carbon-depleted ([C/Fe]$\lsim+0.15$) giants beyond the metal-poor tail ([Fe/H]$\lsim-0.7$) of the thick disk with N enrichment ([N/Fe]$\gsim +0.5$, hereafter N-rich stars) are found throughout the Milky Way, and owe their unusual elemental abundances to rare astrophysical events or nucleosynthetic pathways in different environments \citep[][]{Meszaros2015, Masseron2019, Schiavon2017b, Fernandez-Trincado2018, Fernandez-Trincado2019}. On the other hand, beyond the intrinsic value of identifying the mechanism responsible for the unusual abundance patterns, which are still far from being understood, the exclusive chemistry of such stars are essentials to more broadly improve our understanding of the chain of physical processes experienced by the Milky Way from early. However, the census of chemically anomalous N-rich stars across the Milky Way is still far from complete, especially beyond the bulge regions. Recently, \citet{Martell2016, Fernandez-Trincado2016, Fernandez-Trincado2017L, Fernandez-Trincado2019} reported the discovery of 18 new N-rich stars in the bulge, disk and inner halo of the Milky Way. The present work, reports the discovery of other 31 new N-rich stars in the bulge, disk and inner halo.
 
 Here we revisit the APOGEE spectra to conduct the largest updated census of N-rich stars throughout the Milky Way. Furthermore, the current version of the APOGEE Stellar Parameters and Chemical Abundance Pipeline \citep[ASPCAP,][]{ASPCAP} does not measure the \textit{s}-process elements, but these are measurable from Ce II and Nd II lines \citep{Hasselquist2016, Cunha2017} in the observed spectral window, and carry important information about stellar nucleosynthesis along the RGB and AGB. Here we provide for the first time, measures of the neutron-capture element Ce II \citep[][]{Cunha2017} for some of these candidates. In particular, studies of \textit{s}-process elements provide strong evidence either for or against the uniqueness of the progenitor stars to stellar systems. In Section \ref{section2} we outline our methods to identify candidates enhanced in nitrogen ([N/Fe]$\gsim +0.5$) and depleted in carbon ([C/Fe]$\lsim +0.15$). In Section \ref{section3} we discuss our results, and in Section \ref{section4} we present our concluding remarks.
 
 \section{DATA AND METHODS}
 \label{section2}
 
 \subsection{Targets Analysed}
 \label{sectiontargets}
 
 The sample analysed in this work consist of red giants from the 14th data release of SDSS \citep[DR14,][]{Abolfathi2018, Jonsson2018, Holtzman2018} of the APOGEE-2 survey \citep{Majewski2017}, which has obtained high-resolution (\textit{R}$\sim$22,500) spectra of $\sim$ 270,000 stars in the \textit{H}-band ($\sim\lambda$1.5--1.7$\mu$m), using the 300-fiber cryogenic spectrograph installed on the 2.5m Telescope \citep{Gunn2006} at the Apache Point Observatory, as part of the Sloan Digital Sky Survey IV \citep{Blanton2017}. We refer the reader to \citet{Zasowski2013} and \citet{Zasowski2017} for full details regarding the targeting strategies for APOGEE and APOGEE-2, and \citet{Nidever2015}, \citet{Zamora2015}, \citet{Holtzman2015} and \citet{ASPCAP} for more details concerning data reduction of the APOGEE spectra, determination of radial velocities, atmospheric parameters, and stellar abundances, respectively.
 
 In this work, we have selected a sub-sample of metal-poor ([Fe/H]$\lsim -0.7$) red giants in the first \texttt{Payne} data release of APOGEE abundances \citep[][hereafter Payne-APOGEE]{Payne} that satisfy the following quality cuts to further ensure reliable parameter/abundance derivation:
 
 \begin{itemize}
 	\item[$\circ$] S/N $>$ 70
 	\item[$\circ$]  3000 K $<$T$_{\rm eff}$ $<$ 5500 K
 	\item[$\circ$] log \textit{g} $<$ 3.6
 	\item[$\circ$] \texttt{quality\_flags} $=$ good
 \end{itemize}	
 
 The \texttt{Payne} routine \citep[see][]{Payne} simultaneously derives best-fit values for all atmospheric parameters and abundances using neural networks, with the parameter space of the training set restricted to [Fe/H]$ \gsim -1.5$. For the same reason as in \citet{Fernandez-Trincado2016, Fernandez-Trincado2017L, Fernandez-Trincado2019} and \citet{Schiavon2017a}, we remove stars with [C/Fe] $> +0.15$, because such stars are not typically found in globular clusters, and we want to minimize the contamination by objects such as CH stars \citep[e.g.,][]{Karinkuzhi2015}, leaving us with a total of 6,289 giants with metallicities in the range $-1.5 < $ [Fe/H] $< -0.7$.
 
\begin{center}
 	\begin{figure}
 		\includegraphics[width=105mm]{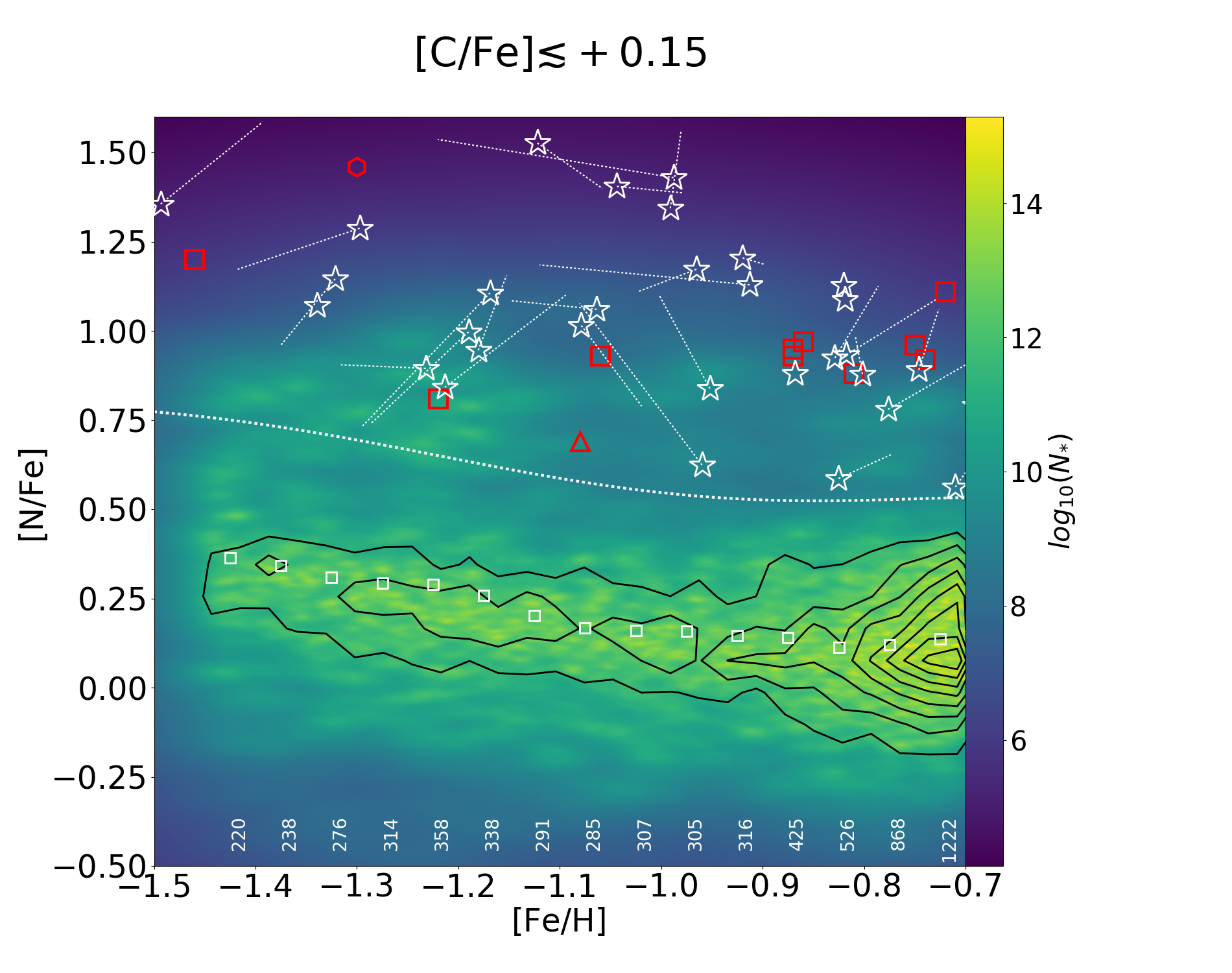}
 		\caption{Kernel Density Estimate (KDE) smoothed distribution of [N/Fe] and [Fe/H] for \texttt{Payne}-APOGEE stars, with the black contours showing the density of objects in the main body of N-normal stars ([N/Fe]$\lsim+0.5$). The number of stars of each bin are shown at the center of the bin at the lower portion of the plot, while the white unfilled square symbols show the mean value of [N/Fe] by bin. The white star symbols are of newly discovered N-rich stars in this study, manually re-analyzed adopting a simple line-by-line approach with the \texttt{BACCHUS} code, with white tiny dotted lines showing the sensitivity  to photometry stellar parameters (see text). The same field sample is compared to a sample of previously identified N-rich stars (red unfilled symbols) from the APOGEE survey and manually inspected line-by-line: \textit{hexagon}--\citet{Fernandez-Trincado2016}, \textit{squares}--\citet{Fernandez-Trincado2017L}, and \textit{triangles}--\citet{Fernandez-Trincado2019}. }
 		\label{Figure1}
 	\end{figure}
 \end{center}

 To search for outliers in the [N/Fe]-[Fe/H] abundance plane (see Figure \ref{Figure1}), we binned in [Fe/H] space (0.05 dex bins), and by fitting a 5$^{th}$ order polynomial to the bulk of the stars, select stars that deviate from the fit by more than $\gsim 2.5\sigma$ from that curve as nitrogen-rich red giants, i.e., we label all stars with nitrogen abundance more than $\sim$0.4 dex above the mean at a fixed metallicity as "N-rich".  The bins were chosen to ensure that at least 200 stars occupied each bin (see inner label numbers in Figure \ref{Figure1}). 
 
 In a similar way as described in \citet{Schiavon2017a}, we note that over a more limited metallicity range, a sixth-order polynomical also captures the mean behavior of our data set well, which includes stars located towards the disk, bulge and halo, simultaneously. This is the first time, that such global feature is evaluated across the Milky Way (disk+halo+bulge) to homogeneously identify such anomalous stars. The initial sample contained about 300 N-rich candidates, according to the [N/Fe] versus [Fe/H] abundance plane from \texttt{Payne}-APOGEE determinations. These stars have high [N/Fe] ratios ($\gsim +0.5$). 
 
There was expected to be some contamination by star clusters and other previously reported outliers, which we remove from our sample. We found that 177 out of 284 stars are confirmed cluster stars from \citet{Masseron2019} and \citep{Fernandez-Trincado2018}, while 31 out of 284 stars are know N-rich stars previously reported in \citet{Martell2016, Fernandez-Trincado2016, Schiavon2017a, Fernandez-Trincado2017L} and \citet{Fernandez-Trincado2019}. This yields 79 stars that are previously overlooked nitrogen-rich objects relative in the final data set. For reasons that are explained below, we remove 48 of those stars from our final analysis. 
 
 Figure \ref{Figure1} shows our final data set in the [N/Fe]-[Fe/H] plane against the giants that have unusually high N abundances. After applying a number of stringent selection criteria, this yielded the discovery of 31 further N-rich stars (bona fide chemically anomalous giants), adding to the $\sim 100$ such objects out of $\sim7,000$ regular disk (thick) and halo stars from previous studies. 
 
\begin{center}
	\begin{figure*}
		\includegraphics[width=190mm]{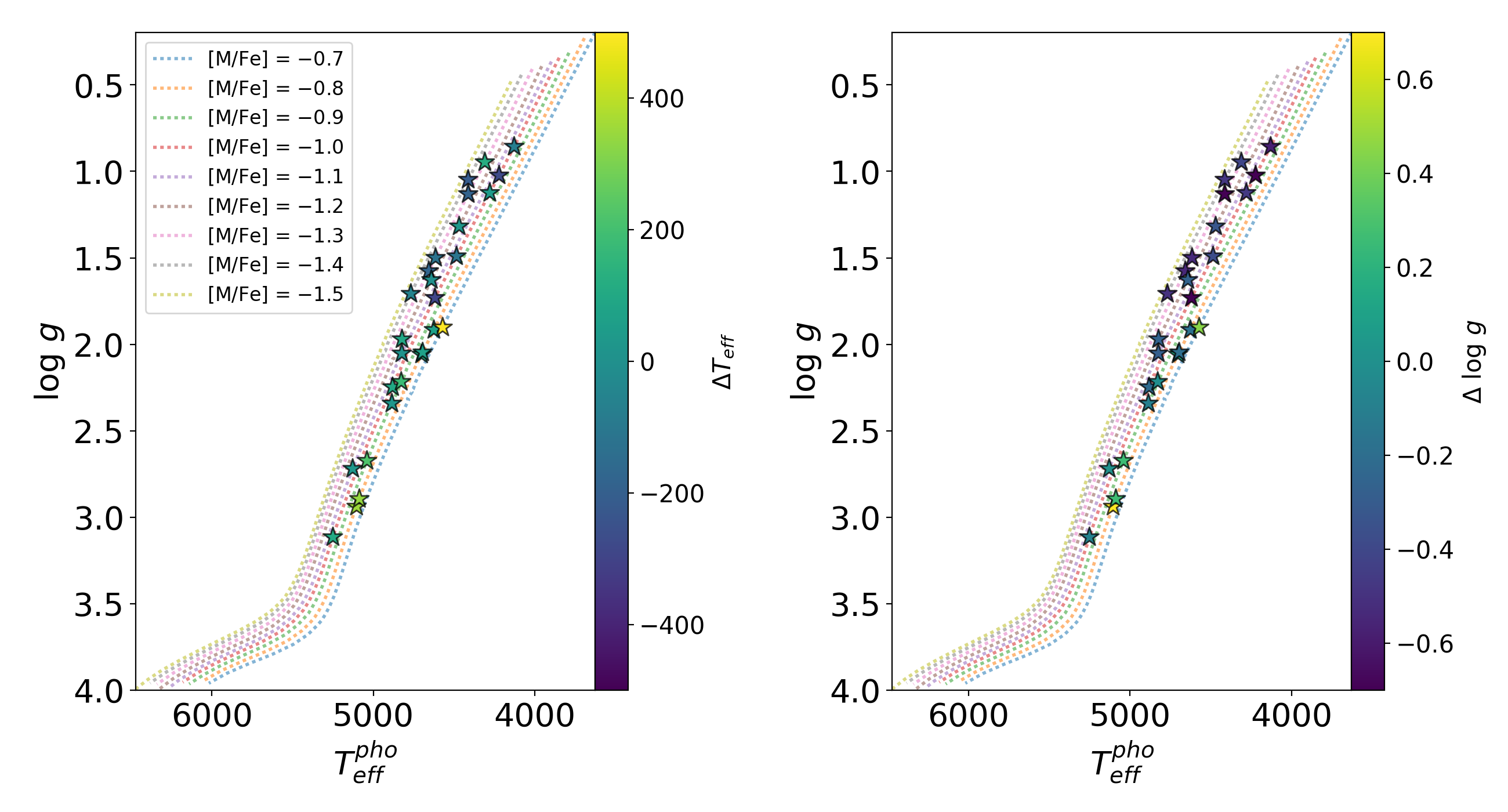}
		\caption{Location of the N-rich stars (star symbols) analyzed in this work in the  log $g$ -- T$_{\rm eff}$ plane. The symbols are color-coded by the differences between the photometric and \texttt{Payne} temperatures, $\Delta$ T$_{\rm eff} = $  T$^{pho}_{\rm eff} - $ T$^{\texttt{Payne}}_{\rm eff} $, and the log \textit{g} from 10 Gyr isochrones and Payne log \textit{g}, $\Delta$log \textit{g} $ = $  log \textit{g} - log \textit{g}$^{\texttt{Payne}}$.}
		\label{Figure2}
	\end{figure*}
\end{center}

\begin{table*}
		\begin{center}
			\setlength{\tabcolsep}{1.0mm}  
			\caption{Adopted atmospheric parameters of our target stars, radial velocity scatter information ($RV_{SCATTER}$), and frequency of observation per object (N$_{\rm visits}$).}
			\begin{tabular}{ccccccccccc}
				\hline
				\hline
				APOGEE$-$ID              &   \texttt{Payne} [Fe/H] & $J_{2MASS}-K_{s,2MASS}$  & $\langle A_{K}^{WISE} \rangle$ &  \textit{E(B-V)} & T$^{pho}_{\rm eff}$ & log$^{\rm iso}$ \textit{g} & \texttt{Payne} T$_{\rm eff}$  & \texttt{Payne} log \textit{g} & \textit{RV$_{SCATTER}$} & N$_{\rm visits}$\\  
				&   K & mag  & mag &  & K & dex & K & dex & km s$^{-1}$ & \\  
				\hline
				\hline
				2M01121802$+$6219193 & $-1.421$ & 1.044 & 0.336 & ...   & ...      & ...     & 4931.82   & 2.253     & 0.08      & 3       \\
				2M02000451$-$0229333 & $-1.289$ & 0.699 & 0.045 & 0.028 & 4614.73  & 1.499   & 4738.89   & 2.032     & 0.19      & 6       \\
				2M06273068$-$0440140 & $-0.797$ & 0.798 & 0.151 & 0.420 & 5103.98  & 2.938   & 4752.06   & 2.160      & 0.07      & 3       \\
				2M11062158$-$0712222 & $-0.893$ & 0.572 & 0.042 & 0.038 & 5039.55  & 2.673   & 4820.19   & 2.416     & 0.14      & 5       \\
				2M11514952$+$2015267 & $-1.225$ & 0.644 & 0.071 & 0.048 & 4822.40  & 1.970   & 4726.44   & 2.235     & 0.06      & 3       \\
				2M12010401$-$0058306 & $-1.009$ & 0.612 & 0.033 & 0.024 & 4880.99  & 2.247   & 4779.53   & 2.461     & 1.52      & 3       \\
				2M12042878$+$1949535 & $-1.384$ & 0.645 & 0.026 & 0.022 & 4767.64  & 1.707   & 4824.94   & 2.208     & 0.15      & 3       \\
				2M12092246$+$0545111 & $-1.289$ & 0.762 & 0.107 & 0.016 & 4412.57  & 1.131   & 4599.00    & 1.830      & 0.44      & 4       \\
				2M12444724$-$0207405 & $-0.931$ & 0.615 & 0.067 & 0.033 & 4885.23  & 2.342   & 4859.61   & 2.439     & 0.15      & 3       \\
				2M13481355$-$0040484 & $-1.014$ & 0.698 & 0.055 & 0.028 & 4618.69  & 1.731   & 4914.06   & 2.445     & 0.21      & 4       \\
				2M13503160$+$4411389 & $-1.082$ & 0.499 & 0.061 & 0.012 & 5250.21  & 3.114   & 5142.27   & 3.201     & 0.17      & 3       \\
				2M14082554$+$4711096 & $-1.118$ & 0.533 & 0.053 & 0.015 & 5128.05  & 2.719   & 5103.23   & 2.716     & 0.00       & 1       \\
				2M14582162$+$4117544 & $-1.012$ & 0.876 & 0.091 & 0.013 & 4128.17  & 0.857   & 4204.34   & 1.434     & 0.00       & 1       \\
				2M15082716$+$6710075 & $-1.302$ & 0.812 & 0.038 & 0.033 & 4309.77  & 0.947   & 4186.79   & 1.355     & 0.80      & 26      \\
				2M15183589$+$0027100 & $-1.217$ & 0.769 & 0.095 & 0.058 & 4468.69  & 1.317   & 4444.12   & 1.643     & 0.35      & 13      \\
				2M15193208$+$0025293 & $-0.760$ & 0.718 & 0.089 & 0.061 & 4625.32  & 1.916   & 4551.84   & 2.092     & 0.09      & 7       \\
				2M15195065$+$0221533 & $-0.777$ & 0.679 & 0.070 & 0.041 & 4701.46  & 2.058   & 4583.81   & 1.918     & 0.00       & 1       \\
				2M15535831$+$4333280 & $-1.256$ & 0.675 & 0.108 & 0.013 & 4656.55  & 1.576   & 4835.53   & 2.118     & 0.34      & 4       \\
				2M16362792$+$3901180 & $-0.842$ & 0.668 & 0.055 & 0.018 & 4694.92  & 2.046   & 4628.36   & 2.247     & 0.04      & 2       \\
				2M16464310$+$4731033 & $-1.013$ & 0.842 & 0.093 & 0.021 & 4221.58  & 1.023   & 4497.21   & 1.729     & 0.37      & 12      \\
				2M17294680$-$2644220 & $-0.912$ & 1.486 & 0.493 & ...   & ...      & ...     & 4486.78   & 1.743     & 0.25      & 2       \\
				2M18022601$-$3106232 & $-0.835$ & 0.977 & 0.326 & ...   & ...      & ...     & 4642.16   & 1.829     & 0.27      & 7       \\
				2M18110406$-$2602142 & $-0.713$ & 1.117 & 0.190 & ...   & ...      & ...     & 4418.24   & 1.752     & 0.00       & 1       \\
				2M18200243$+$0156016 & $-0.736$ & 1.168 & 0.322 & 0.710 & 4570.84  & 1.902   & 4047.75   & 1.449     & 0.16      & 3       \\
				2M18461977$-$3021506 & $-1.002$ & 0.933 & 0.156 & 0.197 & 4278.82  & 1.124   & 4234.47   & 1.584     & 0.00       & 1       \\
				2M18472793$-$3033242 & $-1.156$ & 0.783 & 0.101 & 0.173 & 4640.83  & 1.628   & 4647.83   & 1.891     & 0.00       & 1       \\
				2M18502108$-$2923442 & $-1.023$ & 0.832 & 0.152 & 0.162 & 4484.83  & 1.490   & 4598.75   & 1.866     & 0.00       & 1       \\
				2M19004420$+$4421082 & $-1.079$ & 0.658 & 0.088 & 0.069 & 4823.72  & 2.052   & 4813.71   & 2.332     & 0.16      & 3       \\
				2M20124750$+$1818007 & $-0.863$ & 0.763 & 0.157 & 0.237 & 4827.49  & 2.216   & 4645.88   & 2.231     & 0.22      & 6       \\
				2M21181769$+$0946422 & $-1.357$ & 0.783 & 0.116 & 0.049 & 4412.88  & 1.047   & 4624.36   & 1.525     & 0.00       & 1       \\
				2M22015914$+$1543129 & $-0.795$ & 0.566 & 0.045 & 0.049 & 5086.80  & 2.893   & 4744.41   & 2.631     & 0.26      & 4       \\
				\hline
				\hline
			\end{tabular}  \label{table1}
		\end{center}
\end{table*}


\subsection{Stellar Parameters and Chemical Abundance
	Measurements}

In order to examine the reliability of these high-[N/Fe] outliers, we performed a careful inspection of each APOGEE spectrum. We made use of the Brussels Automatic Stellar Parameter (\texttt{BACCHUS}) code \citep{BACCHUS} to derive the metallicity, broadening parameters, and chemical abundances for the newly identified N-rich sample, making a careful line selection as well as providing abundances based on a line-by-line differential approach. 

\texttt{BACCHUS} relies on the radiative transfer code Turbospectrum \citep{Alvarez1998, Plez2012} and the MARCS model atmosphere grid \citep{Gustafsson2008}. For each element and each line, the abundance determination proceeds as in \citet{Hawkins2016}, \citet{Fernandez-Trincado2017L, Fernandez-Trincado2018}, and \citet{Fernandez-Trincado2019}, and summarized here for guindance: (\textit{i}) a spectrum synthesis, using the full set of (atomic and molecular) lines, is used to find the local continuum level via a linear fit; (\textit{ii}) cosmic and telluric rejections are performed; (\textit{iii}) the local S/N is estimated; (\textit{iv}) a series of flux points contributing to a given absorption line are automatically selected; and (\textit{v}) abundances are then derived by comparing the observed spectrum with a set of convolved synthetic spectra characterised by different abundances. Four different abundance determinations are used: (\textit{i}) line-profile fitting; (\textit{ii}) core line-intensity comparison; (\textit{iii}) global goodness-of-fit estimate; and (\textit{iv}) equivalent width comparison. Each diagnostic yields validation flags. Based on these flags, a decision tree then rejects the line or accepts it, keeping the best-fit abundance. We adopted the $\chi^2$ diagnostic as it is the most robust. However, we store the information from the other diagnostics, including the standard deviation, between all four methods. 

The linelist used in this work is the latest internal DR14 atomic/molecular linelist (linelist.20170418). For a more detailed description of these lines, we refer the reader to a forthcoming paper (Holtzman et al., in preparation). The current version of ASPCAP/DR14 and the \texttt{Payne} routine do not determine the \textit{s}-process elements (Nd II and Ce II). Thus, we determine, for the first time, these elements in our target stars, adopting the linelists provided in \citet{Hasselquist2016} and \citet{Cunha2017}. 

\begin{center}
	\begin{figure*}
		\includegraphics[width=165mm]{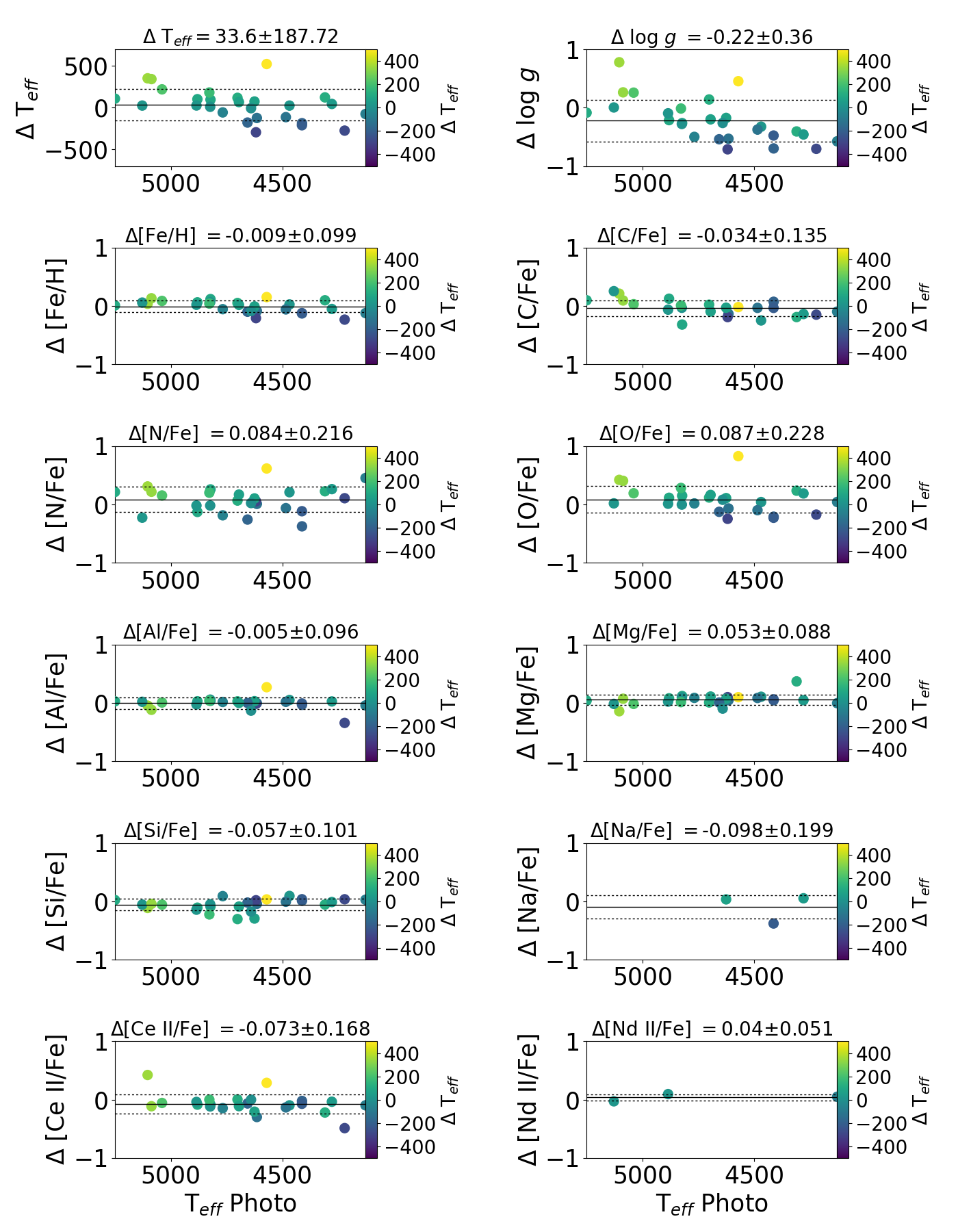}
		\caption{Differences in abundances produced by two runs adopting different temperatures: photometric versus \texttt{Payne} APOGEE temperatures; otherwise the same calculation method was employed. The symbols are color-coded by the differences between the photometric and \texttt{Payne} temperatures. The average and $\pm$ errors give the standard deviation around the mean of the differences, and are listed in the title of each panel.}
		\label{Figuredifferences}
	\end{figure*}
\end{center}

\begin{center}
	\begin{figure*}
		\includegraphics[width=165mm]{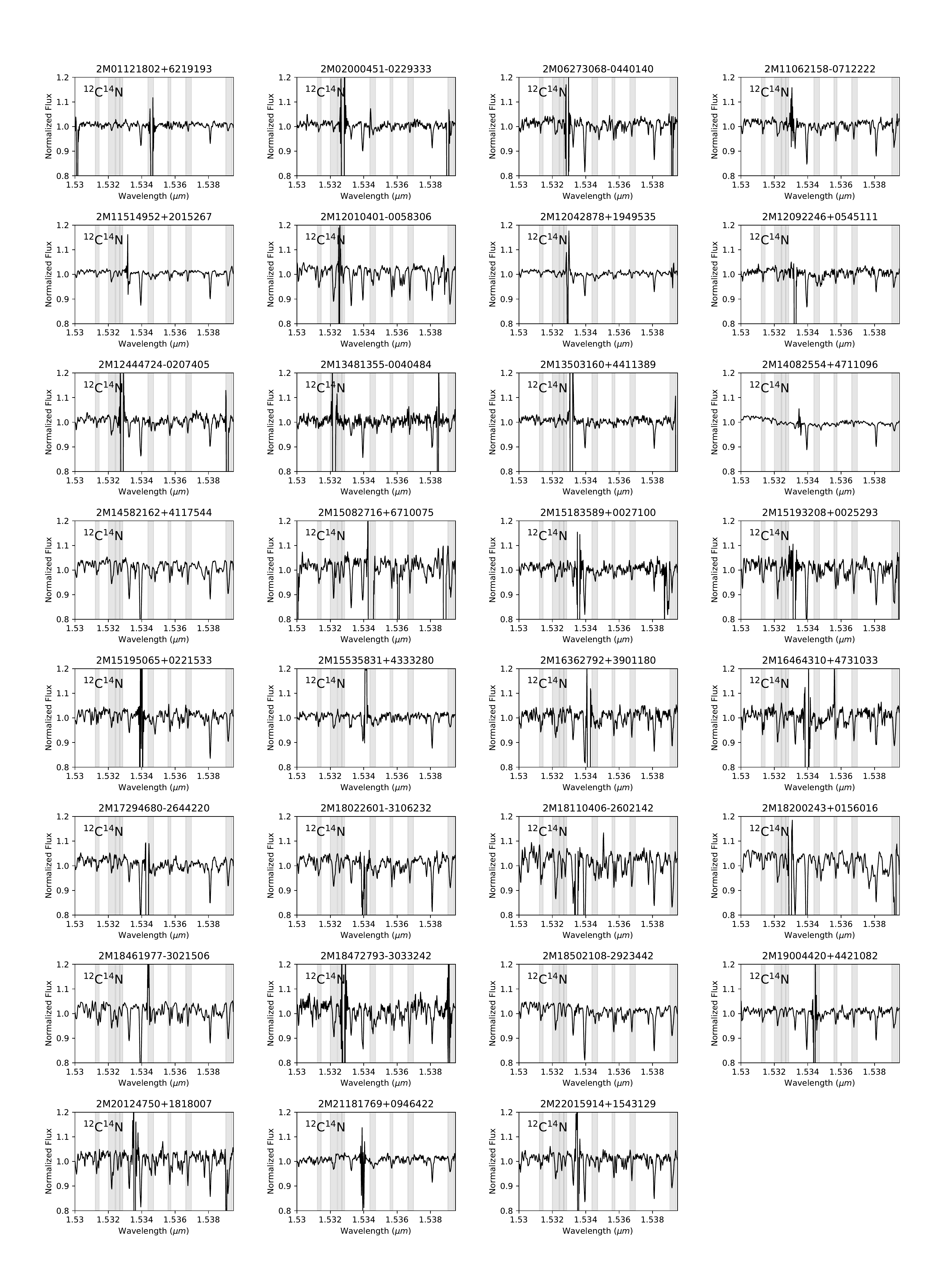}
		\caption{The \textit{H}-band spectra of our N-rich field stars, covering spectral regions around the $^{12}$C$^{14}$N band. The grey vertical bands indicate some of the wavelength regimes of the spectral features used in our re-analysis.}
		\label{spectra}
	\end{figure*}
\end{center}

For the light elements, a mix of heavily CN-cycled and $\alpha$-poor MARCS models were used, as well as the same molecular lines adopted by \citet{Smith2013}, were employed to determine the C, N, and O abundances.  In addition, we have adopted the C, N, and O abundances that satisfy the fitting of all molecular lines consistently; i.e., we first derive $^{16}$O abundances from $^{16}$OH lines, then derive $^{12}$C from $^{12}$C$^{16}$O lines and $^{14}$N from $^{12}$C$^{14}$N lines, and the CNO abundances are derived several times to minimize the OH, CO, and CN dependences \citep[see, e.g.,][]{Smith2013, Fernandez-Trincado2016,  Fernandez-Trincado2017L, Fernandez-Trincado2018, Fernandez-Trincado2019}.

\textit{Atmospheric parameters from spectroscopy}: In order to provide a consistent chemical analysis, we re-determine the chemical abundances assuming as input the effective temperature (T$_{\rm eff}$), surface gravity (log \textit{g}), and metallicity ([Fe/H]) as derived by the Payne-APOGEE runs \citep[see e.g.,][]{Payne}.

\textit{Atmospheric parameters from photometry}: We also applied a simple approach of fixing T$^{pho}_{\rm eff}$ and log \textit{g} to values determined independently of spectroscopy, in order to check for any significant deviation in the chemical abundances. For this, the photometric effective temperatures were calculated from the $J_{2MASS}-K_{s,2MASS}$ color relation using the methodology presented in \citet{Gonzalez2009}. Photometry is extinction corrected using the Rayleigh Jeans Color Excess (RJCE) method \citep[e.g.,][]{Majewski2011}. The results are listed in Table \ref{table1}. We estimate surface gravity from 10 Gyr \texttt{PARSEC} \citep{Bressan2012} isochrones, as illustrated in Figure \ref{Figure2}, since 10 Gyr is the typical age of Galactic GCs \citep{Harris1996}. The mean elemental abundances derived with the \texttt{BACCHUS} code are listed in Table \ref{table2}.

Figure \ref{Figure2} compares the difference between the adopted atmospheric parameters as the stars ascend the giant branch. Figure \ref{Figuredifferences} shows the difference of the derived chemical species from the \texttt{BACCHUS} pipeline for both adopted atmospheric parameters. The same figure indicates that small systematic differences are present between the spectroscopic and atmospheric parameters. However, for a few stars, the \texttt{Payne}-APOGEE raw temperature above $\gsim$ 5000 K and below $\lsim$ 3500 K in metal-poor stars showed significant,  200--400 K offsets, compared to photometry along with significant surface gravity (0.25--0.7 dex) offsets at the same temperature range, and are thus expected to display large scatter in the abundance analysis, as seen in rows 2, 3, 4 and 5 in Figure \ref{Figuredifferences}, which may be the reason for the large scatter in [N/Fe], [O/Fe], [Na/Fe], [Ce II/Fe], and [Nd II/Fe]. It is important to note that \texttt{BACCHUS} recovers the [Fe I/H] abundance ratios of these stars within 0.09 dex for both photometric and spectroscopic temperatures. The adoption of a purely photometry temperature scale enables us to be somewhat idependent of the \texttt{Payne}-APOGEE pipeline and  the APOGEE/ASPCAP pipeline, which gives important comparison data for future pipeline validation. The final results presented in this paper are based on the spectroscopic atmospheric parameters provided by \texttt{Payne}-APOGEE, and are use to estimate our final errors (listed in Table \ref{error1}).

\begin{table*}
	\begin{tiny}
		\begin{center}
			\setlength{\tabcolsep}{3.0mm}  
			\caption{Mean elemental abundances derived for our target stars using the "abund" module in \texttt{BACCHUS} code, adopting the atmospheric parameters from \texttt{Payne}-APOGEE (abundances labeled as [Fe/H]sp and [X/Fe]sp), and atmospheric parameters from photometry and isochrones (abundances labeled as [Fe/H]pho and [X/Fe]pho).}
			\begin{tabular}{cccccccccccc}
				\hline 
				\hline 
				APOGEE$-$ID            & [Fe/H]sp   & [C/Fe]sp   & [N/Fe]sp  &    [O/Fe]sp  & [Mg/Fe]sp &  [Al/Fe]sp   &  [Si/Fe]sp  &        [Ce/Fe]sp  &  [Nd/Fe]sp     &   [Na/Fe]sp    & Classification   \\
				\hline
				\hline
				2M01121802$+$6219193 &  $-$1.34   &   $-$0.03  &   1.07    &      0.64    &    0.12   &     ...      &   ...       &        $<$0.42       & ...        &       ...        & \textit{FG-like/dGs}  \\
				2M02000451$-$0229333 &  $-$1.23   &   $-$0.03  &   0.89    &      0.55    &    0.18   &    $-$0.02   &   0.32      &        0.64       &    ...         &       ...    & \textit{FG-like/dGs}     \\
				2M06273068$-$0440140 &  $-$0.69   &   $-$0.29  &   0.79    &      0.35    &    0.05   &     0.19     &   0.26      &        0.24       &    0.83        &    ...    &  \textit{FG-like/dGs}   \\
				2M11062158$-$0712222 &  $-$0.82   &   $-$0.22  &   0.93    &      0.33    &    0.03   &     0.09     &   0.21      &        0.23       &    ...         &       ...     & \textit{FG-like/dGs}    \\
				2M11514952$+$2015267 &  $-$1.21   &   $-$0.08  &   0.84    &      0.40    &    0.18   &     0.07     &   0.38      &        0.05       &    ...         &       ...    &  \textit{FG-like/dGs}    \\
				2M12010401$-$0058306 &  $-$1.12   &   $-$0.00  &   1.53    &      0.48    &    0.21   &     0.09     &   0.44      &        1.00       &    ...         &       ...      & \textit{FG-like/dGs}    \\
				2M12042878$+$1949535 &  $-$1.32   &$<$$-$0.19  &   1.15    &      0.59    &    0.27   &     0.31     &   0.24      &        0.37       &    1.11        &       ...      & \textit{FG-like/dGs}    \\
				2M12092246$+$0545111 &  $-$1.17   &   $-$0.51  &   1.11    &      0.29    &    0.02   &     0.74     &   0.22      &     $<$0.22       &    ...         &       ...      & \textit{SG-like}    \\
				2M12444724$-$0207405 &  $-$0.92   &   $-$0.27  &   1.20    &      0.42    &    0.09   &     0.39     &   0.36      &        0.32       & $<$0.76        &       ...    & \textit{FG-like/dGs}       \\
				2M13481355$-$0040484 &  $-$0.91   &$<$$-$0.28  &   1.13    &      0.61    &    0.12   &     0.19     &   0.25      &        ...        &    ...         &       ...     & \textit{FG-like/dGs}     \\
				2M13503160$+$4411389 &  $-$0.99   &   $-$0.33  &   1.34    &      ...     &    0.01   &     0.90     &   0.23      &        ...        &    ...         &       ...  & \textit{SG-like}        \\
				2M14082554$+$4711096 &  $-$1.08   &$<$$-$0.26  &   1.01    &      0.51    &    0.18   &    $-$0.01   &   0.32      &    $<$$-$0.02     & $<$1.03   &    ...  &   \textit{FG-like/dGs}      \\
				2M14582162$+$4117544 &  $-$0.96   &   $-$0.42  &   0.62    &      0.24    &    0.08   &     0.16     &   0.17      &        0.24       & $<$0.28        &       ...     & \textit{FG-like/dGs}     \\
				2M15082716$+$6710075 &  $-$1.49   &    0.02    &   1.36    &      0.58    &   $-$0.33 & ...   &   0.16      &        0.48       &    ...         &       ...     & \textit{FG-like/dGs}     \\
				2M15183589$+$0027100 &  $-$1.18   &   $-$0.33  &   0.95    &      0.25    &    0.11   &     0.16     &   0.16      &        0.17       &    ...         &       ...      & \textit{FG-like/dGs}    \\
				2M15193208$+$0025293 &  $-$0.80   &    0.02    &   0.88    &      0.38    &    0.26   &     0.24     &   0.59      &        0.25       &    ...         &    $<$0.30     & \textit{FG-like/dGs}    \\
				2M15195065$+$0221533 &  $-$0.83   &   $-$0.03  &   0.59    &      0.50    &    0.18   &     0.18     &   0.67      &        0.16       &    ...         &       ...       & \textit{FG-like/dGs}   \\
				2M15535831$+$4333280 &  $-$1.19   &     ...    &   0.99    &      0.25    &   $-$0.43 &     1.07     &   0.54      &        0.38       & $<$0.84        &       ...      & \textit{SG-like}    \\
				2M16362792$+$3901180 &  $-$0.75   &   $-$0.22  &   0.89    &      0.27    &   $-$0.02 &     0.07     &   0.27      &        0.25       &    ...         &       ...    & \textit{FG-like/dGs}      \\
				2M16464310$+$4731033 &  $-$0.99   &   $-$0.51  &   1.43    &      0.25    &    0.11   &     0.84     &   0.33      &        1.13       & $<$0.96        &  ...   &  \textit{SG-like}     \\
				2M17294680$-$2644220 &  $-$0.87   &   $-$0.38  &   0.88    &      0.42    &    0.22   &     0.24     &   0.32      &        0.22       &    ...         &       ...       & \textit{FG-like/dGs}   \\
				2M18022601$-$3106232 &  $-$0.82   &   $-$0.41  &   1.13    &      0.24    &    0.34   &     0.74     &   0.67      &        0.21       &    0.75        &       ...     & \textit{SG-like}     \\
				2M18110406$-$2602142 &  $-$0.82   &   $-$0.23  &   1.09    &      0.22    &    0.18   &     0.46     &   0.58      &        0.40       &    ...         &    ...       & \textit{FG-like/dGs}  \\
				2M18200243$+$0156016 &  $-$0.71   &   $-$0.07  &   0.56    &      0.34    &    0.09   &    $-$0.07   &   0.22      &       $-$0.19     &    ...         &       ...   & \textit{FG-like/dGs}       \\
				2M18461977$-$3021506 &  $-$0.95   &   $-$0.27  &   0.84    &      0.26    &    0.04   &     0.13     &   0.22      &        0.35       &    ...         &    $<$0.04   & \textit{FG-like/dGs}      \\
				2M18472793$-$3033242 &  $-$1.06   &    ...     &   1.06    &      0.38    &    0.28   &     0.58     &   0.46      &        0.10       &    ...         &       ...    & \textit{SG-like}      \\
				2M18502108$-$2923442 &  $-$0.97   &   $-$0.27  &   1.17    &      0.46    &    0.30   &     0.66     &   0.53      &        0.37       &    ...         &       ...     & \textit{SG-like}     \\
				2M19004420$+$4421082 &  $-$1.04   &   $-$0.46  &   1.40    &      0.38    &    0.13   &     0.26     &   0.33      &        0.67       &    0.70        &       ...  & \textit{FG-like/dGs}        \\
				2M20124750$+$1818007 &  $-$0.83   &   $-$0.07  &   0.92    &      0.25    &    0.23   &     0.30     &   0.43      &        0.17       &    ...         &       ...       & \textit{FG-like/dGs}   \\
				2M21181769$+$0946422 &  $-$1.30   &   $-$0.52  &   1.29    &      0.57    &    0.14   &     0.26     &   0.32      &        0.25       &    ...         &    $<$0.85    & \textit{FG-like/dGs}     \\
				2M22015914$+$1543129 &  $-$0.78   &   $-$0.03  &   0.78    &      0.22    &    0.20   &     0.30     &   0.32      &     $<$0.21       &    ...         &       ...     & \textit{FG-like/dGs}     \\									
				\hline
				\hline
				&   [Fe/H]pho    &  [C/Fe]pho   & [N/Fe]pho  &   [O/Fe]pho  &   [Mg/Fe]pho  &    [Al/Fe]pho &  [Si/Fe]pho   &   [Ce/Fe]pho &  [Nd/Fe]pho    &  [Na/Fe]pho\\
				\hline
				\hline
				2M01121802$+$6219193 &    ...         &     ...      &    ...     &   ...        &     ...       &     ...       &    ...        &      ...     &     ...        &     ...     \\
				2M02000451$-$0229333 &   $-$1.32      &    $-$0.17   &    0.90    &   0.48       &     0.22      &    $-$0.02    &    0.28       &      0.35    &  ...       &     ...    &  \\
				2M06273068$-$0440140 &   $-$0.65      &    $-$0.08   &    1.10    &   0.77       &    $-$0.09    &     0.15      &    0.14       &      0.67    &     ...        &     ...   &   \\
				2M11062158$-$0712222 &   $-$0.73      &    $-$0.19   &    1.09    &   0.53       &     0.01      &     0.10      &    0.16       &      0.17    &  ...       &     ...   &   \\
				2M11514952$+$2015267 &   $-$1.09      &    $-$0.40   &    1.10    &   0.56       &     0.30      &     0.11      &    0.29       &     $-$0.03  &     ...        &     ...  &    \\
				2M12010401$-$0058306 &   $-$1.06      &     0.12     &    1.40    &   0.59       &     0.29      &     0.12      &    0.33       &      0.92    &     ...        &     ...   &   \\
				2M12042878$+$1949535 &   $-$1.37      &     ...      &    0.96    &   0.61       &     0.35      &     0.33      &    0.33       &   $<$0.22    &     ...        &     ...   &   \\
				2M12092246$+$0545111 &   $-$1.29      &    $-$0.44   &    0.73    &   0.08       &     0.07      &     0.70      &    0.23       &      0.20    &     ...        &     ...   &   \\
				2M12444724$-$0207405 &   $-$0.90      &    $-$0.33   &    1.19    &   0.43       &     0.11      &     0.36      &    0.21       &      0.28    &     $<$0.86         &     ...   &   \\
				2M13481355$-$0040484 &   $-$1.12      & $<$$-$0.47   &    1.18    &   0.36       &  0.22      &     0.17      &    0.27       &      0.32    & ...      &     ...   &   \\
				2M13503160$+$4411389 &   $-$0.98      &    $-$0.23   &    1.56    &$<$1.06       &     0.05      &     0.92      &    0.26       &      ...     &     ...        &     ...   &   \\
				2M14082554$+$4711096 &   $-$1.02      &    $-$0.01   &    0.79    &   0.53       &     0.16      &     0.01      &    0.26       &      ...     &  $<$1.01       &     ...   &   \\
				2M14582162$+$4117544 &   $-$1.08      &    $-$0.51   &    1.08    &   0.28       &     0.08      &     0.11      &    0.19       &      0.14    &  $<$0.33       &     ...  &    \\
				2M15082716$+$6710075 &   $-$1.39      &    $-$0.16   &    1.58    &   0.82       &  0.04     &     ...       &    0.10       &   $<$0.26    &    ...     &     ...    &  \\
				2M15183589$+$0027100 &   $-$1.15      &    $-$0.57   &    1.16    &   0.29       &     0.21      &     0.20      &    0.25       &      0.07    &     ...        &     ... &     \\
				2M15193208$+$0025293 &   $-$0.81      &    $-$0.01   &    0.98    &   0.49       &     0.33      &     0.27      &    0.29       &      0.05    &     ...        &  $<$0.33  &   \\
				2M15195065$+$0221533 &   $-$0.77      &    $-$0.00   &    0.65    &   0.62       &     0.18      &     0.20      &    0.36       &      0.16    &     ...        &     ...    &  \\
				2M15535831$+$4333280 &   $-$1.29      &     ...      &    0.74    &   0.13       &    $-$0.42    &     1.07      &    0.51       &      0.32    &     ...        &     ...   &   \\
				2M16362792$+$3901180 &   $-$0.73      &    $-$0.31   &    1.06    &   0.43       &     0.09      &     0.07      &    0.18       &      0.14    &     ...        &     ...  &    \\
				2M16464310$+$4731033 &   $-$1.22      &    $-$0.66   &    1.54    &   0.07       &     ...       &     0.49      &    0.36       &   $<$0.64    &     ...        &     ...    &  \\
				2M17294680$-$2644220 &   ...          &     ...      &    ...     &   ...        &     ...       &     ...       &    ...        &      ...     &     ...        &     ...   &   \\
				2M18022601$-$3106232 &   ...          &     ...      &    ...     &   ...        &     ...       &     ...       &    ...        &      ...     &     ...        &     ...    &  \\
				2M18110406$-$2602142 &   ...          &     ...      &    ...     &   ...        &     ...       &     ...       &    ...        &      ...     &     ...        &     ...   &   \\
				2M18200243$+$0156016 &   $-$0.56      &    $-$0.08   &    1.18    &   1.17       &     0.19      &     0.20      &    0.25       &      0.10    & ...      &     ...    &  \\
				2M18461977$-$3021506 &   $-$1.00      &    $-$0.41   &    1.10    &   0.45       &     0.08      &     0.15      &    0.21       &      0.31    & ...       &  $<$0.09  &   \\
				2M18472793$-$3033242 &   $-$1.15      &     ...      &    1.08    &   0.47       &     0.18      &     0.45      &    0.28       &      0.10    &     ...        &     ...   &   \\
				2M18502108$-$2923442 &   $-$1.02      &    $-$0.30   &    1.11    &   0.37       &     0.38      &     0.68      &    0.52       &      0.24    &     ...        &     ...   &   \\
				2M19004420$+$4421082 &   $-$0.98      &    $-$0.49   &    1.39    &   0.38       &     0.22      &     0.31      &    0.28       &      0.56    &     ...        &     ...    &  \\
				2M20124750$+$1818007 &   $-$0.79      &    $-$0.06   &    1.13    &   0.54       &     0.25      &     0.34      &    0.20       &      0.17    &     ...        &     ...   &   \\
				2M21181769$+$0946422 &   $-$1.42      &    $-$0.55   &    1.17    &   0.34       &     0.20      &     0.25      &    0.35       &      0.18    &  ...       &  $<$0.47  &   \\
				2M22015914$+$1543129 &   $-$0.64      &     0.07     &    0.99    &   0.63       &     0.27      &  $<$0.19      &    0.28       &   $<$0.10    &     ...        &     ...   &   \\
				\hline
				\hline
			\end{tabular}  \label{table2}\\
		\end{center}
	\end{tiny}
				\raggedright {The Solar reference abundances are from \citet{Asplund2005} and \citet{Grevesse2015} for Ce II and Nd II. The \texttt{BACCHUS} pipeline was used to derive the broadening parameters, metallicity, and chemical abundances.}
\end{table*}

\textit{Incidence of false detections}: A cautionary note is in order before proceeding with the analysis. Once the atmospheric parameters are determined, for each star the selected $^{12}$C$^{14}$N lines were visually inspected to ensure that the spectral fit was adequate. If the lines were not well-reproduced by the synthesis it were rejected. In addition, for a substantial fraction of our target stars, the $^{12}$C$^{14}$N lines were also rejected if they were flagged as problematic by the \texttt{BACCHUS} pipeline, i.e., they were stronly blended or too weak in the spectra of stars with the typical T$_{\rm eff}$ and metallicity of the N-rich sample to deliver reliable [N/Fe] abundances. This indicates, as confirmed by visual inspection of 45 out of 79 stars in our sample, that the spectral fits are of poor quality, or that the stellar parameters are unreliable or both, suggesting that the abundances for these stars cannot be relied on. In this way, 45 stars were identified as false detections, leaving us with a grand total of 31 newly identified reliable N-rich field stars (white unfilled 'star' symbols), as illustrated in Figure \ref{Figure1}. Figure \ref{spectra} show examples for a portion of the observed APOGEE spectra in the region around the strong $^{12}$C$^{14}$N bands for our 31 stars.

\section{Results and discussion}
\label{section3}

\subsection{Chemical Signatures}

In this subsection, we discuss the individual abundance ratios and trends, as function of metallicity, for the 31 newly identified N-rich stars. 

\begin{center}
	\begin{figure*}
		\includegraphics[width=190mm]{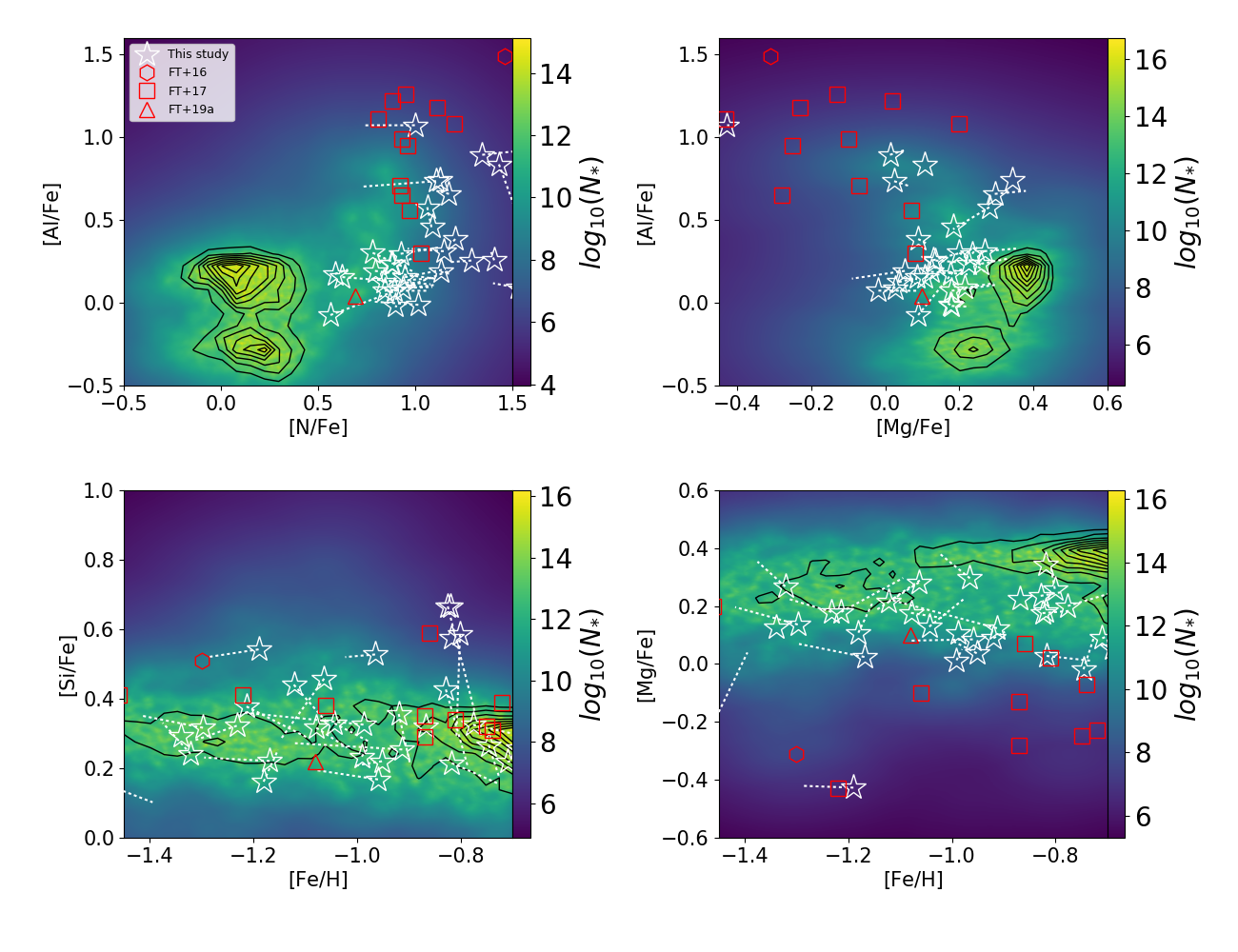}
		\caption{Same as Figure \ref{Figure1}, but for the abundance distribution of the light elements Al, N, Mg, Si, and Fe. The elements for which the newly identified N-rich stars stand out most distinctly from the Milky Way stars are N and Al, where the new stars appear well-separated from the main body of N-normal stars, and less obvious for Si and Mg, with a few exceptions.}
		\label{Figure4}
	\end{figure*}
\end{center}

Figure \ref{Figure4} shows the distribution of [Al/Fe], [N/Fe], [Mg/Fe], [Si/Fe], and [Fe/H] for all stars that made it through the quality criteria discussed in Section \ref{sectiontargets}. These elements were chosen because "migrants" from globular clusters exhibit clear deficiencies and enrichments in these elemental abundances as compared to most stars in Milky Way \citep{Martell2016, Schiavon2017a, Fernandez-Trincado2016, Fernandez-Trincado2017L}. Most obvious in this plot are the high-[N/Fe] outliers with [N/Fe]$\gsim+0.5$, nominally corresponding to a chemically anomalous population, commonly associated with stars possibly removed from a system merged with our Milky Way (unless they are part of a binary system). We did not detect any variation of the radial velocities that would support the hypothesis that these objects formed through the binary channel, however, most of the stars in our sample were observed just once, and those with multiple observations have a short baseline ($<$ 6 months), which makes possible detection of only a small fraction of possible binaries \citep[e.g.,][]{Fernandez-Trincado2019}. The radial velocity dispersion ($RV{scatter}$) of our sample is listed in Table \ref{table1}, which is typically less than 1.5 km s$^{-1}$ over multiple visits, indicating that most of the newly identified N-rich stars unlikely to be variable stars or part of a binary system.

A small fraction ($\sim$29\%) of our giants exhibit large-enrichment in aluminum, [Al/Fe]$\gsim+0.5$, which make them more likely to be \textit{migrants} from globular clusters, as they are not typical giants as seen in dwarf galaxy stars \citep[e.g.,][]{Hasselquist2017, Hasselquist2019}. Such N-/Al-rich stars occupy the same locus as \textit{SG}-like N-/Al-rich giants  found in previous studies \citep{Fernandez-Trincado2016, Fernandez-Trincado2017L}, and the second-generation globular cluster stars \citep{Meszaros2015, Masseron2019}, reinforcing the similarity between such objects and, the globular cluster population. The high levels of Al rule out the possibility that satellite galaxies could have contributed stars to our N-/Al-rich population. These objects exhibit chemical similarities to that the second-generation globular cluster stars, implying that it is possible that a high aluminum and nitrogen abundance could be related to escaped globular cluster stars, or due to metal-poor AGB stars that have experienced strong internal mixing; these chemical properties enable us to classify such objects as \textit{SG-like} (see Table \ref{table2}). This unique collection of N-rich stars significantly contributes to the task of compiling a more thorough census of anomalously high levels of [N/Fe] and [Al/Fe] throughout the Milky Way, and portends the promising results to be expected from future spectroscopic follow-up observations, and to reconsider the question of in situ halo formation \citep{Martell2016}. For other light-element chemical planes, the distinction between the newly identified N-/Al-rich stars and Milky Way population appears to be weaker in Mg and weakest in Si. Here, we show that the [Al/Fe]-[N/Fe] and [Al/Fe]-[Mg/Fe] chemical planes are an especially powerful and reliable diagnostic to identify this unique class of stars among the N-rich population (primarily with metallicities [Fe/H]$\lsim -0.7$). 

At lower [Al/Fe]$\lsim+0.5$ abundances, there is more overlap in [X/Fe] between the chemistry of Milky Way, dwarf galaxy stars \citep{Hasselquist2017, Hasselquist2019}, and the so--called First-Generation of globular cluster stars \citep{Meszaros2015}, i.e., 24 out of 31 N-rich stars in our final data set exhibit chemical abundances that are somewhat distinct in the [Fe/H]-[N/Fe], [Mg/Fe]-[Al/Fe] and [N/Fe]-[Al/Fe] planes, as shown in Figure \ref{Figure1} and the top panels in Figure \ref{Figure4}, but indistinguishable in other chemical planes from stars having chemistry consistent with the halo and thick disk (\textit{left-bottom panel} in Figure \ref{Figure4}), and are distinctly less Mg-enriched (\textit{right-bottom panel} in Figure \ref{Figure4}). Unfortunately, with only few light-/heavy-elements measured, it is not possible to assign the nucleosynthetic origins of these N-rich/Al-normal stars, and disentangle a dwarf galaxy and globular cluster (stars having peculiar chemical composition like the first population) origins; such objects are classified as \textit{FG-like/dGs} in Table \ref{table2}. It is likely that they were contributed by different merger events from a dwarf galaxy such as Gaia-Enceladus \citep{Belokurov2018, Helmi2018} and the Sagittarius dSph \citep{Hasselquist2019}, or massive disrupted globular clusters \citep[e.g.,][]{Kruijssen2015}.

\begin{center}
	\begin{figure*}
		\includegraphics[width=180mm]{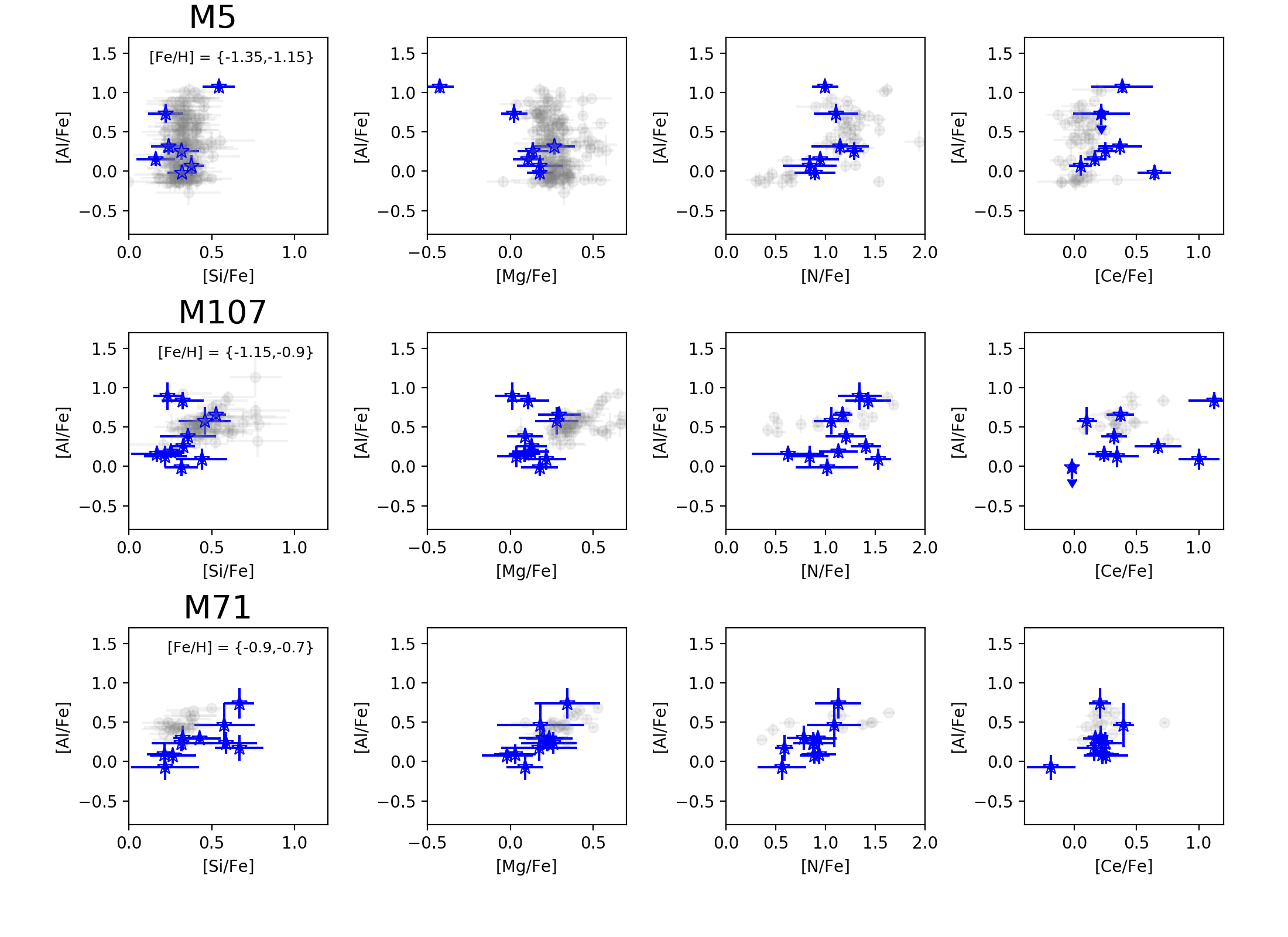}
		\caption{Distribution of [Al/Fe] with light- and heavy-elements ([Si/Fe], [Mg/Fe], [N/Fe], and [Ce/Fe]). The newly identified N-rich stars are highlighted with blue star symbols, and compared with globular cluster stars from \citet{Masseron2019} of similar metallicity. The chemical abundances provided are the average abundance of selected atomic and molecular lines from our manual inspection with the \texttt{BACCHUS} code.}
		\label{Figurecluster}
	\end{figure*}
\end{center}

Figure \ref{Figurecluster} displays our N-rich sample and globular cluster star data together on the [Al/Fe] versus [Si/Fe], [Mg/Fe], [N/Fe], and [Ce/Fe] planes. The mildly metal-poor globular cluster stars of \citet{Masseron2019}, from M5, M107, and M71 are included in this plot, so that the GCs span as wide a metallicity range as our field sample, and have been homogeneously analyzed in the same manner as the N-rich sample presented in this study. Interestingly, the light-/heavy-elements seen in the globular cluster population matches approximately the abundance values determinated for the newly identified N-rich field population at the same [Fe/H]. This result is consistent with findings from other studies, which have characterised SG and FG stars in GCs as having similar chemistry as field stars of same metallicity \citep[e.g.,][]{Lind2015, Martell2016, Fernandez-Trincado2016, Fernandez-Trincado2017L, Fernandez-Trincado2019}. One obvious exception is the star, 2M15535831$+$4333280, extremely Mg-depleted ($<-0.4$) and has high [Al/Fe] and [N/Fe] ratios, similar to large light-element enrichment seen in TYC 5619-109-1 \citep{Fernandez-Trincado2016}, which clearly fall within the extreme limits for [Mg/Fe] of the 'second- generation' globular cluster population. Our sample's [N/Fe] appears definitively anomalous, with a spread in [N/Fe] from around 0.5 to 1.5 dex.

The [Ce/Fe] abundance ratios for stars in our sample also show similar levels to those found for stars residing in globular clusters, with a spread in [Ce/Fe] from around -0.2 to $\sim$1.15. Given the lack of information regarding other neutron-capture elements, it is difficult to assign the nucleosynthetic pathways for such stars, however, we can speculate that the Ce and Nd we have measured are likely to have a pure \textit{s}-process origin.

The [O/Fe] abundance ratio is uncertain in the $T_{\rm eff}$ regime of our objects. The uncertainty arises because \texttt{BACCHUS} determines these abundances from the strengths of $^{12}$C$^{14}$N and $^{12}$C$^{16}$O lines, which become too weak for hotter stars at relatively low metallicities ([Fe/H] $\lsim -0.7$). Moreover, our results show that, C, Na, Ce II, and Nd II abundances have mostly upper limits (see Table \ref{table2}), implying that most of these line
become weak and heavily blended to be accurately measured, and therefore the derived abundances strongly depends on the ability of properly reproducing the blend. APOGEE has 10 Nd II and 9 Ce II features detectable \citep{Hasselquist2016, Cunha2017}, however these lines are highly sensitive to stellar parameters, and therefore, only a few can be used to derive upper limits for a large variety of metal-poor stars. In APOGEE spectra, two Na I lines are visible (1.6373$\mu$m and 1.6388$\mu$m) for a few metal-poor stars, however these lines are weak and heavily blended by telluric features at the typical $T_{\rm eff}$ and metallicity for the stars studied in this work. At this time, we cannot guarantee the quality of the abundances for those elements. 

\subsection{Extra-tidal features around Galactic globular clusters}

For most of the newly identified N-rich stars there are no known globular clusters within an angular separation of one degree, except four giants as listed in Table \ref{separation}. We find for the first time a N-rich star (2M15183589$+$0027100) which appears likely to be a candidate extra-tidal star (within the errors) associated with Pal 5 based on [Fe/H] metallicity, radial velocity and elemental-abundances. We conclude that 2M15183589$+$0027100 is promising an extra-tidal candidate of Pal 5, which appears to be compatible with the globular cluster escapee scenario, as well as supporting spectroscopic evidence that accretion onto the early Milky Way was significant. For the rest of the three N-rich stars with nearby globular clusters, we find a large spread in metallicity and radial velocity, thus making an extra-tidal origin from nearby GCs for these three N-rich stars very unlikely (see Table \ref{separation}). 

\begin{table*}
	\begin{center}
		\setlength{\tabcolsep}{1.0mm}  
		\caption{N-rich giants as tracers of extra-tidal features around Galactic globular clusters}
		\begin{tabular}{ccccccccc}
			\hline
			\hline
			APOGEE$-$ID & [Fe/H] & RV                   & Nearby GC  &  [Fe/H]$_{\rm gc}$ & RV$_{\rm gc}$ & r$_{\rm t, gc}$ & separation & Extra-tidal N-rich giant \\
			& [dex]   & [km s$^{-1}$] &       &    [dex]                    &  [km s$^{-1}$] &  arcmin            & arcmin       &      candidates         \\
			\hline
			\hline
			2M15183589+0027100   &  $-$1.18 &   $-$56.27$\pm$0.34       &   Pal 5    &     $-$1.41    &       $-$58.6$\pm$0.2     &       16.28     &      50.7   &   likely        \\                       
			2M18110406-2602142   &  $-$0.82  &   $-$24.96$\pm$0.01       &  NGC 6553      &     $-$0.18    &        0.7$\pm$0.4           &        8.16            &      25.2      &    unlikely     \\
			2M15195065+0221533   &  $-$0.83 &   $-$158.58$\pm$0.01     &  NGC 5904      &     $-$1.29    &       53.7$\pm$0.3           &       28.4            &     25.8        &   unlikely   \\
			2M16464310+4731033   &  $-$0.99 &   $-$139.80$\pm$0.37     &  NGC 6229      &     $-$1.47    &       $-$138.6$\pm$0.8     &        5.38           &       2.7        &  unlikely  \\
			\hline
			\hline
		\end{tabular}  \label{separation}\\
	\end{center}
\end{table*}

\subsection{Possible evolutionary state of the newly identified N-rich stars}

Here we qualitatively examine the possible evolutionary state of the newly discovered N-rich stars. To accomplish this, the C, N, O, Mg, Al, Si, Na, Ce II, and Nd II abundances are compared with theoretical AGB nucleosynthesis predictions: \texttt{FRUITY}\footnote{FUll-Network Repository of Updated Isotopic Tables and Yields: \url{http://fruity.oa-abruzzo.inaf.it/}} models from \citet{Cristallo2015}, which we have compared to observations with the \texttt{FRUITY} models of metallicity $Z = 2\times{}10^{-3}$, the \texttt{Monash} model of metallicity [Fe/H]$= -1.2$ from \citet{Fishlock2014}, and the \texttt{ATON} model ([Fe/H]$= -1.2$) from \citet{Ventura2016}, as shown in Figure \ref{AGB}. Based on $\chi^2$ fitting of the light-/heavy-elements, we find that the majority of our N-rich stars fit the metal-poor, low-mass (M $<$1.5 -- 5 M$_{\odot}$) AGB yields from \citet{Cristallo2015}, which dominates the production of \textit{s}-process elements (Ce II and Nd II) in most of the cases, but is at odds with the carbon abundance of the \texttt{FRUITY} database. On the other hand, the more massive AGB nucleosynthesis \texttt{ATON} and \texttt{Monash} models in the range of M $> $ 5 -- 7 M$_{\odot}$ is in better agreement with the observations, with the exception of [O/Fe], which shows a significant deviation from the theoretical predictions, perhaps due to the fact that the abundance of [O/Fe] is uncertain in the $T_{\rm eff}$ regime studied here. It is possible that the high nitrogen abundance could be due to a strong internal mixing process, suggesting that the newly identified N-rich stars are probably evolved objects, possibly an in "early-AGB" or AGB phase. More detailed AGB nucleosynthesis models will be necessary to explain the puzzle of the neutron-capture elements observed in these stars, as well more observations for further understanding the chemical peculiarities observed in these stars.

\subsection{Orbits}

In order to provide insight on the origin of our stars across the Milky Way, the positional information of the newly identified N-rich stars was combined with precise proper motions from Gaia DR2 \citep{Lindegren2018, Arenou2018}, radial velocity from the APOGEE-2  survey\citep{Nidever2015, Majewski2017}, and the newly-measured spectro-photometric distances from \citet{Leung2019} as input data for the new state-of-the-art orbital integration package \texttt{GravPot16}\footnote{\url{https://gravpot.utinam.cnrs.fr}}. Orbits are integrated in both an axisymemtric model, and a model including the perturbations due to a realistic (as far as possible) rotating "boxy/peanut" bar, which fits the structural and dynamical parameters of the Galaxy to the best we know of the recent knowledge of our Milky Way (Fernandez-Trincado et al. 2019, in prep.). Figures \ref{orbit1}, \ref{orbit2}, \ref{orbit3}, and \ref{orbit4} show the orbits for each individual star, using as initial conditions the central values, both in the case of the axisymmetric potential in the inertial Galactic frame of reference (\textit{column 1}) and the model with bar (\textit{columns 2, 3 and 4}) in the \textit{noninertial} frame, where the bar is at rest. For each star, we plotted the projection of the orbit on the Galactic plane (X--Y) and on the meridional plane (R-- Z). 

Here, we also provide orbital solutions assuming four different values of the angular velocity of the bar $\Omega_{bar} =$ 35, 40, 45, and 50 km s$^{-1}$ kpc$^{-1}$, with a bar mass of 1.1$\times10^{10}$ M$_{\odot}$, and a present-day angle orientation of 20$^{\circ}$, in the same manner as \citet{Fernandez-Trincado2019}. To model the uncertainty distributions, we sampled a half million orbits using a simple Monte Carlo scheme, assuming a normal distribution for the uncertainties of the input parameters (positions, distance, radial velocity and proper motions). The results are listed in Table \ref{tableorbits1}; the data presented in this table correspond to a backward time integration of 3 Gyr in our dynamical model. Figure \ref{orbitalelements} shows a scatter plot of all the possible combinations among the orbital elements in the non-axisymmetric potential model. We see that, in each panel, most the stars are grouped in two regions, indicating that they likely belong to the same component. In this case, one group is confined to the bulge/bar, and a second group is apparently moving outwards from the co-rotation (CR) region. 

For guidance, the Galactic convention adopted by this study is: $X-$axis is oriented toward $l=$ 0$^{\circ}$ and $b=$ 0$^{\circ}$, and the $Y-$axis is oriented toward $l$ = 90$^{\circ}$ and $b=$0$^{\circ}$, and the disc rotates toward $l=$ 90$^{\circ}$; the velocities are also oriented in these directions. In this convention, the Sun's orbital velocity vector is [U$_{\odot}$,V$_{\odot}$,W$_{\odot}$] = [$11.1$, $12.24$, 7.25] km s$^{-1}$ \citep{Brunthaler2011}. The model has been rescaled to the Sun's Galactocentric distance, 8.3 kpc, and the local rotation velocity of $239$ km s$^{-1}$.

\begin{center}
	\begin{figure}
		\includegraphics[width=90mm]{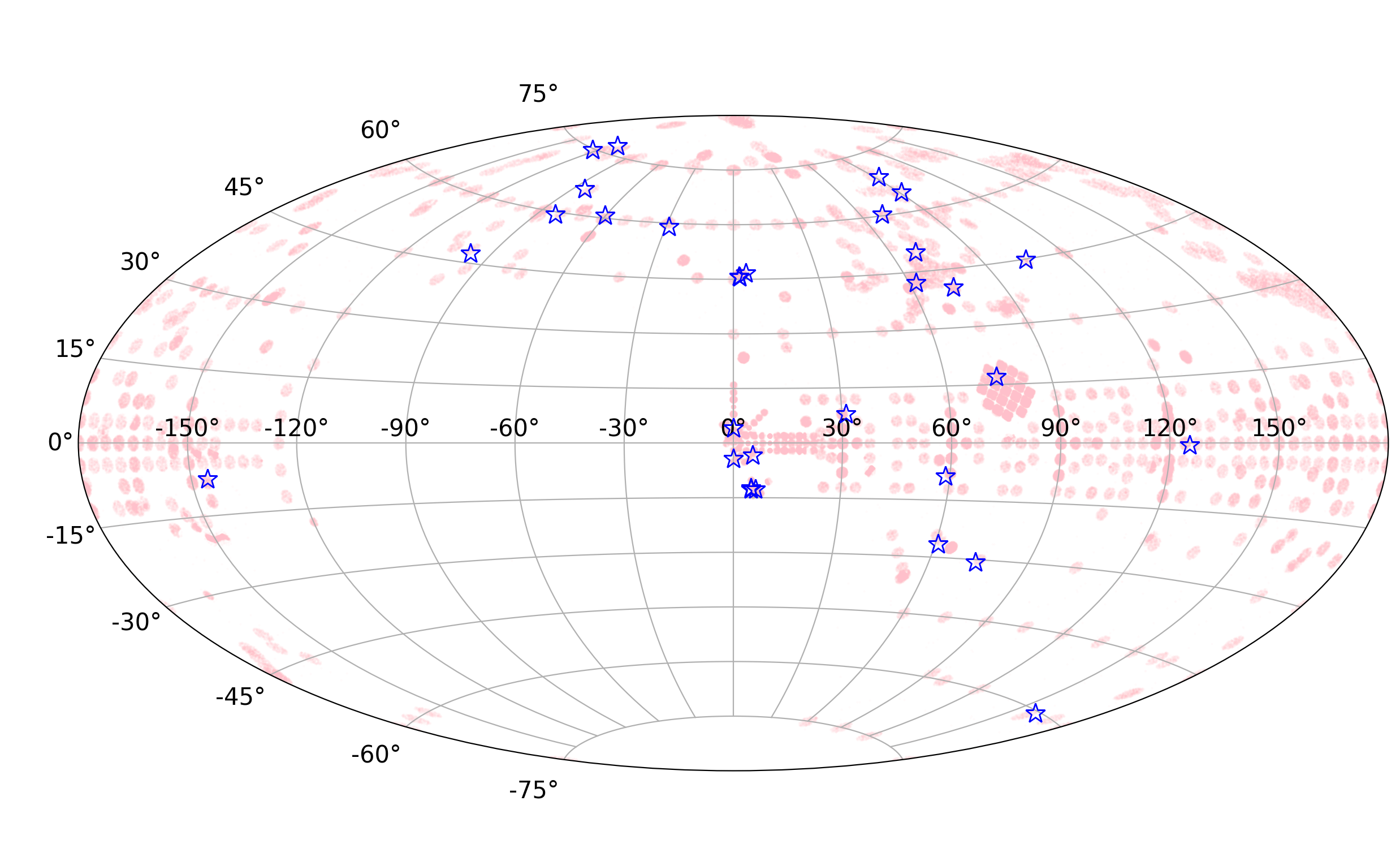}
		\caption{Aitoff projection map in Galactic coordinates for the APOGEE giants used in this work (pink dots). The newly identified N-rich giants are highlighted with blue unfilled ’star’ symbols.}
		\label{aitoff}
	\end{figure}
\end{center}

\begin{center}
	\begin{figure*}
		\includegraphics[width=162mm]{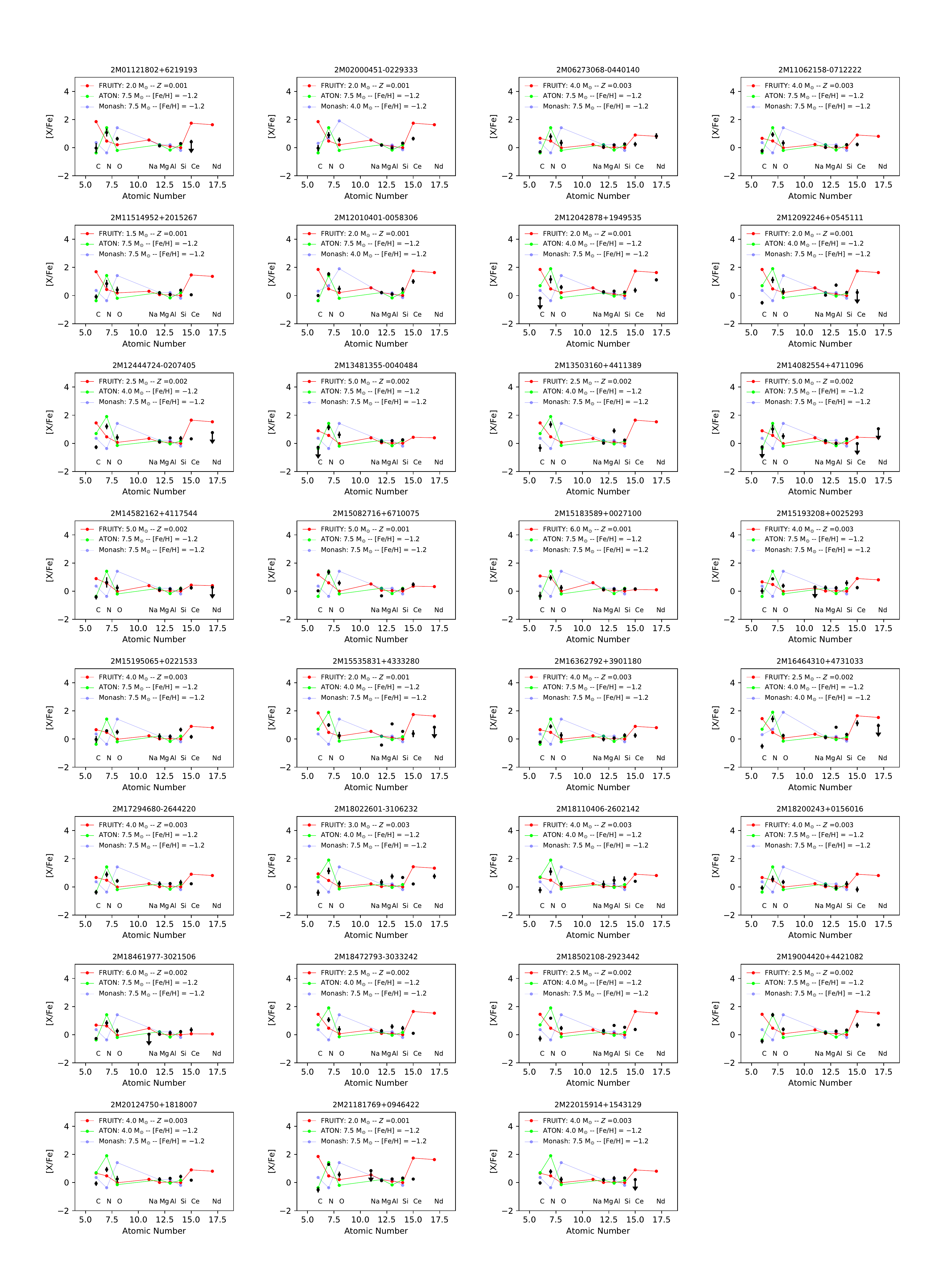}
		\caption{Chemical abundances for the 31 N-rich stars. Each determined abundance is shown as a black dot, solid arrows represent measurements where only an upper limit was possible. The error bars are estimates of the uncertainties in our measurements as listed in Table \ref{error1}. These abundances are compared to three synthetic AGB yields that best fit the observed data; for a selection of the $Z = 0.001$ (\texttt{FRUITY}; red dots and solid line) and $-1.2$ dex (\texttt{ATON}; green dots and solid line, and \texttt{Monash}; blue dots and solid line) models. Note that we have shifted atomic numbers of the heavy-elements Nd II and Ce II by 43 for better illustration of the results.}
		\label{AGB}
	\end{figure*}
\end{center}

Most of the newly identified N-rich stars are found to have radial and prograde orbits, with pericenter values less than 3 kpc (inside the bulge region), apocenter values ranging between 6.5--58 kpc, orbital eccentricities larger than 0.5, and maximum vertical excursions from the Galactic plane ranging between 0.5--58 kpc. The orbital parameters clearly show that a few these stars live in the inner Galaxy, while most are in the inner halo of the Milky Way. Figure \ref{aitoff} shows the Aitoff project of our objects, which reveals that 6 out of 31 of our stars are located toward the bulge region, 2M17294680$-$2644220, 2M18022601$-$3106232,  2M18110406$-$2602142, 2M18461977$-$3021506, 2M18472793$-$3033242, and 2M18502108$-$2923442; from this group we found that the orbital properties of three of them suggest that they are actually halo interlopers into the inner Galaxy (2M17294680$-$2644220, 2M18461977$-$3021506, and 2M18502108$-$2923442), while the other three stars appear to have bulge/bar-like prograde orbits. It is very likely that such N-rich stars trapped into the bulge/bar potential could be linked to merger debris of surviving globular clusters in bulge/bar-like orbits, such as M 62 \citep[see e.g.,][]{Minniti2019}. 

It is also worth mentioning that there are two other N-rich stars, 2M15535831$+$4333280 (\textit{SG-like}), 2M16362792$+$3901180 (\textit{FG-like/dGs}), whose orbits are retrograde with respect to the direction of the Galactic rotation. We found that the abundance patterns of these stars, namely the $\alpha-$elements, neutron-capture elements, and the abundance ratios of [Al/Fe], most resemble the known chemical signature of globular cluster stars. The chemistry and dynamical behaviour of this sub-sample of N-rich stars suggests that they could be the debris of dissipated globular clusters, indicating that there may be a significant population of these peculiar abundance giants residing in the Galactic field. 

Lastly, 26 out of 31 N-rich stars appear to behave as halo-like orbits, intriguingly in the prograde sense with respect to the rotation of the bar, suggesting that they were likely formed during the very early stages of the evolution of the Galaxy \citep[e.g.,][]{Khoperskov2018}, in a similar way as Galactic globular clusters. It is important to note that prograde orbits have been observed before in the inner halo \citep{Bonaca2017, Fernandez-Alvar2018, Hayes2018, Lucey2019}, as well as for other globular clusters \citep{Moreno2014, Perez-Villegas2018}, however this is not yet well-understood. We also found that most of the simulated orbits are situated in the inner Galaxy, which means that most of the N-rich stars are on highly eccentric orbits (with eccentricities greater than 0.6), reaching out to a maximum distance from the Galactic plane larger than 3 kpc. On the other hand, a handful of the N-rich stars have energies allowing the star to move inwards from the bar's corotation radius ($<$ 6.5 kpc). In this region, a class of orbits appears around the Lagrange points on the minor axis of the bar that can be stable, and have a banana-like shape parallel to the bar, as illustrated in a few cases in Figure \ref{orbit1} and \ref{orbit2}, while the orbits liberating around Lagrange points aligned with the bar are unstable and are probably chaotic orbits. Our model naturally predicts trajectories indicating that most of the N-rich stars are confined to the Galactic halo.

Additionally, in Table \ref{tableorbits1} and Figure \ref{orbitalelements} we show the variation of the z-component of the angular momentum ($L_{z}$), as a function of $\Omega_{bar}$. Since this quantity is not conserved in a model like \texttt{GravPot16} (with non-axisymmetric structures), we follow the change, \{$-L_{z}$,$+L_{z}$\}, where negative $L_{z}$ in our reference system means that the cluster orbit is prograde (in the same sense as the disk rotation). Both prograde and prograde-retrograde orbits with respect to the direction of the Galactic rotation are clearly revealed for a few cases, this effect is strongly produced by the presence of the Galactic bar, further indicating a chaotic behavior.

A major limitation of the employed dynamical model is that it ignores secular changes in the Milky Way potential over time, which might be important in understanding of the evolution in the inner Galaxy. An in-depth analysis of such dynamical effects is beyond the scope of this paper, however, the adopted technique shown that the inclusion of a more realistic (as far as possible) bar potential is essential for the description of the dynamical behavior of N-rich stars in the innermost part of the Galaxy.  

\section{Concluding Remarks}
\label{section4}

We have taken advantage of the fist \texttt{Payne} data release of APOGEE abundances, which determine best-fit stellar parameters and abundances using neural networks as an emulator. We applied a series of quality cuts on the initial \texttt{Payne} catalogue, ensuring the data have sufficiently high-quality spectra needed to estimate chemical abundances for a wide range of chemical species, as well as re-examined line-by-line each APOGEE spectrum with the \texttt{BACCHUS} pipeline. Our study presents an unique collection of 31 newly identified N-rich stars towards the bulge and inner halo of the Milky Way that exhibit anomalously high levels of [N/Fe] over a narrow range of metallicities ($-1.5\lsim$[Fe/H]$\lsim-0.7$), below the metal-poor tail of the thick-disk metallicity distribution. Based on their [Al/Fe] abundance ratios, we classified them into two groups, the N-/Al-rich giant-stars ([Al/Fe]$\gsim+0.5$) with chemical signatures similiar to second-generation globular cluster stars, and a second group, the N-rich/Al-normal ([Al/Fe]$<+0.5$) with chemistry similar to the first generation of stars seen in Galactic globular clusters and the population of stars in dwarf galaxies. For many of them, we determined, for the first time the abundances of \textit{s}-process elements (Ce II and Nd II).

We did not detect any significant variation of their radial velocities that would support the hypothesis of mass transfer, however most of the chemical species examined, along with the high nitrogen abundances could supports the idea that the process responsible for the stars can be qualitatively explained by  massive evolved objects, possibly "early-AGB" or AGB stars. We hypothesise that massive (M$>$ 5 --7 M$_{\odot}$) AGB stars may produce a [N/Fe] over-abundance in some of our N-rich giant stars within the Milky Way. However, more detailed AGB nucleosynthesis models, as well as more observations, will be necessary to confirm or refute the scenario related to AGB stars. Combining our  abundance results, orbital analysis, and the absence of radial velocity variations, we conclude that most of the newly identified objects are associated with the bulge/bar structure and the inner Galactic halo, and are likely escaped members of small satellites that were tidally disrupted and captured by the Milky Way.


\section*{Acknowledgments}

 We thank Szabolcs~M{\'e}sz{\'a}ros for helpful support computing the photometry $T_{\rm eff}$, and especially grateful for the technical expertise and assistance provided by the Instituto de Astrof\'isica de Canarias (IAC). This article is based upon work from the “ChETEC” COST Action (CA16117), supported by COST (European Cooperation in Science and Technology). J.G.F-T is supported by FONDECYT No. 3180210. J.G.F-T acknowledges the use of \texttt{TOPCAT} \citep{Taylor2005} through out the course of this investigation. T.C.B. acknowledge partial support for this work from grant PHY 14-30152; Physics Frontier Center / JINA Center for the Evolution of the Elements (JINA-CEE), awarded by the US National Science Foundation. B.T. acknowledges support from the one-hundred-talent project of Sun Yat-Sen University. S.L-M acknowledges funding from the Australian Research Council through Discovery grant DP180101791, and from the UNSW Scientia Fellowship program. Parts of this research were conducted by the Australian Research Council Centre of Excellence for All Sky Astrophysics in 3 Dimensions (ASTRO 3D), through project number CE170100013. A.P-V acknowledges a FAPESP for the postdoctoral fellowship grant no. 2017/15893-1 and the DGAPA-PAPIIT grant IG100319. 
  
 \texttt{BACCHUS} have been executed on computers from the Utinam Institute of the Universit\'e de Franche-Comt\'e, supported by the R\'egion de Franche-Comt\'e and Institut des Sciences de l'Univers (INSU). 
 
 Funding for the \texttt{GravPot16} software has been provided by the Centre national d'\'etudes spatiales (CNES) through grant 0101973 and UTINAM Institute of the Universit\'e de Franche-Comt\'e, supported by the R\'egion de Franche-Comt\'e and Institut des Sciences de l'Univers (INSU). Simulations have been executed on computers from the Utinam Institute of the Universit\'e de Franche-Comt\'e, supported by the R\'egion de Franche-Comt\'e and Institut des Sciences de l'Univers (INSU), and on the supercomputer facilities of the M\'esocentre de calcul de Franche-Comt\'e. 
 
 Funding for the Sloan Digital Sky Survey IV has been provided by the Alfred P. Sloan Foundation, the U.S. Department of Energy Office of Science, and the Participating Institutions. SDSS- IV acknowledges support and resources from the Center for High-Performance Computing at the University of Utah. The SDSS web site is www.sdss.org. SDSS-IV is managed by the Astrophysical Research Consortium for the Participating Institutions of the SDSS Collaboration including the Brazilian Participation Group, the Carnegie Institution for Science, Carnegie Mellon University, the Chilean Participation Group, the French Participation Group, Harvard-Smithsonian Center for Astrophysics, Instituto de Astrof\`{i}sica de Canarias, The Johns Hopkins University, Kavli Institute for the Physics and Mathematics of the Universe (IPMU) / University of Tokyo, Lawrence Berkeley National Laboratory, Leibniz Institut f\"{u}r Astrophysik Potsdam (AIP), Max-Planck-Institut f\"{u}r Astronomie (MPIA Heidelberg), Max-Planck-Institut f\"{u}r Astrophysik (MPA Garching), Max-Planck-Institut f\"{u}r Extraterrestrische Physik (MPE), National Astronomical Observatory of China, New Mexico State University, New York University, University of  Dame, Observat\'{o}rio Nacional / MCTI, The Ohio State University, Pennsylvania State University, Shanghai Astronomical Observatory, United Kingdom Participation Group, Universidad Nacional Aut\'{o}noma de M\'{e}xico, University of Arizona, University of Colorado Boulder, University of Oxford, University of Portsmouth, University of Utah, University of Virginia, University of Washington, University of Wisconsin, Vanderbilt University, and Yale University.


\begin{center}
	\begin{figure*}
		\includegraphics[width=165mm]{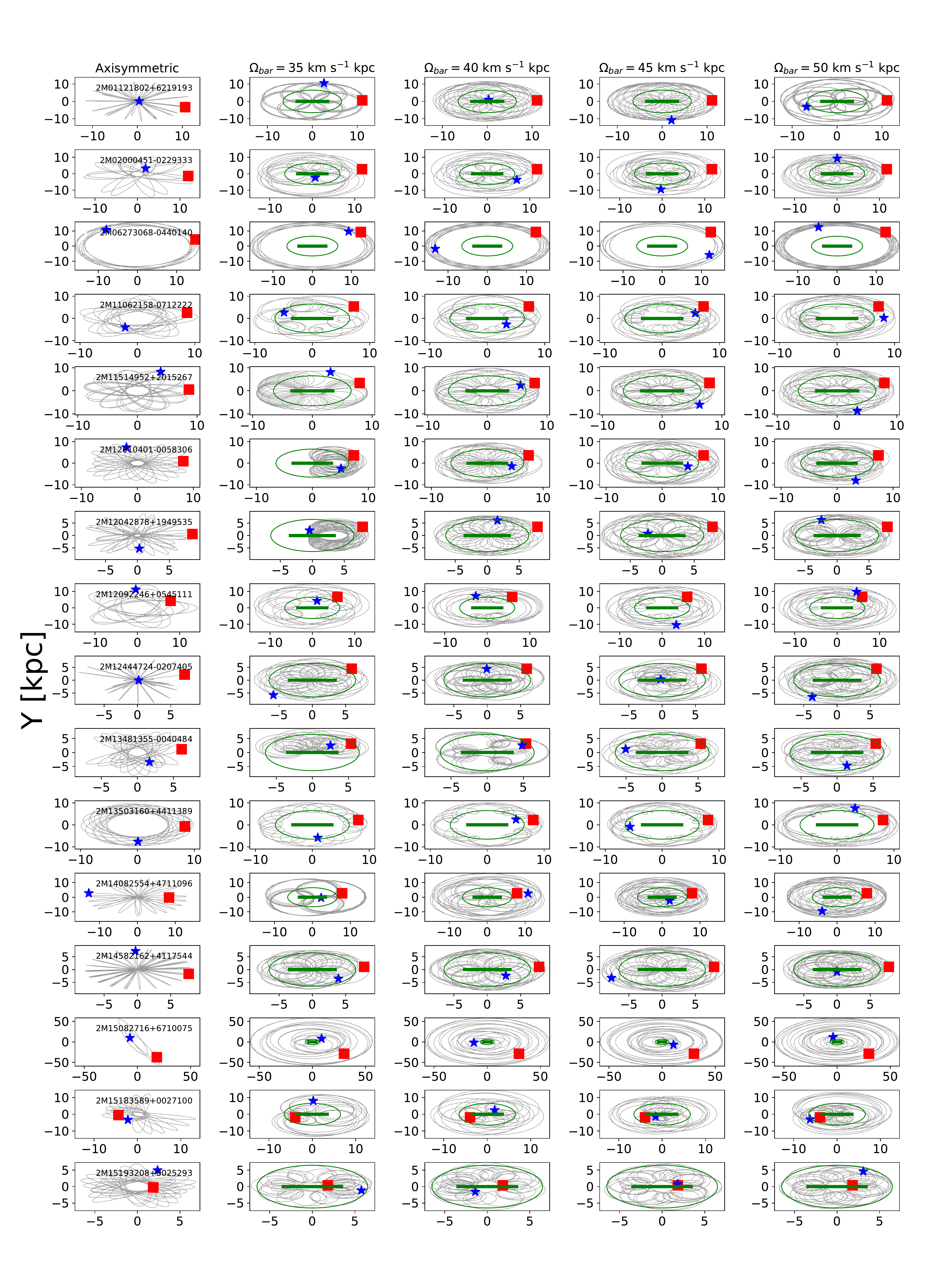}
		\caption{Orbits for the newly identified N-rich stars in an x--y projection, integrated adopting the central values (positions, proper motions, radial velocity and distance) in both an axisymmetric model (column 1) and a model including the Galactic bar potential in the \textit{non-inertial} reference frame where the bar is at rest (columns 2, 3, 4 and 5). The green solid line shows the size of the Galactic bar, and the large green circle the co-rotation radius, CR$\sim$6.5 kpc. The small red square symbol marks the present position of the star, and the blue star symbol marks its final position.}
		\label{orbit1}
	\end{figure*}
\end{center}

\begin{center}
	\begin{figure*}
		\includegraphics[width=165mm]{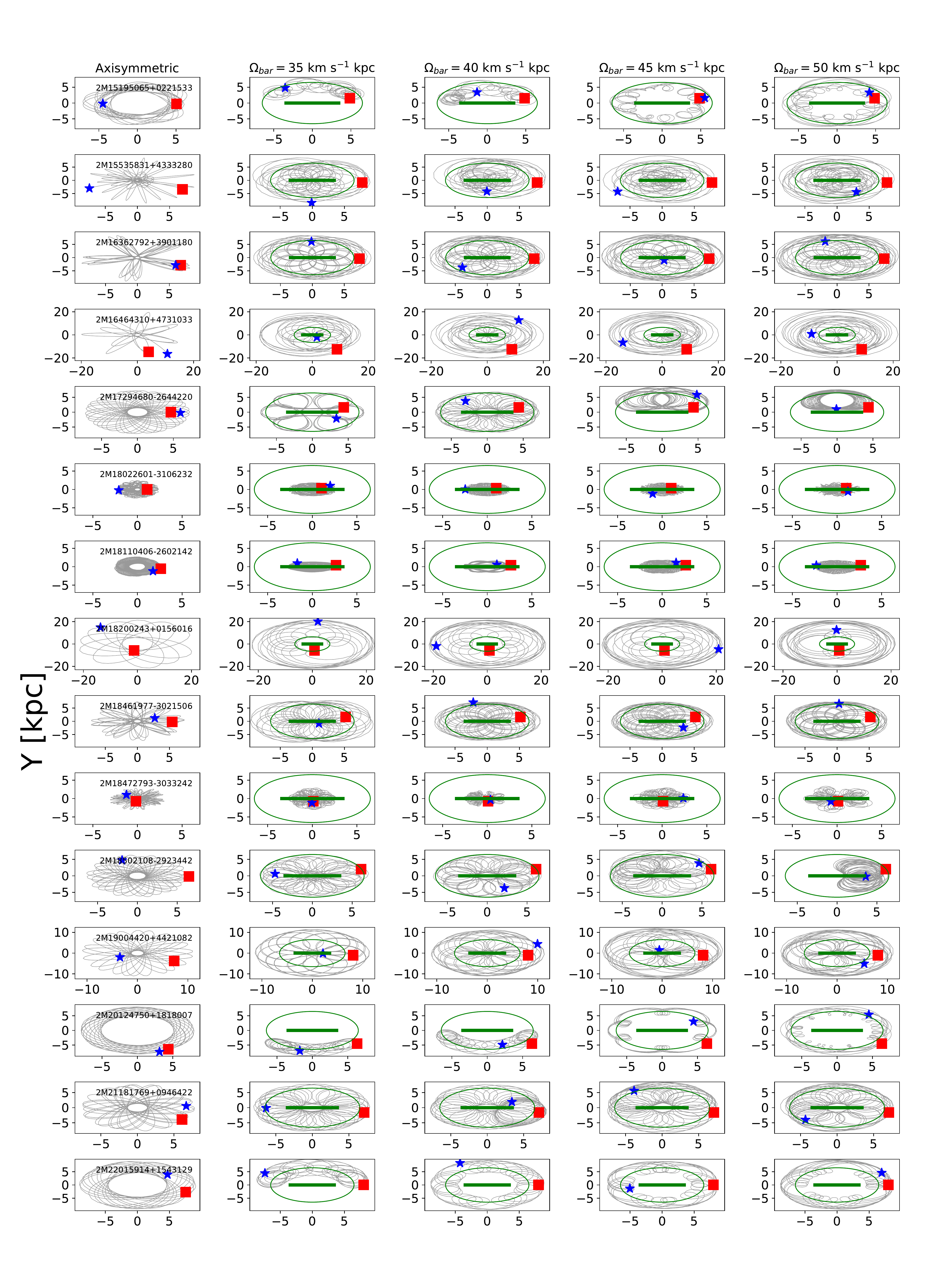}
		\caption{\textit{(continued.)}}
		\label{orbit2}
	\end{figure*}
\end{center}

\begin{center}
	\begin{figure*}
		\includegraphics[width=165mm]{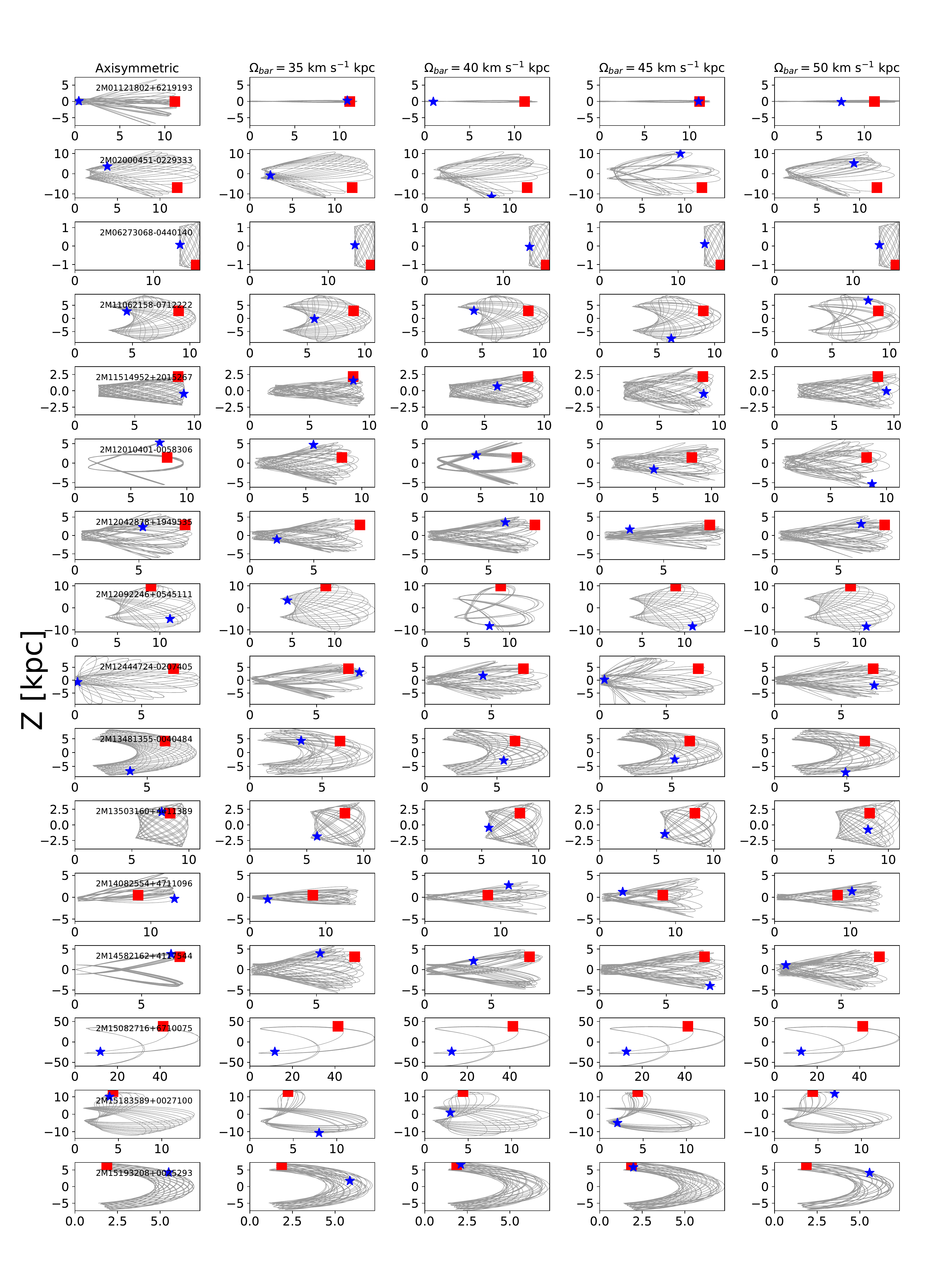}
		\caption{Orbits for the same sample in an R--z projection. The symbols have the same meaning as those in Figure \ref{orbit1}.}
		\label{orbit3}
	\end{figure*}
\end{center}

\begin{center}
	\begin{figure*}
		\includegraphics[width=165mm]{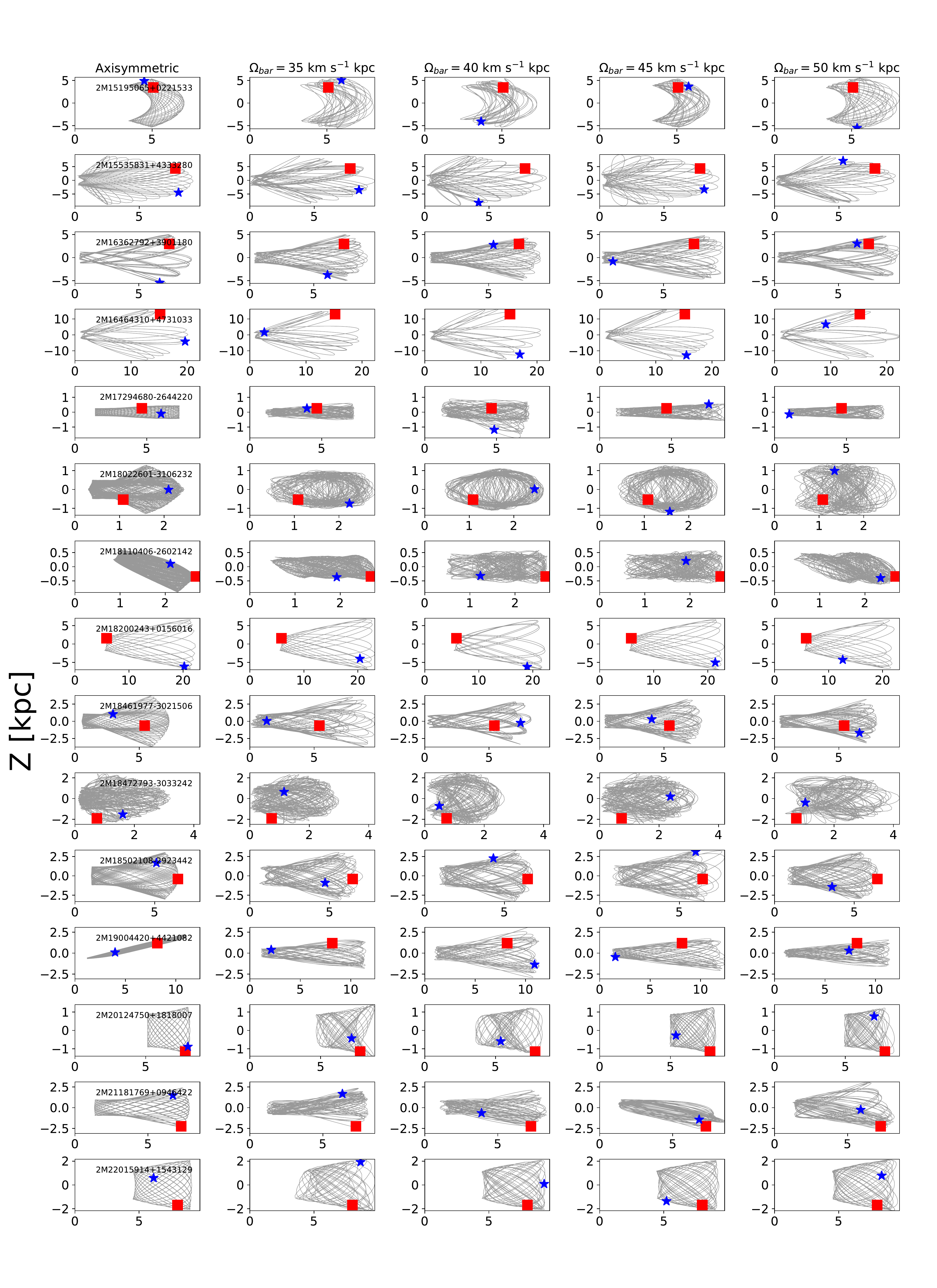}
		\caption{\textit{(continued.)}}
		\label{orbit4}
	\end{figure*}
\end{center}

\begin{center}
	\begin{figure*}
		\includegraphics[width=180mm]{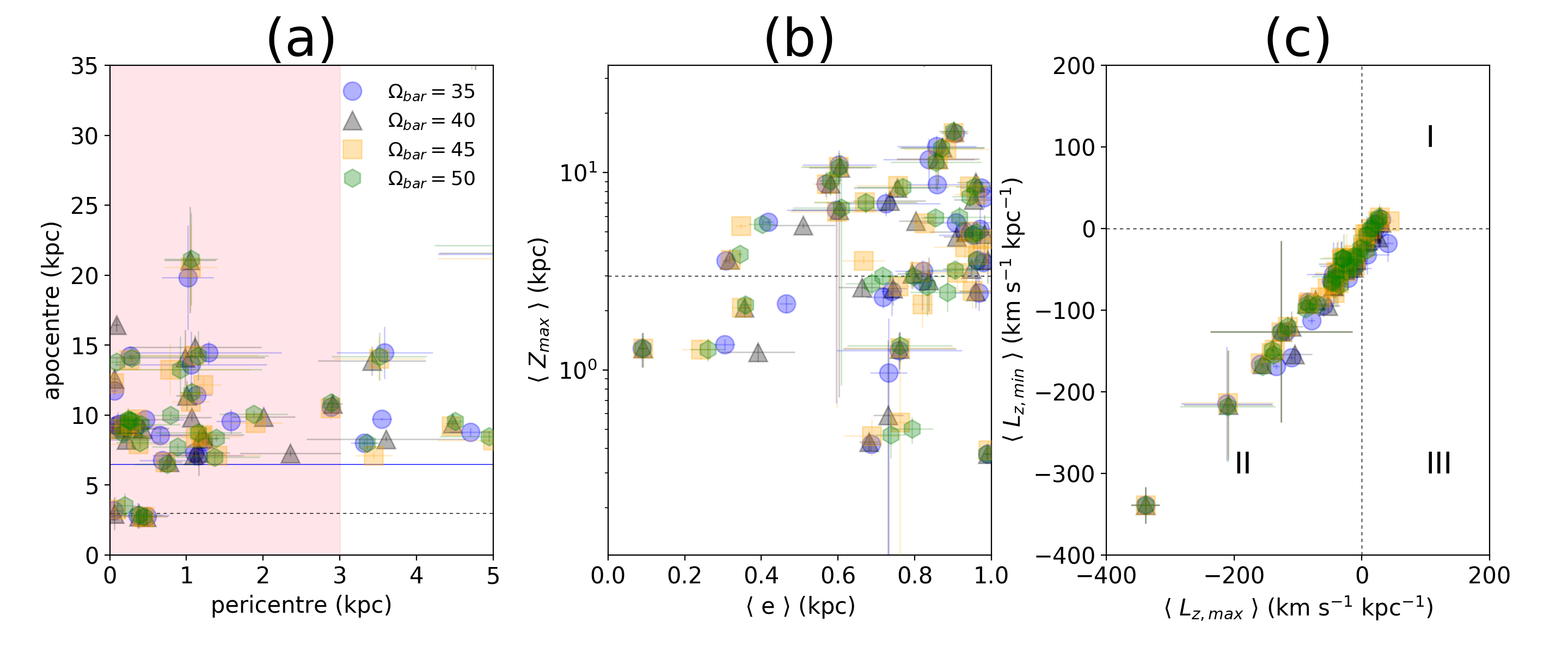}
		\caption{Orbital parameters calculated with different pattern speed of the bar, 35 (blue circles), 40 (grey triangles), 45 (orange squares) and 50 (green hexagons) km s$^{-1}$ kpc$^{-1}$. In panel (a) the shaded region and the black dotted line indicates the radius \citep[i.e., 3 kpc; ][]{Barbuy2018} of the Milky Way bulge, while the blue line indicates the location of the bar’s corotation radius ($CR \sim$6.5 kpc). For panel (a) a star below the black dotted line would have a bulge-like orbit, and in panel (b) the black dotted line represents the edge Zmax of the thick disk \citep[$\sim$3 kpc, ][]{Carollo2010}. In panel (c), the black dotted lines divide the regions with prograde orbits (\textit{region II}) with respect to the direction of the Galactic rotation, retrograde orbits (\textit{region I}), and stars that have prograde-retrograde orbits at the same time (\textit{region III}). The error bars show the uncertainty in the computed orbital parameters.}
		\label{orbitalelements}
	\end{figure*}
\end{center}

\clearpage

\begin{table}
	\begin{tiny}
		\begin{center}
			\setlength{\tabcolsep}{1.0mm}  
			\caption{Abundance determination sensitivity to the stellar parameters from our present measurements.}
			\begin{tabular}{clccccc}
				\hline
				\hline
				APOGEE$-$ID                    &   $X$ &   $\sigma_{[X/H],T_{\rm eff}}$   &    $\sigma_{[X/H],logg}$   &   $\sigma_{[X/H],\xi_{t}}$  &   $\sigma_{mean}$    &   $\sigma_{total}$  \\
				\hline
				\hline
				2M01121802$+$6219193  &   Fe      &   0.048   &    0.010   &   0.015   &   0.102  &   0.114  \\
				2M01121802$+$6219193  &   C       &   0.011   &    0.085   &   0.117   &   0.315  &   0.347  \\
				2M01121802$+$6219193  &   N       &   0.158   &    0.092   &   0.138   &   0.169  &   0.285  \\
				2M01121802$+$6219193  &   O       &   0.134   &    0.027   &   0.012   &   0.051  &   0.147  \\
				2M01121802$+$6219193  &   Mg      &   0.061   &    0.041   &   0.025   &   0.022  &   0.082  \\
				2M01121802$+$6219193  &   Al      &   ..      &    ..      &   ..      &   ..     &   ..     \\
				2M01121802$+$6219193  &   Si      &   ..      &    ..      &   ..      &   ..     &   ..     \\
				2M01121802$+$6219193  &   Ce      &   0.054   &    0.090   &   0.098   &   ..     &   0.144  \\
				2M01121802$+$6219193  &   Nd      &   ..      &    ..      &   ..      &   ..     &   ..     \\
				2M01121802$+$6219193  &   Na      &   ..      &    ..      &   ..      &   ..     &   ..     \\
				\hline
				2M02000451$-$0229333  &   Fe      &   0.040   &    0.010   &   0.005   &   0.101  &   0.109  \\
				2M02000451$-$0229333  &   C       &   0.081   &    0.179   &   0.032   &   0.102  &   0.224  \\
				2M02000451$-$0229333  &   N       &   0.080   &    0.143   &   0.054   &   0.111  &   0.205  \\
				2M02000451$-$0229333  &   O       &   0.123   &    0.021   &   0.005   &   0.131  &   0.181  \\
				2M02000451$-$0229333  &   Mg      &   0.058   &    0.055   &   0.007   &   0.011  &   0.081  \\
				2M02000451$-$0229333  &   Al      &   0.056   &    0.017   &   0.006   &   0.074  &   0.094  \\
				2M02000451$-$0229333  &   Si      &   0.022   &    0.020   &   0.011   &   0.081  &   0.088  \\
				2M02000451$-$0229333  &   Ce      &   0.118   &    0.045   &   0.015   &   0.042  &   0.134  \\
				2M02000451$-$0229333  &   Nd      &   0.0     &    ..      &   ..      &   ..     &   ..     \\
				2M02000451$-$0229333  &   Na      &   ..      &    ..      &   ..      &   ..     &   ..     \\
				\hline
				2M06273068$-$0440140  &   Fe      &   0.047   &    0.024   &   0.024   &   0.087  &   0.105  \\
				2M06273068$-$0440140  &   C       &   0.012   &    0.094   &   0.010   &   0.050  &   0.108  \\
				2M06273068$-$0440140  &   N       &   0.168   &    0.118   &   0.005   &   0.067  &   0.216  \\
				2M06273068$-$0440140  &   O       &   0.102   &    0.034   &   0.017   &   0.185  &   0.215  \\
				2M06273068$-$0440140  &   Mg      &   0.072   &    0.106   &   0.031   &   0.026  &   0.135  \\
				2M06273068$-$0440140  &   Al      &   0.063   &    0.063   &   0.027   &   0.030  &   0.098  \\
				2M06273068$-$0440140  &   Si      &   0.022   &    0.007   &   0.008   &   0.118  &   0.120  \\
				2M06273068$-$0440140  &   Ce      &   0.066   &    0.036   &   0.053   &   0.133  &   0.162  \\
				2M06273068$-$0440140  &   Nd      &   0.041   &    0.124   &   0.019   &   0.154  &   0.203  \\
				2M06273068$-$0440140  &   Na      &   ..      &    ..      &   ..      &   ..     &   ..     \\
				\hline
				2M11062158$-$0712222  &   Fe      &   0.066   &    0.015   &   0.026   &   0.058  &   0.094  \\
				2M11062158$-$0712222  &   C       &   0.015   &    0.025   &   0.044   &   0.137  &   0.147  \\
				2M11062158$-$0712222  &   N       &   0.155   &    0.016   &   0.064   &   0.038  &   0.173  \\
				2M11062158$-$0712222  &   O       &   0.150   &    0.022   &   0.017   &   0.128  &   0.200  \\
				2M11062158$-$0712222  &   Mg      &   0.093   &    0.072   &   0.022   &   0.033  &   0.125  \\
				2M11062158$-$0712222  &   Al      &   0.092   &    0.026   &   0.030   &   0.077  &   0.127  \\
				2M11062158$-$0712222  &   Si      &   0.060   &    0.037   &   0.027   &   0.074  &   0.106  \\
				2M11062158$-$0712222  &   Ce      &   0.018   &    0.057   &   0.013   &   0.115  &   0.130  \\
				2M11062158$-$0712222  &   Nd      &   ..      &    ..      &   ..      &   ..     &   ..     \\
				2M11062158$-$0712222  &   Na      &   ..      &    ..      &   ..      &   ..     &   ..     \\
				\hline
				2M11514952$+$2015267  &   Fe      &   0.053   &    0.035   &   0.046   &   0.084  &   0.115  \\
				2M11514952$+$2015267  &   C       &   0.035   &    0.128   &   0.065   &   0.073  &   0.165  \\
				2M11514952$+$2015267  &   N       &   0.162   &    0.147   &   0.129   &   0.087  &   0.269  \\
				2M11514952$+$2015267  &   O       &   0.138   &    0.073   &   0.072   &   0.135  &   0.219  \\
				2M11514952$+$2015267  &   Mg      &   0.084   &    0.085   &   0.060   &   0.034  &   0.139  \\
				2M11514952$+$2015267  &   Al      &   0.071   &    0.049   &   0.061   &   0.073  &   0.129  \\
				2M11514952$+$2015267  &   Si      &   0.028   &    0.013   &   0.031   &   0.061  &   0.075  \\
				2M11514952$+$2015267  &   Ce      &   0.044   &    0.069   &   0.036   &   0.020  &   0.092  \\
				2M11514952$+$2015267  &   Nd      &   ..      &    ..      &   ..      &   ..     &   ..     \\
				2M11514952$+$2015267  &   Na      &   ..      &    ..      &   ..      &   ..     &   ..     \\
				\hline
				2M12010401$-$0058306  &   Fe      &   0.042   &    5.421   &   0.008   &   0.122  &   0.129  \\
				2M12010401$-$0058306  &   C       &   0.047   &    0.087   &   0.004   &   0.072  &   0.123  \\
				2M12010401$-$0058306  &   N       &   0.082   &    0.054   &   0.016   &   0.087  &   0.133  \\
				2M12010401$-$0058306  &   O       &   0.137   &    0.023   &   0.006   &   0.180  &   0.228  \\
				2M12010401$-$0058306  &   Mg      &   0.086   &    0.072   &   0.008   &   0.044  &   0.121  \\
				2M12010401$-$0058306  &   Al      &   0.066   &    0.017   &   0.011   &   0.112  &   0.132  \\
				2M12010401$-$0058306  &   Si      &   0.034   &    0.035   &   0.015   &   0.145  &   0.154  \\
				2M12010401$-$0058306  &   Ce      &   0.057   &    0.123   &   0.033   &   0.089  &   0.166  \\
				2M12010401$-$0058306  &   Nd      &   ..      &    ..      &   ..      &   ..     &   ..     \\
				2M12010401$-$0058306  &   Na      &   ..      &    ..      &   ..      &   ..     &   ..     \\
				\hline
				2M12042878$+$1949535  &   Fe      &   0.050   &    0.015   &   0.002   &   0.103  &   0.116  \\
				2M12042878$+$1949535  &   C       &   0.036   &    0.084   &   0.001   &   ..     &   0.091  \\
				2M12042878$+$1949535  &   N       &   0.250   &    0.084   &   0.002   &   0.119  &   0.289  \\
				2M12042878$+$1949535  &   O       &   0.132   &    0.029   &   9.994   &   0.087  &   0.161  \\
				2M12042878$+$1949535  &   Mg      &   0.085   &    0.065   &   0.006   &   0.057  &   0.122  \\
				2M12042878$+$1949535  &   Al      &   0.081   &    0.034   &   0.005   &   0.043  &   0.098  \\
				2M12042878$+$1949535  &   Si      &   0.037   &    0.023   &   0.036   &   0.093  &   0.109  \\
				2M12042878$+$1949535  &   Ce      &   0.075   &    0.116   &   0.009   &   0.111  &   0.177  \\
				2M12042878$+$1949535  &   Nd      &   0.067   &    0.092   &   0.022   &   0.024  &   0.118  \\
				2M12042878$+$1949535  &   Na      &   ..      &    ..      &   ..      &   ..     &   ..     \\
				\hline
								2M12092246$+$0545111  &   Fe      &   0.047   &    0.021   &   0.009   &   0.096  &   0.110  \\
								2M12092246$+$0545111  &   C       &   0.064   &    0.110   &   0.030   &   0.016  &   0.132  \\
								2M12092246$+$0545111  &   N       &   0.190   &    0.093   &   0.030   &   0.059  &   0.222  \\
								2M12092246$+$0545111  &   O       &   0.141   &    0.027   &   0.010   &   0.193  &   0.241  \\
								2M12092246$+$0545111  &   Mg      &   0.067   &    0.044   &   0.010   &   0.005  &   0.081  \\
								2M12092246$+$0545111  &   Al      &   0.092   &    0.051   &   0.014   &   0.063  &   0.123  \\
								2M12092246$+$0545111  &   Si      &   0.027   &    0.036   &   0.019   &   0.094  &   0.106  \\
								2M12092246$+$0545111  &   Ce      &   0.164   &    0.158   &   0.004   &   ..     &   0.228  \\
								2M12092246$+$0545111  &   Nd      &   ..      &    ..      &   ..      &   ..     &   ..     \\
								2M12092246$+$0545111  &   Na      &   ..      &    ..      &   ..      &   ..     &   ..     \\
				\hline
				\hline
			\end{tabular}  \label{error1}\\
		\end{center}
	\end{tiny}
				\raggedright{The reported uncertainty for each chemical species in column 7 is: $\sigma_{total}  = \sqrt{\sigma^2_{[X/H], T_{\rm eff}}    + \sigma^2_{[X/H],{\rm log} g} + \sigma^2_{[X/H],\xi_t}  + \sigma^2_{mean}   }$.}
\end{table}

\begin{table}
	\begin{tiny}
		\begin{center}
			\setlength{\tabcolsep}{1.0mm}  
			\caption{\textit{(continued)}}
			\begin{tabular}{clccccc}
				\hline
				\hline
				APOGEE$-$ID                    &   $X$ &   $\sigma_{[X/H],T_{\rm eff}}$   &    $\sigma_{[X/H],logg}$   &   $\sigma_{[X/H],\xi_{t}}$  &   $\sigma_{mean}$    &   $\sigma_{total}$  \\
				\hline
				\hline
				2M12444724$-$0207405  &   Fe      &   0.043   &    0.012   &   0.005   &   0.089  &   0.100  \\
				2M12444724$-$0207405  &   C       &   0.043   &    0.083   &   0.082   &   0.075  &   0.146  \\
				2M12444724$-$0207405  &   N       &   0.106   &    0.053   &   0.089   &   0.140  &   0.204  \\
				2M12444724$-$0207405  &   O       &   0.113   &    0.018   &   0.002   &   0.172  &   0.207  \\
				2M12444724$-$0207405  &   Mg      &   0.065   &    0.078   &   0.006   &   0.035  &   0.108  \\
				2M12444724$-$0207405  &   Al      &   0.073   &    0.035   &   0.013   &   0.063  &   0.104  \\
				2M12444724$-$0207405  &   Si      &   0.070   &    0.066   &   0.073   &   0.119  &   0.170  \\
				2M12444724$-$0207405  &   Ce      &   0.012   &    0.038   &   0.014   &   0.093  &   0.102  \\
				2M12444724$-$0207405  &   Nd      &   0.106   &    0.001   &   0.031   &   ..     &   0.110  \\
				2M12444724$-$0207405  &   Na      &   ..      &    ..      &   ..      &   ..     &   ..     \\
				\hline
				2M13481355$-$0040484  &   Fe      &   0.055   &    0.018   &   0.015   &   0.083  &   0.102  \\
				2M13481355$-$0040484  &   C       &   0.056   &    0.076   &   0.032   &   ...    &   0.099  \\
				2M13481355$-$0040484  &   N       &   0.167   &    0.059   &   0.045   &   0.057  &   0.192  \\
				2M13481355$-$0040484  &   O       &   0.145   &    0.043   &   0.026   &   0.175  &   0.233  \\
				2M13481355$-$0040484  &   Mg      &   0.076   &    0.073   &   0.010   &   0.027  &   0.110  \\
				2M13481355$-$0040484  &   Al      &   0.065   &    0.021   &   0.011   &   0.045  &   0.083  \\
				2M13481355$-$0040484  &   Si      &   0.037   &    0.034   &   0.013   &   0.048  &   0.071  \\
				2M13481355$-$0040484  &   Ce      &   ..      &    ..      &   ..      &   ..     &   ..     \\
				2M13481355$-$0040484  &   Nd      &   ..      &    ..      &   ..      &   ..     &   ..     \\
				2M13481355$-$0040484  &   Na      &   ..      &    ..      &   ..      &   ..     &   ..     \\
				\hline
				2M13503160$+$4411389  &   Fe      &   0.051   &    0.021   &   0.014   &   0.110  &   0.124  \\
				2M13503160$+$4411389  &   C       &   0.049   &    0.220   &   0.100   &   0.089  &   0.263  \\
				2M13503160$+$4411389  &   N       &   0.054   &    0.164   &   0.007   &   0.137  &   0.221  \\
				2M13503160$+$4411389  &   O       &   ..      &    ..      &   ..      &   ..     &   ..     \\
				2M13503160$+$4411389  &   Mg      &   0.073   &    0.077   &   0.006   &   0.010  &   0.108  \\
				2M13503160$+$4411389  &   Al      &   0.085   &    0.095   &   0.006   &   0.120  &   0.176  \\
				2M13503160$+$4411389  &   Si      &   0.033   &    0.012   &   0.025   &   0.075  &   0.087  \\
				2M13503160$+$4411389  &   Ce      &   ..      &    ..      &   ..      &   ..     &   ..     \\
				2M13503160$+$4411389  &   Nd      &   ..      &    ..      &   ..      &   ..     &   ..     \\
				2M13503160$+$4411389  &   Na      &   ..      &    ..      &   ..      &   ..     &   ..     \\
				\hline
				2M14082554$+$4711096  &   Fe      &   0.047   &    0.016   &   0.041   &   0.088  &   0.110  \\
				2M14082554$+$4711096  &   C       &   0.132   &    0.152   &   0.117   &   ..     &   0.134  \\
				2M14082554$+$4711096  &   N       &   0.246   &    0.102   &   0.052   &   0.161  &   0.316  \\
				2M14082554$+$4711096  &   O       &   0.129   &    0.032   &   0.035   &   0.162  &   0.212  \\
				2M14082554$+$4711096  &   Mg      &   0.068   &    0.075   &   0.036   &   0.021  &   0.110  \\
				2M14082554$+$4711096  &   Al      &   0.060   &    0.027   &   0.055   &   0.075  &   0.114  \\
				2M14082554$+$4711096  &   Si      &   0.036   &    0.015   &   0.022   &   0.089  &   0.100  \\
				2M14082554$+$4711096  &   Ce      &   0.001   &    0.001   &   0.009   &   ..     &   0.009  \\
				2M14082554$+$4711096  &   Nd      &   0.043   &    0.043   &   0.049   &   ..     &   0.078  \\
				2M14082554$+$4711096  &   Na      &   ..      &    ..      &   ..      &   ..     &   ..     \\
				\hline
				2M14582162$+$4117544  &   Fe      &   0.031   &    0.056   &   0.022   &   0.083  &   0.107  \\
				2M14582162$+$4117544  &   C       &   0.018   &    0.073   &   0.002   &   0.151  &   0.169  \\
				2M14582162$+$4117544  &   N       &   0.211   &    0.227   &   0.181   &   0.038  &   0.362  \\
				2M14582162$+$4117544  &   O       &   0.164   &    0.039   &   0.025   &   0.119  &   0.208  \\
				2M14582162$+$4117544  &   Mg      &   0.066   &    0.045   &   0.030   &   0.042  &   0.095  \\
				2M14582162$+$4117544  &   Al      &   0.079   &    0.040   &   0.030   &   0.027  &   0.098  \\
				2M14582162$+$4117544  &   Si      &   0.069   &    0.113   &   0.065   &   0.044  &   0.154  \\
				2M14582162$+$4117544  &   Ce      &   0.048   &    0.111   &   0.014   &   0.042  &   0.129  \\
				2M14582162$+$4117544  &   Nd      &   0.027   &    0.006   &   0.008   &   ..     &   0.029  \\
				2M14582162$+$4117544  &   Na      &   ..      &    ..      &   ..      &   ..     &   ..     \\
					\hline
				2M15082716$+$6710075  &   Fe      &   0.040   &    0.048   &   0.018   &   0.167  &   0.180  \\
				2M15082716$+$6710075  &   C       &   0.014   &    0.041   &   0.008   &   0.086  &   0.097  \\
				2M15082716$+$6710075  &   N       &   0.115   &    0.051   &   0.012   &   0.122  &   0.176  \\
				2M15082716$+$6710075  &   O       &   0.146   &    0.042   &   0.012   &   0.087  &   0.176  \\
				2M15082716$+$6710075  &   Mg      &   0.064   &    0.001   &   0.014   &   0.044  &   0.079  \\
				2M15082716$+$6710075  &   Al      &  ..  &    ..   &   ..   &   ..     &   ..  \\
				2M15082716$+$6710075  &   Si      &   0.026   &    0.039   &   0.081   &   0.014  &   0.095  \\
				2M15082716$+$6710075  &   Ce      &   0.044   &    0.089   &   0.025   &   0.100  &   0.144  \\
				2M15082716$+$6710075  &   Nd      &   ..      &    ..      &   ..      &   ..     &   ..     \\
				2M15082716$+$6710075  &   Na      &   ..      &    ..      &   ..      &   ..     &   ..     \\
				\hline
								2M15183589$+$0027100  &   Fe      &   0.030   &    0.009   &   0.009   &   0.095  &   0.101  \\
								2M15183589$+$0027100  &   C       &   0.095   &    0.122   &   0.002   &   0.233  &   0.280  \\
								2M15183589$+$0027100  &   N       &   0.038   &    0.112   &   0.012   &   0.145  &   0.188  \\
								2M15183589$+$0027100  &   O       &   0.148   &    0.016   &   0.001   &   0.111  &   0.186  \\
								2M15183589$+$0027100  &   Mg      &   0.062   &    0.048   &   0.017   &   0.047  &   0.093  \\
								2M15183589$+$0027100  &   Al      &   0.064   &    0.019   &   0.016   &   0.060  &   0.091  \\
								2M15183589$+$0027100  &   Si      &   0.060   &    0.064   &   0.034   &   0.067  &   0.116  \\
								2M15183589$+$0027100  &   Ce      &   0.018   &    0.060   &   0.009   &   0.061  &   0.088  \\
								2M15183589$+$0027100  &   Nd      &   ..      &    ..      &   ..      &   ..     &   ..     \\
								2M15183589$+$0027100  &   Na      &   ..      &    ..      &   ..      &   ..     &   ..     \\
								\hline
								2M15193208$+$0025293  &   Fe      &   0.037   &    0.021   &   0.005   &   0.085  &   0.095  \\
								2M15193208$+$0025293  &   C       &   0.029   &    0.075   &   0.012   &   0.192  &   0.209  \\
								2M15193208$+$0025293  &   N       &   0.070   &    0.049   &   0.014   &   0.053  &   0.103  \\
								2M15193208$+$0025293  &   O       &   0.108   &    0.015   &   0.002   &   0.107  &   0.153  \\
								2M15193208$+$0025293  &   Mg      &   0.086   &    0.108   &   0.009   &   0.031  &   0.142  \\
								2M15193208$+$0025293  &   Al      &   0.086   &    0.061   &   0.010   &   0.090  &   0.139  \\
								2M15193208$+$0025293  &   Si      &   0.078   &    0.083   &   0.115   &   0.091  &   0.187  \\
								2M15193208$+$0025293  &   Ce      &   0.069   &    0.032   &   0.080   &   0.065  &   0.128  \\
								2M15193208$+$0025293  &   Nd      &   ..      &    ..      &   ..      &   ..     &   ..     \\
								2M15193208$+$0025293  &   Na      &   0.021   &    0.014   &   ..      &   ..     &   0.025  \\
								\hline
								2M15195065$+$0221533  &   Fe      &   0.087   &    0.082   &   0.005   &   0.134  &   0.180  \\
								2M15195065$+$0221533  &   C       &   0.052   &    0.082   &   0.007   &   0.217  &   0.238  \\
								2M15195065$+$0221533  &   N       &   0.069   &    0.037   &   0.021   &   0.045  &   0.093  \\
								2M15195065$+$0221533  &   O       &   0.128   &    0.046   &   0.002   &   0.105  &   0.172  \\
								2M15195065$+$0221533  &   Mg      &   0.159   &    0.160   &   0.009   &   0.035  &   0.229  \\
								2M15195065$+$0221533  &   Al      &   0.107   &    0.097   &   0.010   &   0.070  &   0.162  \\
								2M15195065$+$0221533  &   Si      &   0.018   &    0.018   &   0.127   &   0.065  &   0.145  \\
								2M15195065$+$0221533  &   Ce      &   0.043   &    0.129   &   8.498   &   0.009  &   0.136  \\
								2M15195065$+$0221533  &   Nd      &   ..      &    ..      &   ..      &   ..     &   ..     \\
								2M15195065$+$0221533  &   Na      &   ..      &    ..      &   ..      &   ..     &   ..     \\
				\hline
			\end{tabular}  \label{error2}
		\end{center}
	\end{tiny}
\end{table}

\begin{table}
	\begin{tiny}
		\begin{center}
			\setlength{\tabcolsep}{1.0mm}  
			\caption{\textit{(continued)}}
			\begin{tabular}{clccccc}
				\hline
				\hline
				APOGEE$-$ID                    &   $X$ &   $\sigma_{[X/H],T_{\rm eff}}$   &    $\sigma_{[X/H],logg}$   &   $\sigma_{[X/H],\xi_{t}}$  &   $\sigma_{mean}$    &   $\sigma_{total}$  \\
				\hline
				\hline
				2M15535831$+$4333280  &   Fe      &   0.025   &    0.002   &   0.017   &   0.084  &   0.090  \\
				2M15535831$+$4333280  &   C       &   ..      &    ..      &   ..      &   ..     &   ..     \\
				2M15535831$+$4333280  &   N       &   0.037   &    0.058   &   0.034   &   0.109  &   0.134  \\
				2M15535831$+$4333280  &   O       &   0.069   &    0.086   &   0.047   &   0.225  &   0.255  \\
				2M15535831$+$4333280  &   Mg      &   0.026   &    0.017   &   0.026   &   0.074  &   0.085  \\
				2M15535831$+$4333280  &   Al      &   0.054   &    0.051   &   0.022   &   0.016  &   0.079  \\
				2M15535831$+$4333280  &   Si      &   0.011   &    0.032   &   0.010   &   0.091  &   0.098  \\
				2M15535831$+$4333280  &   Ce      &   0.013   &    0.233   &   0.080   &   0.015  &   0.247  \\
				2M15535831$+$4333280  &   Nd      &   0.011   &    0.066   &   0.041   &   ..     &   0.078  \\
				2M15535831$+$4333280  &   Na      &   ..      &    ..      &   ..      &   ..     &   ..     \\
				\hline
				2M16362792$+$3901180  &   Fe      &   0.039   &    0.016   &   0.018   &   0.074  &   0.087  \\
				2M16362792$+$3901180  &   C       &   0.022   &    0.093   &   0.024   &   0.053  &   0.112  \\
				2M16362792$+$3901180  &   N       &   0.084   &    0.080   &   0.060   &   0.070  &   0.149  \\
				2M16362792$+$3901180  &   O       &   0.105   &    0.042   &   0.049   &   0.178  &   0.216  \\
				2M16362792$+$3901180  &   Mg      &   0.052   &    0.074   &   0.032   &   0.121  &   0.155  \\
				2M16362792$+$3901180  &   Al      &   0.077   &    0.039   &   0.006   &   0.048  &   0.099  \\
				2M16362792$+$3901180  &   Si      &   0.047   &    0.048   &   0.011   &   0.121  &   0.139  \\
				2M16362792$+$3901180  &   Ce      &   0.042   &    0.160   &   0.045   &   0.051  &   0.179  \\
				2M16362792$+$3901180  &   Nd      &   ..      &    ..      &   ..      &   ..     &   ..     \\
				2M16362792$+$3901180  &   Na      &   ..      &    ..      &   ..      &   ..     &   ..     \\
				\hline
				2M16464310$+$4731033  &   Fe      &   0.036   &    0.031   &   0.010   &   0.132  &   0.141  \\
				2M16464310$+$4731033  &   C       &   0.069   &    0.089   &   0.119   &   0.046  &   0.170  \\
				2M16464310$+$4731033  &   N       &   0.118   &    0.066   &   0.146   &   0.112  &   0.229  \\
				2M16464310$+$4731033  &   O       &   0.110   &    0.041   &   0.020   &   0.071  &   0.139  \\
				2M16464310$+$4731033  &   Mg      &   0.071   &    0.033   &   0.016   &   0.098  &   0.127  \\
				2M16464310$+$4731033  &   Al      &   0.098   &    0.038   &   0.018   &   0.039  &   0.114  \\
				2M16464310$+$4731033  &   Si      &   0.021   &    0.065   &   0.028   &   0.104  &   0.128  \\
				2M16464310$+$4731033  &   Ce      &   0.046   &    0.154   &   0.029   &   0.131  &   0.210  \\
				2M16464310$+$4731033  &   Nd      &   0.058   &    0.060   &   0.100   &   ..     &   0.130  \\
				2M16464310$+$4731033  &   &   ..      &    ..      &   ..      &   ..     &   ..     \\
				\hline
				2M17294680$-$2644220  &   Fe      &   0.031   &    0.040   &   0.020   &   0.065  &   0.086  \\
				2M17294680$-$2644220  &   C       &   0.016   &    0.075   &   0.007   &   0.151  &   0.170  \\
				2M17294680$-$2644220  &   N       &   0.144   &    0.056   &   0.037   &   0.123  &   0.202  \\
				2M17294680$-$2644220  &   O       &   0.115   &    0.040   &   0.022   &   0.069  &   0.142  \\
				2M17294680$-$2644220  &   Mg      &   0.080   &    0.126   &   0.036   &   0.036  &   0.158  \\
				2M17294680$-$2644220  &   Al      &   0.062   &    0.062   &   0.019   &   0.054  &   0.105  \\
				2M17294680$-$2644220  &   Si      &   0.075   &    0.089   &   0.005   &   0.140  &   0.182  \\
				2M17294680$-$2644220  &   Ce      &   0.023   &    0.071   &   0.010   &   0.065  &   0.099  \\
				2M17294680$-$2644220  &   Nd      &   ..      &    ..      &   ..      &   ..     &   ..     \\
				2M17294680$-$2644220  &   Na      &   ..      &    ..      &   ..      &   ..     &   ..     \\
				\hline
				2M18022601$-$3106232  &   Fe      &   0.077   &    0.065   &   0.052   &   0.070  &   0.134  \\
				2M18022601$-$3106232  &   C       &   0.031   &    0.160   &   0.022   &   0.141  &   0.216  \\
				2M18022601$-$3106232  &   N       &   0.099   &    0.177   &   0.086   &   0.070  &   0.231  \\
				2M18022601$-$3106232  &   O       &   0.084   &    0.079   &   0.077   &   0.140  &   0.197  \\
				2M18022601$-$3106232  &   Mg      &   0.120   &    0.123   &   0.061   &   0.075  &   0.198  \\
				2M18022601$-$3106232  &   Al      &   0.112   &    0.091   &   0.050   &   0.116  &   0.192  \\
				2M18022601$-$3106232  &   Si      &   0.012   &    0.007   &   0.015   &   0.088  &   0.090  \\
				2M18022601$-$3106232  &   Ce      &   0.019   &    0.069   &   0.022   &   0.048  &   0.089  \\
				2M18022601$-$3106232  &   Nd      &   0.080   &    0.001   &   0.044   &   0.154  &   0.179  \\
				2M18022601$-$3106232  &   Na      &   0.027   &    ..      &   0.009   &   ..     &   0.028  \\
				\hline
				2M18110406$-$2602142  &   Fe      &   0.115   &    0.085   &   0.005   &   0.102  &   0.176  \\
				2M18110406$-$2602142  &   C       &   0.049   &    0.143   &   0.035   &   0.145  &   0.213  \\
				2M18110406$-$2602142  &   N       &   0.235   &    0.103   &   0.049   &   0.075  &   0.272  \\
				2M18110406$-$2602142  &   O       &   0.141   &    0.027   &   0.014   &   0.040  &   0.150  \\
				2M18110406$-$2602142  &   Mg      &   0.213   &    0.150   &   0.011   &   0.025  &   0.262  \\
				2M18110406$-$2602142  &   Al      &   0.225   &    0.163   &   0.011   &   0.040  &   0.281  \\
				2M18110406$-$2602142  &   Si      &   0.076   &    0.063   &   0.097   &   0.116  &   0.181  \\
				2M18110406$-$2602142  &   Ce      &   0.030   &    0.050   &   0.005   &   0.060  &   0.084  \\
				2M18110406$-$2602142  &   Nd      &   ..      &    ..      &   ..      &   ..     &   ..     \\
				2M18110406$-$2602142  &   Na      &   ..      &    ..      &   ..      &   ..     &   ..     \\
				\hline
				2M18200243$+$0156016  &   Fe      &   0.028   &    0.045   &   0.038   &   0.092  &   0.113  \\
				2M18200243$+$0156016  &   C       &   0.053   &    0.126   &   0.017   &   0.093  &   0.167  \\
				2M18200243$+$0156016  &   N       &   0.190   &    0.112   &   0.086   &   0.054  &   0.244  \\
				2M18200243$+$0156016  &   O       &   0.130   &    0.045   &   0.038   &   0.068  &   0.158  \\
				2M18200243$+$0156016  &   Mg      &   0.061   &    0.070   &   0.051   &   0.034  &   0.111  \\
				2M18200243$+$0156016  &   Al      &   0.101   &    0.007   &   0.077   &   0.098  &   0.161  \\
				2M18200243$+$0156016  &   Si      &   0.048   &    0.057   &   0.135   &   0.133  &   0.204  \\
				2M18200243$+$0156016  &   Ce      &   0.137   &    0.112   &   0.026   &   0.081  &   0.197  \\
				2M18200243$+$0156016  &   Nd      &   ..      &    ..      &   ..      &   ..     &   ..     \\
				2M18200243$+$0156016  &   Na      &   ..      &    ..      &   ..      &   ..     &   ..     \\
				\hline
								2M18461977$-$3021506  &   Fe      &   0.041   &    0.044   &   0.022   &   0.098  &   0.117  \\
								2M18461977$-$3021506  &   C       &   0.051   &    0.074   &   0.022   &   0.035  &   0.100  \\
								2M18461977$-$3021506  &   N       &   0.114   &    0.077   &   0.053   &   0.119  &   0.190  \\
								2M18461977$-$3021506  &   O       &   0.152   &    0.023   &   0.013   &   0.094  &   0.181  \\
								2M18461977$-$3021506  &   Mg      &   0.069   &    0.063   &   0.025   &   0.065  &   0.117  \\
								2M18461977$-$3021506  &   Al      &   0.087   &    0.047   &   0.026   &   0.094  &   0.139  \\
								2M18461977$-$3021506  &   Si      &   0.071   &    0.078   &   0.046   &   0.056  &   0.128  \\
								2M18461977$-$3021506  &   Ce      &   0.019   &    0.168   &   0.028   &   0.024  &   0.174  \\
								2M18461977$-$3021506  &   Nd      &   ..      &    ..      &   ..      &   ..     &   ..     \\
								2M18461977$-$3021506  &   Na      &   0.006   &    0.009   &   0.004   &   ..     &   0.012  \\
								\hline
								2M18472793$-$3033242  &   Fe      &   0.036   &    0.015   &   0.028   &   0.156  &   0.163  \\
								2M18472793$-$3033242  &   C       &   ..      &    ..      &   ..      &   ..     &   ..     \\
								2M18472793$-$3033242  &   N       &   0.126   &    0.036   &   0.026   &   0.114  &   0.176  \\
								2M18472793$-$3033242  &   O       &   0.124   &    0.021   &   0.022   &   0.175  &   0.217  \\
								2M18472793$-$3033242  &   Mg      &   0.078   &    0.067   &   0.027   &   0.073  &   0.130  \\
								2M18472793$-$3033242  &   Al      &   0.107   &    0.092   &   0.019   &   0.102  &   0.176  \\
								2M18472793$-$3033242  &   Si      &   0.070   &    0.079   &   0.032   &   0.111  &   0.157  \\
								2M18472793$-$3033242  &   Ce      &   0.018   &    0.075   &   0.012   &   0.018  &   0.081  \\
								2M18472793$-$3033242  &   Nd      &   ..      &    ..      &   ..      &   ..     &   ..     \\
								2M18472793$-$3033242  &   Na      &   ..      &    ..      &   ..      &   ..     &   ..     \\
								\hline
			\end{tabular}  \label{error4}
		\end{center}
	\end{tiny}
\end{table}

\begin{table}
	\begin{tiny}
		\begin{center}
			\setlength{\tabcolsep}{1.0mm}  
			\caption{\textit{(continued)}}
			\begin{tabular}{clccccc}
				\hline
				\hline
				APOGEE$-$ID                    &   $X$ &   $\sigma_{[X/H],T_{\rm eff}}$   &    $\sigma_{[X/H],logg}$   &   $\sigma_{[X/H],\xi_{t}}$  &   $\sigma_{mean}$    &   $\sigma_{total}$  \\
				\hline
				\hline
				2M18502108$-$2923442  &   Fe      &   0.037   &    0.004   &   0.005   &   0.073  &   0.082  \\
				2M18502108$-$2923442  &   C       &   0.027   &    0.069   &   0.001   &   0.184  &   0.199  \\
				2M18502108$-$2923442  &   N       &   0.077   &    0.036   &   0.002   &   0.044  &   0.096  \\
				2M18502108$-$2923442  &   O       &   0.127   &    0.018   &   0.001   &   0.083  &   0.153  \\
				2M18502108$-$2923442  &   Mg      &   0.084   &    0.081   &   0.011   &   0.054  &   0.129  \\
				2M18502108$-$2923442  &   Al      &   0.079   &    0.041   &   0.012   &   0.025  &   0.093  \\
				2M18502108$-$2923442  &   Si      &   0.013   &    0.016   &   0.018   &   0.050  &   0.057  \\
				2M18502108$-$2923442  &   Ce      &   0.032   &    0.058   &   0.013   &   0.090  &   0.112  \\
				2M18502108$-$2923442  &   Nd      &   ..      &    ..      &   ..      &   ..     &   ..     \\
				2M18502108$-$2923442  &   Na      &   ..      &    ..      &   ..      &   ..     &   ..     \\
				\hline
				2M19004420$+$4421082  &   Fe      &   0.046   &    0.014   &   0.018   &   0.093  &   0.106  \\
				2M19004420$+$4421082  &   C       &   0.074   &    0.108   &   0.044   &   0.074  &   0.157  \\
				2M19004420$+$4421082  &   N       &   0.098   &    0.107   &   0.023   &   0.030  &   0.151  \\
				2M19004420$+$4421082  &   O       &   0.115   &    0.060   &   0.012   &   0.050  &   0.140  \\
				2M19004420$+$4421082  &   Mg      &   0.064   &    0.063   &   0.022   &   0.009  &   0.093  \\
				2M19004420$+$4421082  &   Al      &   0.073   &    0.025   &   0.028   &   0.052  &   0.098  \\
				2M19004420$+$4421082  &   Si      &   0.024   &    0.018   &   0.010   &   0.062  &   0.070  \\
				2M19004420$+$4421082  &   Ce      &   0.037   &    0.091   &   0.023   &   0.158  &   0.188  \\
				2M19004420$+$4421082  &   Nd      &   0.075   &    0.001   &   0.001   &   0.021  &   0.078  \\
				2M19004420$+$4421082  &   Na      &   ..      &    ..      &   ..      &   ..     &   ..     \\
				\hline
				2M20124750$+$1818007  &   Fe      &   0.039   &    0.009   &   0.007   &   0.118  &   0.125  \\
				2M20124750$+$1818007  &   C       &   0.007   &    0.069   &   0.001   &   0.153  &   0.168  \\
				2M20124750$+$1818007  &   N       &   0.169   &    0.042   &   0.012   &   0.064  &   0.186  \\
				2M20124750$+$1818007  &   O       &   0.133   &    0.016   &   0.009   &   0.180  &   0.225  \\
				2M20124750$+$1818007  &   Mg      &   0.082   &    0.106   &   0.011   &   0.038  &   0.140  \\
				2M20124750$+$1818007  &   Al      &   0.054   &    0.016   &   0.020   &   0.052  &   0.080  \\
				2M20124750$+$1818007  &   Si      &   0.044   &    0.052   &   0.056   &   0.094  &   0.129  \\
				2M20124750$+$1818007  &   Ce      &   0.024   &    0.084   &   0.001   &   0.044  &   0.098  \\
				2M20124750$+$1818007  &   Nd      &   ..      &    ..      &   ..      &   ..     &   ..     \\
				2M20124750$+$1818007  &   Na      &   ..      &    ..      &   ..      &   ..     &   ..     \\
				\hline
				2M21181769$+$0946422  &   Fe      &   0.042   &    0.014   &   0.018   &   0.095  &   0.107  \\
				2M21181769$+$0946422  &   C       &   0.083   &    0.061   &   0.007   &   0.151  &   0.183  \\
				2M21181769$+$0946422  &   N       &   0.052   &    0.053   &   0.023   &   0.050  &   0.093  \\
				2M21181769$+$0946422  &   O       &   0.148   &    0.029   &   0.018   &   0.140  &   0.207  \\
				2M21181769$+$0946422  &   Mg      &   0.079   &    0.029   &   0.027   &   0.028  &   0.093  \\
				2M21181769$+$0946422  &   Al      &   0.055   &    0.018   &   0.028   &   0.090  &   0.111  \\
				2M21181769$+$0946422  &   Si      &   0.028   &    0.045   &   0.027   &   0.088  &   0.107  \\
				2M21181769$+$0946422  &   Ce      &   0.010   &    0.064   &   0.009   &   0.065  &   0.092  \\
				2M21181769$+$0946422  &   Nd      &   ..      &    ..      &   ..      &   ..     &   ..     \\
				2M21181769$+$0946422  &   Na      &   ..      &    ..      &   ..      &   ..     &   ..     \\
				\hline
				2M22015914$+$1543129  &   Fe      &   0.032   &    0.010   &   0.004   &   0.090  &   0.096  \\
				2M22015914$+$1543129  &   C       &   0.038   &    0.116   &   0.009   &   0.036  &   0.127  \\
				2M22015914$+$1543129  &   N       &   0.129   &    0.086   &   0.009   &   0.070  &   0.170  \\
				2M22015914$+$1543129  &   O       &   0.090   &    0.026   &   0.002   &   0.164  &   0.189  \\
				2M22015914$+$1543129  &   Mg      &   0.083   &    0.113   &   0.005   &   0.045  &   0.148  \\
				2M22015914$+$1543129  &   Al      &   0.054   &    0.046   &   0.011   &   0.140  &   0.157  \\
				2M22015914$+$1543129  &   Si      &   0.023   &    0.018   &   0.007   &   0.048  &   0.057  \\
				2M22015914$+$1543129  &   Ce      &   0.015   &    0.062   &   0.004   &   ..     &   0.064  \\
				2M22015914$+$1543129  &   Nd      &   ..      &    ..      &   ..      &   ..     &   ..     \\
				2M22015914$+$1543129  &   Na      &   ..      &    ..      &   ..      &   ..     &   ..     \\
				\hline
				\hline
			\end{tabular}  \label{error5}
		\end{center}
	\end{tiny}
\end{table}

\begin{table*}
	      \centering
	\setlength{\tabcolsep}{1.3mm}  
	\label{tableorbits1}
	\begin{tiny}
		\caption{Orbital elements obtained using a simple Monte Carlo approach for the 31 N-rich stars. The average value of the orbital elements (pericentric and apocentric radii, the eccentricity, the maximum distance the orbit reaches above/below the Galactic plane, as well as the maximum and minimum of the \textit{z}-component of the angular momentum in the inertial frame, $L_{z}$) was found for the half million realizations, with uncertainty ranges given by the 16th \textit{(subscript)} and 84th \textit{(superscript)} percentile values.}
		\begin{tabular}{cccccccccccccc}
			\hline
			\hline
			$\Omega_{bar}$ &  APOGEE$-$ID             &  RA          &  DEC           &    $RV\pm{}\Delta$                &  $d\pm{}\Delta$                &       $\mu_{\alpha}\pm{}\Delta$            &       $\mu_{\delta}\pm{}\Delta$           &    $\langle r_{min} \rangle$              &      $\langle r_{max} \rangle$              &       $\langle e \rangle$            &        $\langle Z_{max} \rangle$               &   $\langle L_{z, min} \rangle$                                &  $\langle L_{z, max} \rangle$      \\    
			km s$^{-1}$ kpc$^{-1}$ & & ($^{\circ}$)  &      ($^{\circ}$)  &  km s$^{-1}$ &  kpc & mas yr$^{-1}$  & mas yr$^{-1}$ & kpc & kpc &  & kpc & km s$^{-1}$ kpc$^{-1}$ & km s$^{-1}$ kpc$^{-1}$  \\               
			\hline
			\hline
			35 &  2M01121802$+$6219193  &  18.075099   &    62.322044   & $-$256.06$\pm$0.07   &  4.02$\pm$0.39    &      5.369$\pm$0.031   &   $-$1.376$\pm$0.035   &    0.05$^{ 0.10}_{ 0.02}$   &  11.73$^{ 12.04}_{ 11.51}$  &   0.99$^{0.99}_{0.98}$  &     0.37$^{ 0.40}_{ 0.35}$  &    $-$9.0$^{      0.0}_{  -15.0}$  &     11.0$^{    21.0}_{      7.0}$ \\
			40 &                        &              &                &                      &                   &                        &                        &    0.06$^{ 0.10}_{ 0.02}$   &  12.59$^{ 12.70}_{ 11.48}$  &   0.99$^{0.99}_{0.98}$  &     0.37$^{ 0.40}_{ 0.35}$  &   $-$15.0$^{     -5.0}_{  -20.0}$  &      9.0$^{    18.0}_{      3.0}$ \\
			45 &                        &              &                &                      &                   &                        &                        &    0.05$^{ 0.12}_{ 0.02}$   &  12.29$^{ 13.49}_{ 11.86}$  &   0.99$^{0.99}_{0.97}$  &     0.39$^{ 0.44}_{ 0.36}$  &   $-$12.0$^{    -10.0}_{  -24.0}$  &      4.0$^{    12.0}_{     -4.0}$ \\
			50 &                        &              &                &                      &                   &                        &                        &    0.09$^{ 0.18}_{ 0.04}$   &  13.81$^{ 14.02}_{ 13.50}$  &   0.98$^{0.99}_{0.97}$  &     0.37$^{ 0.42}_{ 0.36}$  &   $-$26.0$^{ -13.0}_{ -35.0}$  &        7.0$^{  21.0}_{  -1.2}$ \\
			35 &  2M02000451$-$0229333  &  30.018804   &  $-$2.492584   &     68.68$\pm$0.19   &  7.89$\pm$1.13    &      3.701$\pm$0.076   &   $-$4.492$\pm$0.048   &    1.28$^{ 2.27}_{ 0.36}$   &  14.45$^{ 15.38}_{ 13.69}$  &   0.83$^{0.95}_{0.71}$  &    11.62$^{15.23}_{ 8.79}$  &   $-$56.0$^{    -14.0}_{  -93.0}$  &  $-$45.0$^{    -2.0}_{    -84.0}$ \\
			40 &                        &              &                &                      &                   &                        &                        &    1.11$^{ 2.13}_{ 0.39}$   &  14.84$^{ 15.68}_{ 13.98}$  &   0.86$^{0.94}_{0.73}$  &    11.74$^{15.05}_{ 8.99}$  &   $-$56.0$^{    -17.0}_{  -96.0}$  &  $-$39.0$^{    -7.0}_{    -78.0}$ \\
			45 &                        &              &                &                      &                   &                        &                        &    1.09$^{ 2.00}_{ 0.33}$   &  14.25$^{ 15.13}_{ 13.76}$  &   0.85$^{0.95}_{0.74}$  &    11.64$^{14.68}_{ 8.83}$  &   $-$54.0$^{    -14.0}_{  -94.0}$  &  $-$37.0$^{     1.0}_{    -72.0}$ \\
			50 &                        &              &                &                      &                   &                        &                        &    1.15$^{ 2.20}_{ 0.38}$   &  14.25$^{ 17.16}_{ 13.64}$  &   0.85$^{0.95}_{0.72}$  &    11.30$^{15.38}_{ 9.17}$  &   $-$59.0$^{ -20.8}_{ -93.0}$  &    $-$41.0$^{  -5.8}_{ -81.0}$ \\
			35 &  2M06273068$-$0440140  &  96.877869   &  $-$4.67057    &     66.63$\pm$0.06   &  7.94$\pm$1.05    &   $-$0.260$\pm$0.046   &   $-$0.057$\pm$0.045   &   13.27$^{14.19}_{12.38}$   &  15.86$^{ 17.10}_{ 14.67}$  &   0.08$^{0.09}_{0.08}$  &     1.28$^{ 1.56}_{ 1.05}$  &  $-$339.0$^{   -317.0}_{ -362.0}$  & $-$339.0$^{  -317.0}_{   -361.0}$ \\
			40 &                        &              &                &                      &                   &                        &                        &   13.24$^{14.18}_{12.40}$   &  15.89$^{ 17.15}_{ 14.65}$  &   0.09$^{0.10}_{0.08}$  &     1.28$^{ 1.56}_{ 1.05}$  &  $-$339.0$^{   -317.0}_{ -362.0}$  & $-$339.0$^{  -317.0}_{   -362.0}$ \\
			45 &                        &              &                &                      &                   &                        &                        &   13.26$^{14.16}_{12.34}$   &  15.89$^{ 17.19}_{ 14.69}$  &   0.09$^{0.09}_{0.08}$  &     1.28$^{ 1.56}_{ 1.05}$  &  $-$339.0$^{   -317.0}_{ -362.0}$  & $-$339.0$^{  -317.0}_{   -362.0}$ \\
			50 &                        &              &                &                      &                   &                        &                        &   13.25$^{14.18}_{12.35}$   &  15.90$^{ 17.15}_{ 14.68}$  &   0.09$^{0.10}_{0.08}$  &     1.28$^{ 1.56}_{ 1.05}$  &  $-$339.0$^{-317.0}_{-362.0}$  &   $-$339.0$^{-317.0}_{-362.0}$ \\
			35 &  2M11062158$-$0712222  &  166.589953  &  $-$7.206176   &     43.25$\pm$0.13   &  3.86$\pm$0.36    &  $-$13.358$\pm$0.057   &   $-$6.870$\pm$0.041   &    2.88$^{ 2.95}_{ 2.78}$   &  10.56$^{ 11.24}_{ 10.04}$  &   0.56$^{0.60}_{0.54}$  &     8.78$^{ 10.0}_{ 7.54}$  &   $-$90.0$^{    -84.0}_{  -96.0}$  &  $-$85.0$^{   -80.0}_{    -88.0}$ \\
			40 &                        &              &                &                      &                   &                        &                        &    2.90$^{ 3.03}_{ 2.78}$   &  10.83$^{ 11.47}_{ 10.33}$  &   0.57$^{0.59}_{0.54}$  &     8.83$^{ 9.93}_{ 7.75}$  &   $-$92.0$^{    -83.0}_{ -102.0}$  &  $-$86.0$^{   -78.0}_{    -92.0}$ \\
			45 &                        &              &                &                      &                   &                        &                        &    2.87$^{ 2.95}_{ 2.77}$   &  10.48$^{ 11.72}_{ 10.02}$  &   0.57$^{0.61}_{0.54}$  &     8.73$^{ 10.3}_{ 7.41}$  &   $-$90.0$^{    -85.0}_{  -95.0}$  &  $-$83.0$^{   -78.0}_{    -87.0}$ \\
			50 &                        &              &                &                      &                   &                        &                        &    2.88$^{ 2.97}_{ 2.79}$   &  10.88$^{ 11.31}_{ 10.01}$  &   0.57$^{0.60}_{0.54}$  &     9.09$^{10.25}_{ 7.43}$  &   $-$92.0$^{ -84.0}_{ -98.0}$  &    $-$84.0$^{ -80.0}_{ -87.0}$ \\
			35 &  2M11514952$+$2015267  &  177.956372  &    20.257444   &     10.85$\pm$0.05   &  2.25$\pm$0.20    &   $-$3.249$\pm$0.083   &  $-$15.064$\pm$0.061   &    1.57$^{ 1.75}_{ 1.37}$   &   9.55$^{ 10.96}_{  9.23}$  &   0.71$^{0.76}_{0.67}$  &     2.33$^{ 2.64}_{ 2.05}$  &   $-$94.0$^{    -78.0}_{ -106.0}$  &  $-$60.0$^{   -53.0}_{    -68.0}$ \\
			40 &                        &              &                &                      &                   &                        &                        &    2.00$^{ 2.37}_{ 1.54}$   &   9.87$^{  9.96}_{  9.77}$  &   0.66$^{0.73}_{0.60}$  &     2.61$^{ 3.04}_{ 2.51}$  &   $-$88.0$^{    -75.0}_{ -101.0}$  &  $-$73.0$^{   -57.0}_{    -84.0}$ \\
			45 &                        &              &                &                      &                   &                        &                        &    1.90$^{ 2.22}_{ 1.51}$   &   9.43$^{  9.68}_{  9.33}$  &   0.66$^{0.72}_{0.61}$  &     3.58$^{ 3.71}_{ 3.37}$  &   $-$84.0$^{    -71.0}_{  -95.0}$  &  $-$65.0$^{   -54.0}_{    -75.0}$ \\
			50 &                        &              &                &                      &                   &                        &                        &    1.87$^{ 2.32}_{ 1.43}$   &  10.06$^{ 10.29}_{  9.73}$  &   0.68$^{0.75}_{0.61}$  &     2.74$^{ 3.14}_{ 2.41}$  &   $-$93.0$^{ -81.0}_{-101.0}$  &    $-$69.0$^{ -54.0}_{ -83.0}$ \\
			35 &  2M12010401$-$0058306  &  180.26672   &  $-$0.975172   &    145.41$\pm$1.52   &  1.69$\pm$0.10    &      3.765$\pm$0.077   &  $-$20.473$\pm$0.039   &    0.46$^{ 1.09}_{ 0.08}$   &   9.70$^{  9.78}_{  9.58}$  &   0.90$^{0.98}_{0.79}$  &     5.56$^{ 5.75}_{ 5.39}$  &   $-$49.0$^{    -42.0}_{  -57.0}$  &  $-$13.0$^{    20.0}_{    -36.2}$ \\
			40 &                        &              &                &                      &                   &                        &                        &    1.07$^{ 1.36}_{ 0.85}$   &   9.84$^{ 10.23}_{  9.71}$  &   0.80$^{0.84}_{0.75}$  &     5.68$^{ 5.85}_{ 5.59}$  &   $-$52.0$^{    -47.0}_{  -61.0}$  &  $-$39.0$^{   -32.0}_{    -47.0}$ \\
			45 &                        &              &                &                      &                   &                        &                        &    1.05$^{ 1.29}_{ 0.79}$   &  10.94$^{ 11.20}_{ 10.82}$  &   0.82$^{0.86}_{0.78}$  &     5.53$^{ 6.19}_{ 5.31}$  &   $-$67.0$^{    -58.0}_{  -74.0}$  &  $-$37.0$^{   -29.0}_{    -45.0}$ \\
			50 &                        &              &                &                      &                   &                        &                        &    0.79$^{ 0.97}_{ 0.53}$   &   9.97$^{ 10.27}_{  9.68}$  &   0.85$^{0.89}_{0.82}$  &     5.90$^{ 6.45}_{ 5.35}$  &   $-$52.0$^{ -46.0}_{ -63.0}$  &    $-$25.0$^{ -18.0}_{ -34.0}$ \\
			35 &  2M12042878$+$1949535  &  181.119949  &    19.831541   &     83.82$\pm$0.14   &  2.93$\pm$0.31    &   $-$6.604$\pm$0.072   &  $-$13.721$\pm$0.053   &    0.20$^{ 1.01}_{ 0.05}$   &   9.18$^{  9.55}_{  8.77}$  &   0.95$^{0.98}_{0.80}$  &     4.83$^{ 5.64}_{ 4.22}$  &   $-$32.0$^{     -7.8}_{  -56.0}$  &      8.5$^{    33.0}_{    -36.0}$ \\
			40 &                        &              &                &                      &                   &                        &                        &    0.30$^{ 0.99}_{ 0.05}$   &   9.05$^{  9.35}_{  8.84}$  &   0.93$^{0.98}_{0.79}$  &     5.02$^{ 5.39}_{ 4.28}$  &   $-$28.0$^{     -9.0}_{  -49.0}$  &   $-$8.0$^{     8.0}_{    -37.0}$ \\
			45 &                        &              &                &                      &                   &                        &                        &    0.35$^{ 0.99}_{ 0.11}$   &   9.71$^{  9.91}_{  9.47}$  &   0.93$^{0.97}_{0.81}$  &     4.18$^{ 4.83}_{ 3.79}$  &   $-$32.0$^{    -12.0}_{  -57.0}$  &  $-$12.0$^{     2.0}_{    -37.0}$ \\
			50 &                        &              &                &                      &                   &                        &                        &    0.41$^{ 1.07}_{ 0.05}$   &   9.33$^{ 10.01}_{  8.96}$  &   0.91$^{0.98}_{0.80}$  &     5.96$^{ 6.26}_{ 4.66}$  &   $-$34.0$^{  -7.0}_{ -60.0}$  &    $-$13.0$^{   7.0}_{ -38.0}$ \\
			35 &  2M12092246$+$0545111  &  182.34361   &    5.7531      &      1.44$\pm$0.43   & 10.80$\pm$1.93    &   $-$1.320$\pm$0.060   &   $-$1.700$\pm$0.031   &    3.58$^{ 4.22}_{ 2.97}$   &  14.47$^{ 16.38}_{ 12.66}$  &   0.60$^{0.69}_{0.50}$  &    10.94$^{12.99}_{ 8.82}$  &  $-$126.0$^{   -114.8}_{ -137.0}$  & $-$119.0$^{  -106.0}_{   -133.0}$ \\
			40 &                        &              &                &                      &                   &                        &                        &    3.41$^{ 4.27}_{ 2.86}$   &  13.86$^{ 15.59}_{ 13.49}$  &   0.60$^{0.68}_{0.52}$  &    10.61$^{12.50}_{ 9.41}$  &  $-$120.0$^{   -107.0}_{ -144.1}$  & $-$110.0$^{   -99.0}_{   -134.0}$ \\
			45 &                        &              &                &                      &                   &                        &                        &    3.51$^{ 4.12}_{ 2.93}$   &  14.03$^{ 15.50}_{ 12.39}$  &   0.60$^{0.67}_{0.49}$  &    10.75$^{12.49}_{ 8.68}$  &  $-$120.0$^{   -108.8}_{ -134.0}$  & $-$116.0$^{  -102.0}_{   -126.0}$ \\
			50 &                        &              &                &                      &                   &                        &                        &    3.51$^{ 4.16}_{ 2.92}$   &  14.19$^{ 15.79}_{ 12.36}$  &   0.60$^{0.69}_{0.49}$  &    10.69$^{12.56}_{ 8.81}$  &  $-$120.0$^{-110.0}_{-135.0}$  &   $-$116.0$^{-102.0}_{-132.0}$ \\
			35 &  2M12444724$-$0207405  &  191.196857  &  $-$2.127943   &    105.54$\pm$0.15   &  5.06$\pm$0.56    &   $-$1.419$\pm$0.051   &   $-$8.712$\pm$0.030   &    0.08$^{ 0.47}_{ 0.03}$   &   9.26$^{  9.98}_{  8.86}$  &   0.98$^{0.99}_{0.90}$  &     7.49$^{ 8.57}_{ 6.69}$  &    $-$4.0$^{      9.0}_{  -23.0}$  &     13.0$^{    30.0}_{      8.0}$ \\
			40 &                        &              &                &                      &                   &                        &                        &    0.22$^{ 0.57}_{ 0.06}$   &   9.26$^{  9.80}_{  8.98}$  &   0.95$^{0.98}_{0.88}$  &     7.28$^{ 7.58}_{ 6.75}$  &    $-$3.5$^{     13.0}_{  -26.0}$  &     15.0$^{    30.0}_{     -9.0}$ \\
			45 &                        &              &                &                      &                   &                        &                        &    0.18$^{ 0.60}_{ 0.04}$   &   9.23$^{  9.67}_{  8.88}$  &   0.96$^{0.99}_{0.87}$  &     7.77$^{ 9.02}_{ 6.54}$  &    $-$7.0$^{     11.0}_{  -23.2}$  &      5.0$^{    22.0}_{     -6.0}$ \\
			50 &                        &              &                &                      &                   &                        &                        &    0.27$^{ 0.53}_{ 0.05}$   &   9.60$^{ 10.35}_{  8.99}$  &   0.94$^{0.98}_{0.89}$  &     7.58$^{ 8.52}_{ 6.89}$  &    $-$8.0$^{  13.0}_{ -34.0}$  &        7.0$^{  28.0}_{ -10.0}$ \\
			35 &  2M13481355$-$0040484  &  207.056489  &  $-$0.680118   &    210.41$\pm$0.20   &  4.81$\pm$0.70    &   $-$6.905$\pm$0.056   &   $-$2.524$\pm$0.045   &    0.65$^{ 1.07}_{ 0.35}$   &   8.56$^{  9.19}_{  8.23}$  &   0.85$^{0.92}_{0.76}$  &     8.69$^{ 9.33}_{ 8.03}$  &   $-$41.0$^{    -27.0}_{  -55.0}$  &  $-$12.0$^{    -4.0}_{    -25.0}$ \\
			40 &                        &              &                &                      &                   &                        &                        &    1.20$^{ 1.81}_{ 0.72}$   &   8.59$^{  9.35}_{  8.01}$  &   0.75$^{0.85}_{0.62}$  &     8.36$^{ 9.47}_{ 7.83}$  &   $-$36.0$^{    -27.0}_{  -50.0}$  &  $-$30.0$^{   -17.0}_{    -44.2}$ \\
			45 &                        &              &                &                      &                   &                        &                        &    1.18$^{ 1.80}_{ 0.74}$   &   8.48$^{  9.17}_{  7.99}$  &   0.75$^{0.85}_{0.63}$  &     8.58$^{ 9.30}_{ 7.84}$  &   $-$36.0$^{    -27.0}_{  -49.0}$  &  $-$31.0$^{   -19.0}_{    -45.0}$ \\
			50 &                        &              &                &                      &                   &                        &                        &    1.15$^{ 1.79}_{ 0.66}$   &   8.73$^{  9.33}_{  7.87}$  &   0.76$^{0.86}_{0.62}$  &     8.43$^{ 9.22}_{ 7.56}$  &   $-$36.0$^{ -25.0}_{ -48.0}$  &    $-$29.0$^{ -15.0}_{ -43.0}$ \\
			35 &  2M13503160$+$4411389  &  207.6317    &    44.194141   &     20.54$\pm$0.16   &  2.00$\pm$0.05    &      2.079$\pm$0.017   &  $-$13.203$\pm$0.019   &    5.35$^{ 5.41}_{ 5.28}$   &  10.09$^{ 10.17}_{ 10.05}$  &   0.30$^{0.31}_{0.30}$  &     3.58$^{ 3.66}_{ 3.41}$  &  $-$166.0$^{   -164.0}_{ -167.0}$  & $-$159.0$^{  -157.0}_{   -160.0}$ \\
			40 &                        &              &                &                      &                   &                        &                        &    5.32$^{ 5.37}_{ 5.19}$   &  10.29$^{ 10.32}_{ 10.27}$  &   0.31$^{0.33}_{0.31}$  &     3.62$^{ 3.77}_{ 3.44}$  &  $-$166.0$^{   -165.0}_{ -168.0}$  & $-$156.0$^{  -154.0}_{   -157.0}$ \\
			45 &                        &              &                &                      &                   &                        &                        &    5.31$^{ 5.35}_{ 5.24}$   &  10.30$^{ 10.36}_{ 10.26}$  &   0.31$^{0.32}_{0.31}$  &     3.63$^{ 3.71}_{ 3.45}$  &  $-$165.0$^{   -163.0}_{ -166.0}$  & $-$159.0$^{  -157.0}_{   -160.0}$ \\
			50 &                        &              &                &                      &                   &                        &                        &    5.34$^{ 5.42}_{ 5.28}$   &  10.95$^{ 10.97}_{ 10.93}$  &   0.34$^{0.34}_{0.33}$  &     3.84$^{ 3.96}_{ 3.71}$  &  $-$169.0$^{-168.0}_{-171.0}$  &   $-$155.0$^{-153.0}_{-157.0}$ \\ 
						35 &  2M14082554$+$4711096  &  212.106431  &    47.186001   &  $-$62.76$\pm$0.00   &  0.49$\pm$0.01    & $-$151.479$\pm$0.033   &  $-$10.543$\pm$0.034   &    0.26$^{ 0.30}_{ 0.20}$   &  14.22$^{ 14.34}_{ 13.93}$  &   0.96$^{0.97}_{0.95}$  &     3.59$^{ 4.04}_{ 2.45}$  &       6.0$^{      8.0}_{    3.0}$  &     21.0$^{    24.0}_{     18.0}$ \\
						40 &                        &              &                &                      &                   &                        &                        &    0.08$^{ 0.13}_{ 0.04}$   &  16.45$^{ 16.75}_{ 15.80}$  &   0.98$^{0.99}_{0.98}$  &     3.59$^{ 4.06}_{ 2.34}$  &   $-$11.0$^{     -4.0}_{  -19.2}$  &     25.0$^{    30.0}_{     21.0}$ \\
						45 &                        &              &                &                      &                   &                        &                        &    0.24$^{ 0.33}_{ 0.15}$   &  13.97$^{ 14.32}_{ 13.75}$  &   0.96$^{0.97}_{0.95}$  &     3.83$^{ 4.21}_{ 3.37}$  &       9.0$^{     12.0}_{    5.0}$  &     43.0$^{    50.0}_{     34.0}$ \\
						50 &                        &              &                &                      &                   &                        &                        &    0.28$^{ 0.31}_{ 0.21}$   &  14.06$^{ 14.29}_{ 13.88}$  &   0.96$^{0.97}_{0.95}$  &     3.60$^{ 3.97}_{ 3.29}$  &      11.0$^{  13.0}_{   6.0}$  &       25.0$^{  27.0}_{  24.0}$ \\
						35 &  2M14582162$+$4117544  &  224.590099  &    41.298454   &  $-$73.89$\pm$0.00   &  3.57$\pm$0.33    &  $-$13.150$\pm$0.057   &   $-$6.297$\pm$0.064   &    0.13$^{ 0.39}_{ 0.05}$   &   9.05$^{  9.45}_{  8.69}$  &   0.97$^{0.98}_{0.92}$  &     5.17$^{ 6.58}_{ 3.78}$  &       0.0$^{      9.0}_{  -28.0}$  &     26.0$^{    40.0}_{     18.8}$ \\
						40 &                        &              &                &                      &                   &                        &                        &    0.08$^{ 0.32}_{ 0.03}$   &   9.03$^{  9.32}_{  8.74}$  &   0.98$^{0.99}_{0.93}$  &     4.87$^{ 6.11}_{ 3.66}$  &    $-$6.0$^{      9.0}_{  -20.0}$  &     15.0$^{    27.0}_{      9.0}$ \\
						45 &                        &              &                &                      &                   &                        &                        &    0.09$^{ 0.49}_{ 0.02}$   &   9.04$^{  9.38}_{  8.64}$  &   0.97$^{0.99}_{0.89}$  &     4.61$^{ 5.13}_{ 3.79}$  &    $-$7.0$^{     16.2}_{  -18.0}$  &      8.0$^{    28.0}_{     -2.0}$ \\
						50 &                        &              &                &                      &                   &                        &                        &    0.17$^{ 0.43}_{ 0.04}$   &   8.72$^{  9.20}_{  8.45}$  &   0.96$^{0.98}_{0.90}$  &     4.97$^{ 5.75}_{ 4.39}$  &       1.0$^{  13.0}_{ -19.0}$  &       17.0$^{  27.0}_{  -2.0}$ \\
						35 &  2M15082716$+$6710075  &  227.113188  &    67.168762   & $-$233.12$\pm$0.79   & 54.34$\pm$4.20    &   $-$0.203$\pm$0.089   &      0.112$\pm$0.085   &    4.77$^{ 7.54}_{ 2.14}$   &  58.43$^{ 82.86}_{ 35.40}$  &   0.82$^{0.93}_{0.70}$  &    58.14$^{81.48}_{34.06}$  &  $-$126.5$^{     27.2}_{ -195.0}$  & $-$126.0$^{    27.2}_{   -195.0}$ \\
						40 &                        &              &                &                      &                   &                        &                        &    4.76$^{ 7.46}_{ 2.18}$   &  58.43$^{ 82.86}_{ 35.40}$  &   0.82$^{0.93}_{0.70}$  &    58.12$^{81.30}_{34.10}$  &  $-$126.0$^{     27.2}_{ -195.0}$  & $-$126.0$^{    27.2}_{   -195.0}$ \\
						45 &                        &              &                &                      &                   &                        &                        &    4.76$^{ 7.43}_{ 2.14}$   &  58.43$^{ 82.86}_{ 35.40}$  &   0.82$^{0.93}_{0.70}$  &    58.13$^{81.30}_{34.10}$  &  $-$126.5$^{     27.2}_{ -195.0}$  & $-$126.0$^{    27.2}_{   -195.0}$ \\
						50 &                        &              &                &                      &                   &                        &                        &    4.71$^{ 7.43}_{ 2.15}$   &  58.43$^{ 82.86}_{ 35.39}$  &   0.82$^{0.93}_{0.70}$  &    58.13$^{81.32}_{34.10}$  &  $-$126.5$^{  27.2}_{-195.0}$  &   $-$126.0$^{  27.2}_{-195.0}$ \\		
						35 &  2M15183589$+$0027100  &  229.649545  &    0.452789    &  $-$56.27$\pm$0.34   & 18.16$\pm$2.53    &   $-$2.689$\pm$0.071   &   $-$2.806$\pm$0.063   &    1.05$^{ 2.21}_{ 0.23}$   &  13.59$^{ 16.27}_{ 11.51}$  &   0.85$^{0.95}_{0.75}$  &    13.58$^{15.07}_{11.38}$  &   $-$46.0$^{    -17.0}_{  -77.0}$  &  $-$33.0$^{    -3.0}_{    -69.0}$ \\
						40 &                        &              &                &                      &                   &                        &                        &    0.98$^{ 2.29}_{ 0.20}$   &  14.09$^{ 15.79}_{ 11.75}$  &   0.87$^{0.96}_{0.74}$  &    13.37$^{14.86}_{11.46}$  &   $-$43.5$^{    -20.0}_{  -76.2}$  &  $-$32.0$^{    -3.0}_{    -75.0}$ \\
						45 &                        &              &                &                      &                   &                        &                        &    0.78$^{ 2.26}_{ 0.17}$   &  13.26$^{ 15.91}_{ 12.16}$  &   0.88$^{0.97}_{0.74}$  &    13.11$^{14.82}_{11.59}$  &   $-$35.5$^{    -19.0}_{  -76.2}$  &  $-$24.0$^{    -1.0}_{    -74.2}$ \\
						50 &                        &              &                &                      &                   &                        &                        &    0.92$^{ 2.30}_{ 0.16}$   &  13.27$^{ 16.37}_{ 11.65}$  &   0.86$^{0.97}_{0.74}$  &    13.33$^{15.16}_{11.85}$  &   $-$37.0$^{ -17.0}_{ -78.2}$  &    $-$28.0$^{  -1.0}_{ -76.0}$ \\			
			\hline
			\hline
		\end{tabular}  
	\end{tiny}
\end{table*}

\begin{table*}
	\setlength{\tabcolsep}{1.3mm}  
	\label{tableorbits2}
	\begin{tiny}
		\caption{\textit{(continued)}}
		\begin{tabular}{cccccccccccccc}
			\hline
			\hline
			$\Omega_{bar}$ &  APOGEE$-$ID             &  RA          &  DEC           &    $RV\pm{}\Delta$                &  $d\pm{}\Delta$                &       $\mu_{\alpha}\pm{}\Delta$            &       $\mu_{\delta}\pm{}\Delta$           &    $\langle r_{min} \rangle$              &      $\langle r_{max} \rangle$              &       $\langle e \rangle$            &        $\langle Z_{max} \rangle$               &   $\langle L_{z, min} \rangle$                                &  $\langle L_{z, max} \rangle$      \\    
			km s$^{-1}$ kpc$^{-1}$ & & ($^{\circ}$)  &      ($^{\circ}$)  &  km s$^{-1}$ &  kpc & mas yr$^{-1}$  & mas yr$^{-1}$ & kpc & kpc &  & kpc & km s$^{-1}$ kpc$^{-1}$ & km s$^{-1}$ kpc$^{-1}$  \\               
			\hline
			\hline
			35 &  2M15193208$+$0025293  &  229.883704  &    0.424806    &     16.79$\pm$0.08   &  9.21$\pm$0.97    &   $-$1.542$\pm$0.126   &   $-$0.608$\pm$0.163   &    1.15$^{ 1.73}_{ 0.54}$   &   7.13$^{  8.11}_{  6.61}$  &   0.72$^{0.87}_{0.58}$  &     6.97$^{ 8.10}_{ 6.16}$  &   $-$42.0$^{    -31.0}_{  -54.0}$  &  $-$26.0$^{   -10.0}_{    -40.0}$ \\
			40 &                        &              &                &                      &                   &                        &                        &    1.09$^{ 1.52}_{ 0.68}$   &   7.15$^{  7.68}_{  6.84}$  &   0.73$^{0.83}_{0.64}$  &     7.18$^{ 7.84}_{ 6.45}$  &   $-$41.0$^{    -24.8}_{  -59.0}$  &  $-$23.5$^{   -14.0}_{    -34.0}$ \\
			45 &                        &              &                &                      &                   &                        &                        &    1.40$^{ 1.98}_{ 0.84}$   &   7.11$^{  7.62}_{  6.64}$  &   0.66$^{0.80}_{0.54}$  &     7.09$^{ 7.75}_{ 6.34}$  &   $-$38.0$^{    -23.0}_{  -53.0}$  &  $-$32.0$^{   -19.0}_{    -46.0}$ \\
			50 &                        &              &                &                      &                   &                        &                        &    1.36$^{ 2.00}_{ 0.83}$   &   6.98$^{  7.56}_{  6.56}$  &   0.67$^{0.80}_{0.53}$  &     7.04$^{ 7.89}_{ 6.25}$  &   $-$37.0$^{ -23.8}_{ -51.0}$  &    $-$32.0$^{ -19.0}_{ -46.0}$ \\
			35 &  2M15195065$+$0221533  &  229.961053  &    2.364827    &  $-$158.58$\pm$0.01  &  4.71$\pm$0.39    &   $-$2.500$\pm$0.072   &   $-$1.367$\pm$0.065   &    3.31$^{ 3.41}_{ 3.06}$   &   8.02$^{  8.15}_{  7.83}$  &   0.41$^{0.43}_{0.40}$  &     5.63$^{ 5.73}_{ 5.46}$  &  $-$113.0$^{   -109.0}_{ -117.0}$  &  $-$79.0$^{   -74.0}_{    -85.0}$ \\
			40 &                        &              &                &                      &                   &                        &                        &    2.35$^{ 3.50}_{ 2.19}$   &   7.26$^{  7.46}_{  7.05}$  &   0.50$^{0.52}_{0.36}$  &     5.40$^{ 5.55}_{ 5.22}$  &   $-$95.0$^{    -88.0}_{ -103.0}$  &  $-$53.0$^{   -50.0}_{    -88.0}$ \\
			45 &                        &              &                &                      &                   &                        &                        &    3.44$^{ 3.67}_{ 3.18}$   &   7.12$^{  7.41}_{  6.88}$  &   0.34$^{0.36}_{0.33}$  &     5.36$^{ 5.56}_{ 5.20}$  &   $-$93.0$^{    -86.0}_{ -100.0}$  &  $-$85.0$^{   -77.0}_{    -93.0}$ \\
			50 &                        &              &                &                      &                   &                        &                        &    3.35$^{ 3.55}_{ 3.23}$   &   8.00$^{  8.17}_{  7.33}$  &   0.40$^{0.42}_{0.35}$  &     5.48$^{ 5.77}_{ 5.18}$  &   $-$98.0$^{ -90.0}_{-101.0}$  &    $-$88.0$^{ -80.0}_{ -92.0}$ \\
			35 &  2M15535831$+$4333280  &  238.492995  &    43.557785   &  $-$182.35$\pm$0.34  &  5.69$\pm$0.76    &   $-$6.000$\pm$0.041   &   $-$0.476$\pm$0.051   &    0.12$^{ 0.60}_{ 0.04}$   &   9.35$^{  9.87}_{  8.78}$  &   0.97$^{0.99}_{0.88}$  &     8.39$^{ 8.87}_{ 7.68}$  &       0.0$^{     16.0}_{  -13.0}$  &     21.0$^{     41.0}_{     9.0}$ \\
			40 &                        &              &                &                      &                   &                        &                        &    0.19$^{ 0.55}_{ 0.05}$   &   9.39$^{  9.90}_{  8.73}$  &   0.95$^{0.98}_{0.89}$  &     8.93$^{ 9.60}_{ 7.96}$  &       0.0$^{     14.0}_{  -15.0}$  &     15.0$^{     31.0}_{     1.0}$ \\
			45 &                        &              &                &                      &                   &                        &                        &    0.27$^{ 0.76}_{ 0.05}$   &   9.25$^{  9.51}_{  8.96}$  &   0.94$^{0.98}_{0.84}$  &     8.56$^{ 9.33}_{ 7.68}$  &       0.0$^{     22.0}_{  -14.0}$  &     14.0$^{     32.0}_{    -3.0}$ \\
			50 &                        &              &                &                      &                   &                        &                        &    0.23$^{ 0.56}_{ 0.04}$   &   9.68$^{ 10.71}_{  9.07}$  &   0.95$^{0.99}_{0.89}$  &     8.46$^{ 9.07}_{ 7.57}$  &       0.0$^{  16.0}_{ -23.0}$  &       13.0$^{  34.0}_{  -1.0}$ \\
			35 &  2M16362792$+$3901180  &  249.11635   &    39.021675   &  $-$191.99$\pm$0.03  &  4.44$\pm$0.27    &   $-$8.336$\pm$0.023   &      1.191$\pm$0.026   &    0.33$^{ 0.49}_{ 0.13}$   &   9.16$^{  9.26}_{  9.07}$  &   0.92$^{0.97}_{0.89}$  &     5.04$^{ 5.32}_{ 4.90}$  &      11.0$^{     17.0}_{    3.0}$  &     30.0$^{     35.0}_{    26.0}$ \\
			40 &                        &              &                &                      &                   &                        &                        &    0.42$^{ 0.56}_{ 0.22}$   &   9.01$^{  9.13}_{  8.89}$  &   0.90$^{0.95}_{0.88}$  &     4.72$^{ 4.98}_{ 4.42}$  &      14.0$^{     20.2}_{    7.0}$  &     28.0$^{     34.0}_{    24.0}$ \\
			45 &                        &              &                &                      &                   &                        &                        &    0.28$^{ 0.45}_{ 0.13}$   &   9.50$^{  9.79}_{  9.06}$  &   0.94$^{0.97}_{0.90}$  &     5.03$^{ 5.69}_{ 4.48}$  &       9.0$^{     14.0}_{    1.0}$  &     25.0$^{     32.0}_{    18.0}$ \\
			50 &                        &              &                &                      &                   &                        &                        &    0.24$^{ 0.42}_{ 0.12}$   &   9.64$^{  9.89}_{  9.38}$  &   0.95$^{0.97}_{0.91}$  &     4.84$^{ 5.06}_{ 4.68}$  &       6.0$^{  15.0}_{   0.8}$  &       21.0$^{  30.0}_{  15.0}$ \\
			35 &  2M16464310$+$4731033  &  251.679603  &    47.517605   &  $-$139.80$\pm$0.37  & 20.02$\pm$3.61    &   $-$1.218$\pm$0.055   &   $-$0.654$\pm$0.101   &    1.01$^{ 1.38}_{ 0.71}$   &  19.82$^{ 24.38}_{ 16.89}$  &   0.90$^{0.93}_{0.86}$  &    15.78$^{17.71}_{14.47}$  &   $-$53.0$^{    -38.0}_{  -68.0}$  &  $-$39.0$^{    -27.0}_{   -54.0}$ \\
			40 &                        &              &                &                      &                   &                        &                        &    1.04$^{ 1.39}_{ 0.72}$   &  21.09$^{ 24.31}_{ 16.73}$  &   0.90$^{0.93}_{0.86}$  &    16.08$^{17.66}_{14.35}$  &   $-$53.0$^{    -39.0}_{  -69.0}$  &  $-$40.0$^{    -26.0}_{   -56.2}$ \\
			45 &                        &              &                &                      &                   &                        &                        &    1.05$^{ 1.40}_{ 0.70}$   &  20.58$^{ 23.90}_{ 17.74}$  &   0.90$^{0.93}_{0.86}$  &    15.93$^{17.82}_{14.62}$  &   $-$53.0$^{    -38.0}_{  -68.0}$  &  $-$40.0$^{    -25.0}_{   -56.0}$ \\
			50 &                        &              &                &                      &                   &                        &                        &    1.05$^{ 1.39}_{ 0.68}$   &  21.14$^{ 23.30}_{ 16.68}$  &   0.90$^{0.93}_{0.86}$  &    16.25$^{17.95}_{14.40}$  &   $-$53.0$^{ -38.8}_{ -67.0}$  &    $-$40.0$^{ -25.0}_{ -55.0}$ \\
			35 &  2M17294680$-$2644220  &  262.445037  &  $-$26.739462  &  $-$217.76$\pm$0.25  &  3.65$\pm$0.49    &   $-$3.679$\pm$0.174   &   $-$6.375$\pm$0.114   &    1.10$^{ 1.58}_{ 0.83}$   &   7.31$^{  8.20}_{  6.33}$  &   0.73$^{0.77}_{0.67}$  &     0.96$^{ 2.20}_{ 0.50}$  &   $-$71.0$^{    -57.0}_{  -84.0}$  &  $-$42.0$^{    -31.0}_{   -61.0}$ \\
			40 &                        &              &                &                      &                   &                        &                        &    1.16$^{ 1.56}_{ 0.75}$   &   7.29$^{  9.55}_{  6.33}$  &   0.73$^{0.78}_{0.70}$  &     0.58$^{ 2.13}_{ 0.48}$  &   $-$70.0$^{    -55.0}_{ -101.0}$  &  $-$44.0$^{    -27.0}_{   -60.0}$ \\
			45 &                        &              &                &                      &                   &                        &                        &    1.11$^{ 1.24}_{ 0.82}$   &   8.40$^{  8.71}_{  6.36}$  &   0.76$^{0.78}_{0.74}$  &     0.54$^{ 1.77}_{ 0.46}$  &   $-$79.0$^{    -54.0}_{  -88.0}$  &  $-$43.0$^{    -31.0}_{   -48.0}$ \\
			50 &                        &              &                &                      &                   &                        &                        &    0.88$^{ 1.40}_{ 0.74}$   &   7.72$^{  8.16}_{  7.59}$  &   0.79$^{0.81}_{0.70}$  &     0.50$^{ 0.60}_{ 0.45}$  &   $-$69.0$^{ -66.0}_{ -77.2}$  &    $-$34.0$^{ -28.0}_{ -53.0}$ \\
					35 &  2M18022601$-$3106232  &  270.608394  &  $-$31.106453  &  $-$170.48$\pm$0.26  &  7.23$\pm$1.15    &      2.127$\pm$0.126   &   $-$5.193$\pm$0.104   &    0.36$^{ 0.86}_{ 0.04}$   &   2.82$^{  3.97}_{  2.16}$  &   0.75$^{0.96}_{0.63}$  &     1.25$^{ 1.65}_{ 1.06}$  &   $-$43.0$^{    -21.0}_{  -57.0}$  &   $-$8.0$^{      6.0}_{   -25.0}$ \\
					40 &                        &              &                &                      &                   &                        &                        &    0.38$^{ 0.78}_{ 0.04}$   &   2.77$^{  4.13}_{  2.35}$  &   0.76$^{0.96}_{0.66}$  &     1.28$^{ 1.65}_{ 1.11}$  &   $-$39.0$^{    -20.0}_{  -59.0}$  &   $-$9.0$^{      6.0}_{   -22.0}$ \\
					45 &                        &              &                &                      &                   &                        &                        &    0.41$^{ 0.69}_{ 0.04}$   &   2.75$^{  4.06}_{  2.40}$  &   0.76$^{0.96}_{0.67}$  &     1.29$^{ 1.65}_{ 1.17}$  &   $-$36.0$^{    -20.0}_{  -65.0}$  &   $-$9.5$^{      7.0}_{   -19.0}$ \\
					50 &                        &              &                &                      &                   &                        &                        &    0.38$^{ 0.66}_{ 0.04}$   &   2.88$^{  3.92}_{  2.34}$  &   0.76$^{0.96}_{0.69}$  &     1.33$^{ 1.67}_{ 1.19}$  &   $-$35.0$^{ -20.0}_{ -64.0}$  &     $-$8.0$^{   8.0}_{ -18.0}$ \\
					35 &  2M18110406$-$2602142  &  272.766955  &  $-$26.037289  &   $-$24.96$\pm$0.01  &  5.70$\pm$0.54    &   $-$1.181$\pm$0.100   &   $-$5.176$\pm$0.079   &    0.48$^{ 0.65}_{ 0.42}$   &   2.77$^{  3.26}_{  2.21}$  &   0.68$^{0.71}_{0.65}$  &     0.42$^{ 0.45}_{ 0.39}$  &   $-$47.0$^{    -28.0}_{  -57.0}$  &  $-$13.0$^{    -10.0}_{   -19.0}$ \\
					40 &                        &              &                &                      &                   &                        &                        &    0.48$^{ 0.65}_{ 0.41}$   &   2.71$^{  3.27}_{  2.23}$  &   0.68$^{0.70}_{0.65}$  &     0.43$^{ 0.49}_{ 0.40}$  &   $-$48.0$^{    -33.0}_{  -55.0}$  &  $-$13.0$^{    -10.0}_{   -18.0}$ \\
					45 &                        &              &                &                      &                   &                        &                        &    0.44$^{ 0.60}_{ 0.29}$   &   2.73$^{  3.36}_{  2.24}$  &   0.68$^{0.81}_{0.66}$  &     0.46$^{ 0.61}_{ 0.41}$  &   $-$46.0$^{    -36.0}_{  -56.0}$  &  $-$11.0$^{     -6.9}_{   -17.0}$ \\
					50 &                        &              &                &                      &                   &                        &                        &    0.44$^{ 0.52}_{ 0.26}$   &   2.74$^{  3.37}_{  2.25}$  &   0.73$^{0.78}_{0.70}$  &     0.46$^{ 0.65}_{ 0.42}$  &   $-$47.0$^{ -37.0}_{ -53.1}$  &    $-$11.0$^{  -6.0}_{ -14.1}$ \\
					35 &  2M18200243$+$0156016  &  275.010145  &    1.933787    &     73.81$\pm$0.16   & 11.20$\pm$1.45    &   $-$2.277$\pm$0.084   &   $-$9.631$\pm$0.075   &    5.45$^{ 6.74}_{ 4.40}$   &  21.52$^{ 45.92}_{ 12.08}$  &   0.59$^{0.74}_{0.48}$  &     6.44$^{ 15.0}_{ 3.47}$  &  $-$215.0$^{   -158.0}_{ -298.0}$  & $-$212.0$^{   -152.0}_{  -294.2}$ \\
					40 &                        &              &                &                      &                   &                        &                        &    5.42$^{ 6.73}_{ 4.46}$   &  21.58$^{ 45.81}_{ 12.39}$  &   0.60$^{0.74}_{0.49}$  &     6.47$^{ 14.9}_{ 3.43}$  &  $-$216.0$^{   -161.8}_{ -297.2}$  & $-$209.0$^{   -155.0}_{  -295.2}$ \\
					45 &                        &              &                &                      &                   &                        &                        &    5.42$^{ 6.68}_{ 4.37}$   &  21.20$^{ 45.36}_{ 12.96}$  &   0.59$^{0.74}_{0.49}$  &     6.32$^{ 14.8}_{ 3.48}$  &  $-$213.5$^{   -165.0}_{ -295.3}$  & $-$210.0$^{   -149.0}_{  -294.2}$ \\
					50 &                        &              &                &                      &                   &                        &                        &    5.41$^{ 6.65}_{ 4.28}$   &  22.14$^{ 44.89}_{ 12.35}$  &   0.60$^{0.74}_{0.49}$  &     6.62$^{14.85}_{ 3.26}$  &  $-$218.0$^{-159.0}_{-295.2}$  &   $-$210.0$^{-144.0}_{-293.3}$ \\
					35 &  2M18461977$-$3021506  &  281.582414  &  $-$30.364075  &  $-$224.99$\pm$0.00  &  2.97$\pm$0.35    &   $-$8.525$\pm$0.077   &  $-$17.357$\pm$0.066   &    0.10$^{ 0.24}_{ 0.04}$   &   9.28$^{ 10.19}_{  8.29}$  &   0.97$^{0.98}_{0.94}$  &     3.51$^{ 3.70}_{ 3.07}$  &   $-$18.0$^{      8.0}_{  -38.0}$  &     41.0$^{     52.0}_{     36.0}$ \\
					40 &                        &              &                &                      &                   &                        &                        &    0.21$^{ 0.67}_{ 0.04}$   &   8.24$^{  8.94}_{  7.65}$  &   0.94$^{0.98}_{0.83}$  &     3.25$^{ 3.44}_{ 2.86}$  &       7.0$^{     25.1}_{  -14.0}$  &     30.0$^{     44.0}_{     15.0}$ \\
					45 &                        &              &                &                      &                   &                        &                        &    0.37$^{ 0.85}_{ 0.06}$   &   7.93$^{  8.37}_{  7.37}$  &   0.91$^{0.98}_{0.79}$  &     3.11$^{ 3.94}_{ 2.77}$  &      14.0$^{     31.0}_{   -4.0}$  &     28.0$^{     43.0}_{     12.0}$ \\
					50 &                        &              &                &                      &                   &                        &                        &    0.39$^{ 0.68}_{ 0.06}$   &   8.03$^{  8.51}_{  7.67}$  &   0.90$^{0.98}_{0.83}$  &     3.20$^{ 3.61}_{ 3.00}$  &      14.0$^{  25.0}_{  -3.2}$  &       27.0$^{  41.0}_{  10.0}$ \\
								35 &  2M18472793$-$3033242  &  281.866403  &  $-$30.556734  &     87.59$\pm$0.05   &  8.69$\pm$1.15    &      0.421$\pm$0.060   &   $-$6.731$\pm$0.052   &    0.04$^{ 0.28}_{ 0.02}$   &   3.19$^{  3.96}_{  2.52}$  &   0.96$^{0.98}_{0.86}$  &     2.45$^{ 3.07}_{ 2.15}$  &   $-$16.0$^{    -11.0}_{  -26.0}$  &     15.0$^{     23.0}_{     -4.0}$ \\
								40 &                        &              &                &                      &                   &                        &                        &    0.05$^{ 0.24}_{ 0.02}$   &   2.95$^{  4.56}_{  2.18}$  &   0.95$^{0.98}_{0.89}$  &     2.50$^{ 3.07}_{ 2.16}$  &   $-$17.0$^{    -12.0}_{  -34.0}$  &      3.0$^{     14.0}_{     -2.0}$ \\
								45 &                        &              &                &                      &                   &                        &                        &    0.07$^{ 0.26}_{ 0.02}$   &   3.26$^{  4.15}_{  2.43}$  &   0.94$^{0.98}_{0.87}$  &     2.52$^{ 3.05}_{ 2.10}$  &   $-$23.0$^{    -14.0}_{  -37.0}$  &      2.0$^{     11.0}_{     -4.0}$ \\
								50 &                        &              &                &                      &                   &                        &                        &    0.19$^{ 0.33}_{ 0.04}$   &   3.54$^{  3.90}_{  2.12}$  &   0.88$^{0.97}_{0.77}$  &     2.48$^{ 3.09}_{ 2.07}$  &   $-$24.0$^{ -15.0}_{ -33.0}$  &     $-$2.0$^{   6.0}_{  -5.0}$ \\
								35 &  2M18502108$-$2923442  &  282.587861  &  $-$29.395628  &  $-$31.22$\pm$0.00   &  1.90$\pm$0.11    &  $-$16.159$\pm$0.072   &  $-$15.883$\pm$0.062   &    0.68$^{ 0.98}_{ 0.37}$   &   6.75$^{  7.05}_{  6.41}$  &   0.82$^{0.88}_{0.74}$  &     3.16$^{ 3.28}_{ 2.86}$  &   $-$61.0$^{    -39.0}_{  -68.0}$  &  $-$22.0$^{    -11.0}_{    -32.0}$ \\
								40 &                        &              &                &                      &                   &                        &                        &    0.77$^{ 0.90}_{ 0.57}$   &   6.63$^{  6.82}_{  6.47}$  &   0.79$^{0.83}_{0.76}$  &     3.06$^{ 3.70}_{ 2.72}$  &   $-$51.0$^{    -43.0}_{  -56.0}$  &  $-$27.0$^{    -19.0}_{    -32.0}$ \\
								45 &                        &              &                &                      &                   &                        &                        &    0.72$^{ 0.86}_{ 0.57}$   &   6.65$^{  7.92}_{  6.48}$  &   0.81$^{0.83}_{0.79}$  &     3.15$^{ 3.32}_{ 2.69}$  &   $-$52.0$^{    -41.0}_{  -74.2}$  &  $-$25.0$^{    -18.0}_{    -29.0}$ \\
								50 &                        &              &                &                      &                   &                        &                        &    0.74$^{ 1.28}_{ 0.44}$   &   6.49$^{  7.39}_{  6.38}$  &   0.79$^{0.87}_{0.70}$  &     2.98$^{ 3.56}_{ 2.78}$  &   $-$47.0$^{ -37.0}_{ -69.0}$  &    $-$20.0$^{ -10.0}_{ -40.0}$ \\
			35 &  2M19004420$+$4421082  &  285.184205  &    44.352283   & $-$219.58$\pm$0.16   &  4.04$\pm$0.33    &   $-$5.337$\pm$0.036   &  $-$11.575$\pm$0.035   &    1.13$^{ 1.28}_{ 0.89}$   &  11.44$^{ 12.14}_{ 10.71}$  &   0.81$^{0.84}_{0.80}$  &     2.80$^{ 3.45}_{ 1.85}$  &   $-$64.0$^{    -59.0}_{  -71.0}$  &  $-$48.0$^{    -38.0}_{    -57.0}$ \\
			40 &                        &              &                &                      &                   &                        &                        &    1.00$^{ 1.16}_{ 0.87}$   &  11.38$^{ 12.85}_{ 10.71}$  &   0.83$^{0.85}_{0.82}$  &     2.84$^{ 3.79}_{ 2.07}$  &   $-$64.0$^{    -58.0}_{  -82.0}$  &  $-$42.0$^{    -37.0}_{    -51.0}$ \\
			45 &                        &              &                &                      &                   &                        &                        &    1.21$^{ 1.30}_{ 0.83}$   &  12.19$^{ 12.49}_{ 11.44}$  &   0.81$^{0.86}_{0.80}$  &     2.15$^{ 2.91}_{ 1.86}$  &   $-$73.0$^{    -67.0}_{  -77.0}$  &  $-$54.0$^{    -35.0}_{    -57.0}$ \\
			50 &                        &              &                &                      &                   &                        &                        &    1.06$^{ 1.21}_{ 0.91}$   &  11.62$^{ 12.37}_{ 10.92}$  &   0.83$^{0.85}_{0.82}$  &     2.66$^{ 3.31}_{ 1.76}$  &   $-$66.0$^{ -59.0}_{ -74.0}$  &    $-$47.0$^{ -39.7}_{ -53.0}$ \\
			35 &  2M20124750$+$1818007  &  303.197945  &    18.300213   &  $-$58.21$\pm$0.22   &  7.63$\pm$0.59    &   $-$2.890$\pm$0.036   &   $-$5.879$\pm$0.030   &    4.70$^{ 4.85}_{ 4.30}$   &   8.79$^{  9.15}_{  8.54}$  &   0.30$^{0.33}_{0.28}$  &     1.34$^{ 1.42}_{ 1.23}$  &  $-$169.0$^{   -167.0}_{ -173.0}$  & $-$134.0$^{   -127.0}_{   -137.0}$ \\
			40 &                        &              &                &                      &                   &                        &                        &    3.60$^{ 5.63}_{ 3.56}$   &   8.30$^{  8.55}_{  8.21}$  &   0.39$^{0.39}_{0.20}$  &     1.22$^{ 1.42}_{ 1.17}$  &  $-$154.0$^{   -147.0}_{ -167.0}$  & $-$105.0$^{   -103.0}_{   -158.0}$ \\
			45 &                        &              &                &                      &                   &                        &                        &    4.99$^{ 5.71}_{ 4.26}$   &   8.20$^{  8.66}_{  7.98}$  &   0.24$^{0.30}_{0.20}$  &     1.27$^{ 1.43}_{ 1.13}$  &  $-$152.0$^{   -142.0}_{ -170.0}$  & $-$142.0$^{   -127.0}_{   -159.0}$ \\
			50 &                        &              &                &                      &                   &                        &                        &    4.94$^{ 5.70}_{ 4.41}$   &   8.48$^{  8.64}_{  7.84}$  &   0.26$^{0.29}_{0.20}$  &     1.27$^{ 1.43}_{ 1.10}$  &  $-$155.0$^{-139.0}_{-167.0}$  &   $-$138.0$^{-129.0}_{-160.0}$ \\
			35 &  2M21181769$+$0946422  &  319.573711  &    9.7784      & $-$115.86$\pm$0.02   &  5.04$\pm$0.65    &      1.345$\pm$0.057   &   $-$5.096$\pm$0.058   &    1.21$^{ 1.34}_{ 1.10}$   &   8.10$^{  8.22}_{  7.78}$  &   0.73$^{0.76}_{0.70}$  &     2.51$^{ 3.33}_{ 2.07}$  &   $-$67.0$^{    -61.0}_{  -71.0}$  &  $-$46.0$^{    -42.0}_{    -51.0}$ \\
			40 &                        &              &                &                      &                   &                        &                        &    1.23$^{ 1.42}_{ 1.05}$   &   8.31$^{  9.30}_{  8.01}$  &   0.74$^{0.78}_{0.70}$  &     2.59$^{ 3.02}_{ 2.10}$  &   $-$71.0$^{    -63.0}_{  -91.0}$  &  $-$45.0$^{    -39.0}_{    -49.0}$ \\
			45 &                        &              &                &                      &                   &                        &                        &    1.21$^{ 1.33}_{ 0.81}$   &   8.60$^{  8.81}_{  8.48}$  &   0.75$^{0.82}_{0.73}$  &     2.64$^{ 2.91}_{ 2.41}$  &   $-$68.0$^{    -63.0}_{  -77.0}$  &  $-$44.0$^{    -30.0}_{    -49.0}$ \\
			50 &                        &              &                &                      &                   &                        &                        &    1.38$^{ 1.56}_{ 1.20}$   &   8.37$^{  8.64}_{  8.15}$  &   0.71$^{0.75}_{0.67}$  &     2.99$^{ 3.31}_{ 2.84}$  &   $-$64.0$^{ -59.0}_{ -70.0}$  &    $-$47.0$^{ -44.0}_{ -50.0}$ \\
			35 &  2M22015914$+$1543129  &  330.49643   &    15.720271   &  $-$58.32$\pm$0.25   &  3.29$\pm$0.26    &      1.341$\pm$0.050   &   $-$0.273$\pm$0.047   &    3.54$^{ 3.71}_{ 3.45}$   &   9.71$^{  9.78}_{  9.62}$  &   0.46$^{0.47}_{0.44}$  &     2.16$^{ 2.35}_{ 1.92}$  &  $-$158.0$^{   -154.0}_{ -161.0}$  & $-$110.0$^{   -106.0}_{   -115.0}$ \\
			40 &                        &              &                &                      &                   &                        &                        &    4.47$^{ 4.56}_{ 4.37}$   &   9.40$^{  9.57}_{  9.27}$  &   0.35$^{0.37}_{0.34}$  &     2.07$^{ 2.32}_{ 1.85}$  &  $-$149.0$^{   -147.0}_{ -152.0}$  & $-$140.0$^{   -138.0}_{   -143.0}$ \\
			45 &                        &              &                &                      &                   &                        &                        &    4.45$^{ 4.57}_{ 4.32}$   &   9.22$^{  9.85}_{  9.13}$  &   0.34$^{0.38}_{0.33}$  &     2.06$^{ 2.35}_{ 1.89}$  &  $-$148.0$^{   -146.0}_{ -151.0}$  & $-$137.0$^{   -134.0}_{   -141.0}$ \\
			50 &                        &              &                &                      &                   &                        &                        &    4.50$^{ 4.59}_{ 4.41}$   &   9.57$^{  9.76}_{  9.39}$  &   0.35$^{0.37}_{0.34}$  &     2.13$^{ 2.36}_{ 1.91}$  &  $-$149.0$^{-148.0}_{-152.0}$  &   $-$141.0$^{-139.0}_{-143.0}$ \\
			\hline
			\hline
		\end{tabular}  
	\end{tiny}
\end{table*}

\end{document}